\documentclass[12pt,onecolumn,journal,compsoc]{IEEEtran}
\usepackage[utf8]{inputenc}
\usepackage{comment}
\usepackage{subfigure}
\usepackage{balance}
\usepackage{CJKutf8}
\usepackage{ifpdf}
\usepackage{amsmath} 
\usepackage[linesnumbered,ruled]{algorithm2e} 
\usepackage{algorithmic}
\usepackage{array}
\usepackage{url}
\usepackage{graphicx}
\usepackage{hhline}
\usepackage{lineno}
\usepackage{enumitem}
\usepackage{amssymb}
\usepackage{tabularx}
\usepackage{tipa}
\usepackage[nolist]{acronym}
\usepackage{multirow}
\usepackage{rotating}
\usepackage{adjustbox,lipsum}
\usepackage{url}
\usepackage{todonotes}
\usepackage{algorithmic}
\usepackage{hyperref}
\usepackage{titlesec}
\usepackage{xcolor}
\usepackage{soul}

\usepackage{rotating}
\usepackage{graphicx}
\usepackage{wrapfig} 
\usepackage{fancyhdr}

\graphicspath{ {./Figures/} } 

\titleclass{\subsubsubsection}{straight}[\subsection]

\newcounter{subsubsubsection}[subsubsection]
\renewcommand\thesubsubsubsection{\thesubsubsection.\arabic{subsubsubsection}}

\titleformat{\subsubsubsection}
  {\normalfont\normalsize\itshape}{\thesubsubsubsection}{1em}{} 
\titlespacing*{\subsubsubsection}
{0pt}{3.25ex plus 1ex minus .2ex}{1.5ex plus .2ex}

\makeatletter
\renewcommand\paragraph{\@startsection{paragraph}{5}{\z@}%
  {3.25ex \@plus1ex \@minus.2ex}%
  {-1em}%
  {\normalfont\normalsize\itshape}} 
\renewcommand\subparagraph{\@startsection{subparagraph}{6}{\parindent}%
  {3.25ex \@plus1ex \@minus .2ex}%
  {-1em}%
  {\normalfont\normalsize\itshape}} 
\def\toclevel@subsubsubsection{4}
\def\toclevel@paragraph{5}
\def\toclevel@paragraph{6}
\def\l@subsubsubsection{\@dottedtocline{4}{7em}{4em}}
\def\l@paragraph{\@dottedtocline{5}{10em}{5em}}
\def\l@subparagraph{\@dottedtocline{6}{14em}{6em}}
\makeatother

\titleformat{\section}
  {\normalfont\Large\bfseries}{\thesection}{1em}{}
 
\titleformat{\subsection}
  {\normalfont\large\scshape}{\thesubsection}{1em}{}

\begin{document}

\title{Hybrid SDN Evolution: A Comprehensive Survey of the State-of-the-Art}
\author{
	Sajad Khorsandroo$^\alpha$,
	Adrián Gallego Sánchez,
$^\beta$
	Ali Saman Tosun$^\gamma$,
	José Manuel Arco Rodríguez$^\delta$,
	Roberto Doriguzzi-Corin$^\epsilon$\\
	
	\small{$^\alpha$Department of Computer Science, North Carolina A\&T State University, USA}\\
	\small{$^\beta$NETCOM Research Group, University Carlos III of Madrid, Spain}
\\
	\small{$^\gamma$Department of Computer Science, The University of Texas at San Antonio, USA}\\
	\small{$^\delta$Automatica Department, The University of Alcalá,  Spain}\\
	\small{$^\epsilon$Cybersecurity, Fondazione Bruno Kessler - Italy}\\
	
}

\maketitle

\thispagestyle{fancy}
\renewcommand{\headrulewidth}{0pt}
\chead{\scriptsize This is the authors' version of an article that has been published in Elsevier Computer Networks. Changes were made to\\this version by the publisher prior to publication. The final version of record is available at {\color{blue}https://doi.org/10.1016/j.comnet.2021.107981}.} 

\begin{abstract}
\ac{sdn} is an evolutionary networking paradigm which has been adopted by large network and cloud providers, among which are Tech Giants. However, embracing a new and futuristic paradigm as an alternative to well-established and mature legacy networking paradigm requires a lot of time along with considerable financial resources and technical expertise. Consequently, many enterprises can not afford it. A compromise solution then is a hybrid networking environment (a.k.a. \ac{hsdn}) in which \ac{sdn} functionalities are leveraged while existing traditional network infrastructures are acknowledged. 

Recently, \ac{hsdn} has been seen as a viable networking solution for a diverse range of businesses and organizations. Accordingly, the body of literature on \ac{hsdn} research has improved remarkably. On this account, we present this paper as a comprehensive state-of-the-art survey which expands upon \ac{hsdn} from many different perspectives. \\

\end{abstract}




\section{Introduction} \label{sec:Introduction}

Computer networks are typically built using different devices such as switches, routers, firewalls, and load balancers which communicate through various standard protocols. Network administrators are responsible for setting appropriate policies and managing all network devices in order to respond to a wide range of network events. Usually, these challenging tasks are performed manually with a rather limited number of tools available. Consequently, network management and configuration along with tuning network performance are quite tedious and potentially error-prone tasks. "Internet ossification" \cite{1577741} is another serious challenge network operators are faced with. Internet is considered as one of the critical infrastructures in today's world and it has a huge and very complex deployment base. Therefore, it is extremely difficult, or sometimes impossible, for the Internet to be updated in terms of underlying protocols as well as physical infrastructures. Network programmability \cite{van2007network} is a proposed solution to tackle these challenges and help the Internet be updated based on the emerging applications and services. Particularly, \acf{sdn} \cite{mckeown2009software} is an evolutionary networking paradigm in which control plane has been decoupled from data plane. This leads to considerably simplified network configuration and management along with enhanced flexibility and agility. The main idea behind \ac{sdn} is to allow a logically centralized software-based controller (i.e. control plane) takes care of network intelligence and decision making, while data plane is responsible for traffic forwarding tasks. Such tasks can then be programmed through either an open standard interface/protocol (such as \ac{of} \cite{mckeown2008openflow}, or ForCES \cite{yang2004forwarding}) or a domain-specific language (such as \ac{p4} \cite{bosshart2014p4}).
Consequently, major network industry parties have set up the \ac{onf} \cite{web:ONF} to promote \ac{sdn} and to standardize the \ac{of} protocol. This has caused an intense adoption of \ac{sdn} in almost every field of networking, from \ac{dc} \cite {wang2017efficient, velasco2014towards, liu2014sdn} and cloud networks \cite{banikazemi2013meridian, mambretti2015next} to \ac{wan} \cite{Branch:B4, jin2016optimizing, Branch:B4_After}, wireless \cite{costanzo2012software, riggio2015programming} and recently 5G \cite{akyildiz2015wireless, trivisonno2015sdn, yousaf2017nfv}. Hence, both industry and academia are spending considerable amount of time and money to embrace \ac{sdn} as a prevailing networking paradigm. Accordingly, \ac{sdn} market is expected to witness a substantial growth from USD 8.8 billion in 2018 to USD 28.9 billion by 2023, at a Compound Annual Growth Rate (CAGR) of 26.8\% during this period \cite{web:SDNmarket}.

Although \ac{sdn} is a fast growing networking paradigm, the traditional IP networks (a.k.a legacy networks) are still widely in place for good reasons. Despite the fact that initial \ac{sdn} models assumes that a logically centralized controller manages the network, larger and more sophisticated \ac{sdn} deployments need several controllers collaborating in a distributed manner. However, control plane scalability, resiliency, and fault tolerance are still under active investigations. Besides, distributed \ac{sdn} solutions such as controller locality \cite{schmid2013exploiting}, controller placement \cite{jimenez2014controller}, and network state consistency \cite{aslan2016adaptive} still largely rely on the solutions existed in the legacy networking domain.\\
\ac{sdn} security is another issue which needs to be addressed. Although \ac{sdn} market is flourishing and Tech Giants such as Google \cite{Branch:B4} embraced it as the early adopters, security is still a major challenge for small to medium-size enterprises. While \ac{sdn} is getting mature and being widely deployed, it definitely becomes an attractive target too. In addition, \ac{sdn} introduces new attack vectors which did not exist in traditional networks \cite{khorsandroo2019white, khorsandroo2018time}.\\
Complexity is another concern. Although \ac{sdn} is an absolutely competitive advantage for Big Techs \cite{web:googlesdn, web:googlesdn2}, it can impose unnecessary burden on technical staffs in terms of operational complexities. These complexities possibly arise in implementation, deployment, and even administration of the networks \cite{sezer2013we}. Top-tier providers, however, benefit from vast technical resources to tackle these difficulties while their smaller rivals simply cannot.\\
Strictly speaking, not all the possible challenges in adoption of \ac{sdn} are technical. There are indeed business and financial risks as well. When existing legacy network infrastructures work smoothly, organizations are usually reluctant to pay for new equipment and retrain their technical staffs.
Neither are there well-studied, production-level, uniform solutions present to support network migration process incrementally. The situation is exacerbated by potential risk of service interruptions and \ac{sla} violations during the transition from legacy to \ac{sdn}.\\
Despite the fact that transition from legacy network to \ac{sdn} can be associated with some potential challenges (e.g. financial barriers, possible technical difficulties, and lack of standards for a seamless migration, to name a few), \ac{sdn} still has obvious advantages which cannot be overlooked. Therefore, a hybrid deployment (a.k.a \acf{hsdn}) \cite{sinha2017survey, Surveys:HSDN_approaches}, where \ac{sdn} and traditional network nodes coexist, can be a compromise solution. To get the most out of a hybrid networking solution where heterogeneous network equipment, protocols, and paradigms interact, seamless coordination between logically centralized \ac{sdn} control plane and legacy distributed \ac{rib} should be guaranteed \cite{wang2017boosting}.

\begin{figure}[t]
\includegraphics[width=1\columnwidth]{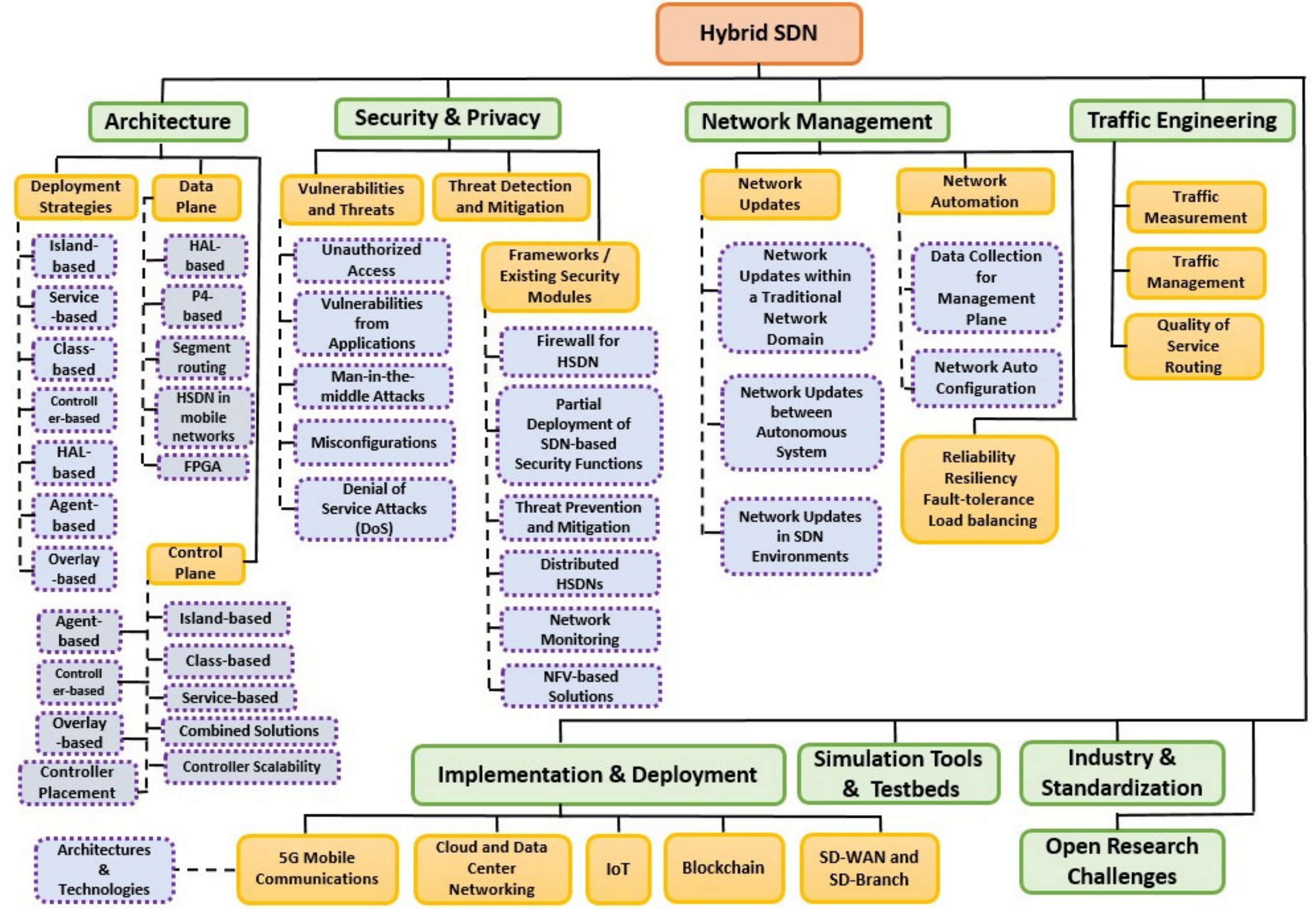}
\caption{An overview of discussed \ac{hsdn} topics}
\label{fig:fig1}
\end{figure}
In this paper, we comprehensively survey state-of-the-art \ac{hsdn}. 
Comprehensively reviewing a large body of related literature, we investigated \ac{hsdn} from multiple perspectives as illustrated in Figure \ref{fig:fig1}. More specifically, Section \ref{sec:Hybrid_Net_Arch} investigates the existing \ac{hsdn} models and  organized them in control and data planes with great details. Security and privacy of \ac{hsdn} are addressed in Section \ref{sec:HSDN_Sec_and_Privacy}. We discuss potential vulnerabilities and possible threats in a hybrid network environment while we also study how they can be detected and mitigated. This section concludes the  discussion with a detailed review of existing security frameworks and solutions in the field. Section \ref{sec:HSDN_Net_Mgmt} explores network management in terms of network data collection, network automation, and network updates. It also investigates existing literature in regards to \ac{hsdn} reliability, resiliency, fault tolerance and load balancing. Traffic engineering with reference to traffic management and measurement along with \ac{qos} and \ac{qos} routing schemes in a hybrid environment are presented in Section \ref{sec:Traffic_Eng}. We elaborate on the implementation and deployment of \ac{hsdn} solutions in the context of emergent networks environments in Section \ref{sec:HSDN_Emergent_Networks}. We specifically focus on the fifth generation cellular networks (a.k.a 5G), cloud and data center networks, \ac{iot}, Blockchain, Software Defined WAN (SD-\ac{wan}) and Software Defined Branch (SD-Branch) use cases. Section \ref{sec:simulation-testbed} provides a run-through of existing simulation tools and testbeds while Section \ref{sec:HSDN_Standardization} overviews existing \ac{hsdn} solutions from business and standardization groups perspective. We explore \ac{hsdn} open issues and potential future research directions in Section \ref{sec:open_issues} while Section \ref{sec:Conclusions} concludes the paper.

\section{Hybrid SDN Architecture} \label{sec:Hybrid_Net_Arch}

The benefits of \ac{sdn} are clear: (i) network programmability, fostering network automation, (ii) network management reducing operational costs by simplifying the management tasks and (iii) network virtualization. These aspects are spurring network operators and enterprises to upgrade their networks with \ac{sdn}-enabled switches and servers. However, introducing \ac{sdn} technologies in a legacy network requires capital and operational investments and might raise security and reliability concerns. The latter aspects, discussed in this paper, are not only due to the \ac{sdn} technology itself, but they are also correlated to the ability of the network architects and administrators to define the deployment strategy that best suits the network scenario, and to ensure a smooth coexistence between legacy and \ac{sdn} devices. 

As also discussed in other works \cite{SANDHYA201735,amin,Huang19survey,Vissicchio14Opportunities}, there exists a variety of deployment strategies for \ac{hsdn} networks. The choice on which strategy to adopt depends on many factors, such as type of the network (either enterprise, telecom operator, data center, etc.), type of network services offered to the users, required performance, capital and operational budgets. In this Section we briefly review the \ac{hsdn} architecture and then we discuss a range of different strategies for \ac{hsdn} deployment.      

\subsection{\acs{hsdn} deployment strategies}\label{sec:deployment_strategies}

\begin{figure}[ht]
\includegraphics[width=1\linewidth]{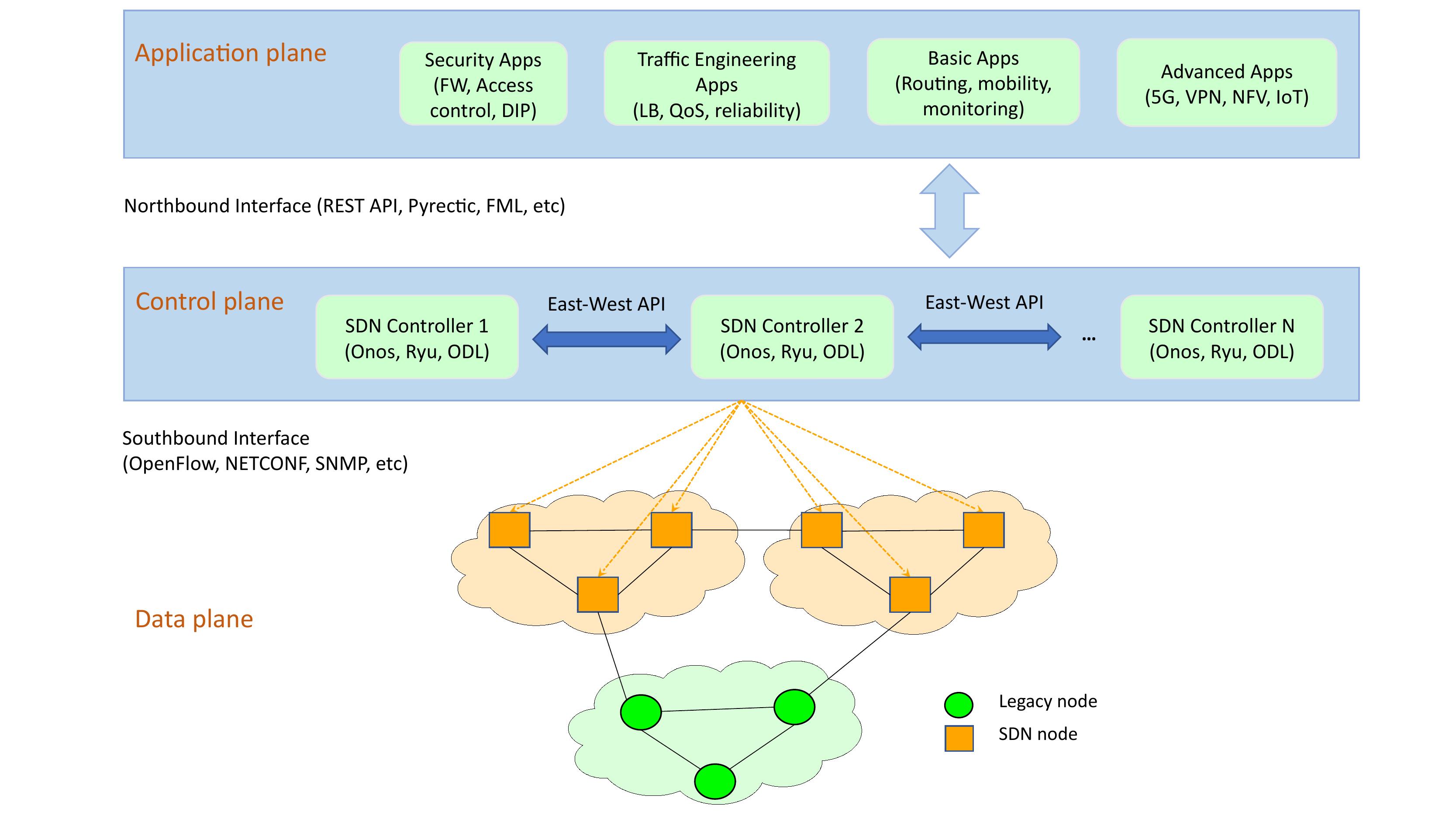}
\centering
\caption{\ac{hsdn} architecture}
\label{fig:SDN_Views}
\end{figure}

Here we introduce the \ac{hsdn} concept starting from the classic plane-oriented view of the \ac{sdn} architecture as depicted in Figure \ref{fig:SDN_Views}. Bottom-up, the \ac{sdn} data plane, also known as forwarding data plane, is built with \ac{sdn}-enabled devices (orange boxes in the figure) either virtual or physical controlled by the \ac{sdn} controller via an open vendor-agnostic \ac{sbi}. A popular common \ac{sbi} is \acf{of} \cite{openflow151}, with other well-known candidate protocols that will be mentioned in this paper. The control plane uses the \ac{sbi} to program the forwarding behaviour of the data plane. A \ac{hsdn} network comprises traditional networking, in which the traffic is forwarded based on the decisions of distributed routing/switching mechanisms, as represented in the figure with green elements.

The \ac{sdn} control plane consists of a centralized software controller that translates the requirements of the applications down to the data plane nodes and gives relevant information up to the \ac{sdn} applications. It supports the network control logic and provides the application layer with an abstracted view of the \ac{sdn}-enabled network resources. The control plane is logically centralized and usually implemented as a physically distributed system for scalability and reliability purposes (see Section \ref{subsec:cp-scalab} for more details). A set of common controllers are reported in Figure \ref{fig:SDN_Views}.    

The \ac{nbi} of the \ac{sdn} controller allows \ac{sdn} applications to send configuration commands to the \ac{sdn} devices and to retrieve information about network topology and data plane state.
Unfortunately, the \ac{nbi} is not yet standardized and a variety of different interfaces are implemented across open-source and commercial \ac{sdn} controllers. This limits the portability of \ac{sdn} applications from one controller to another. Some examples of such interfaces are shown in the \ac{nbi} block of the figure.

As represented in Figure \ref{fig:SDN_Views}, in the \ac{hsdn} networking architecture, both legacy and \ac{sdn} devices coexist. While in the latter the \ac{fib} is populated by the \ac{sdn} controller by translating the high-level routing policies issued by the \ac{sdn} applications, in the former the task is accomplished by distributed routing protocols such as \ac{ospf}, \ac{bgp} and \ac{isis}.
In both cases, the \ac{fib}, also known as forwarding table, is used by the routing or bridging function to identify the proper output port of the device for each packet. The forwarding tables are implemented using \acp{cam} or, in the case of \ac{sdn}-enabled switches, using more complex and expensive \acp{tcam}. \acp{tcam} can store three bit states (0, 1, and ``don’t care``) and work very well in conjunction with OpenFlow, where the flow table entries foresee a wildcard bit, used to inform the switch to ignore the value of the specified header field. As \ac{sdn} enables a much finer traffic control compared to traditional L2 and L3 forwarding, \ac{sdn} forwarding rules requires more memory space (up to 773 bits with recent \ac{of} versions \cite{openflow151}, compared to 60 bit of a L2 forwarding entry). This might limit the application space of \ac{sdn} when hardware enhancements are not available or feasible due to the cost and power consumption of \ac{tcam} components. 

\begin{figure}[ht]
\includegraphics[width=1\linewidth]{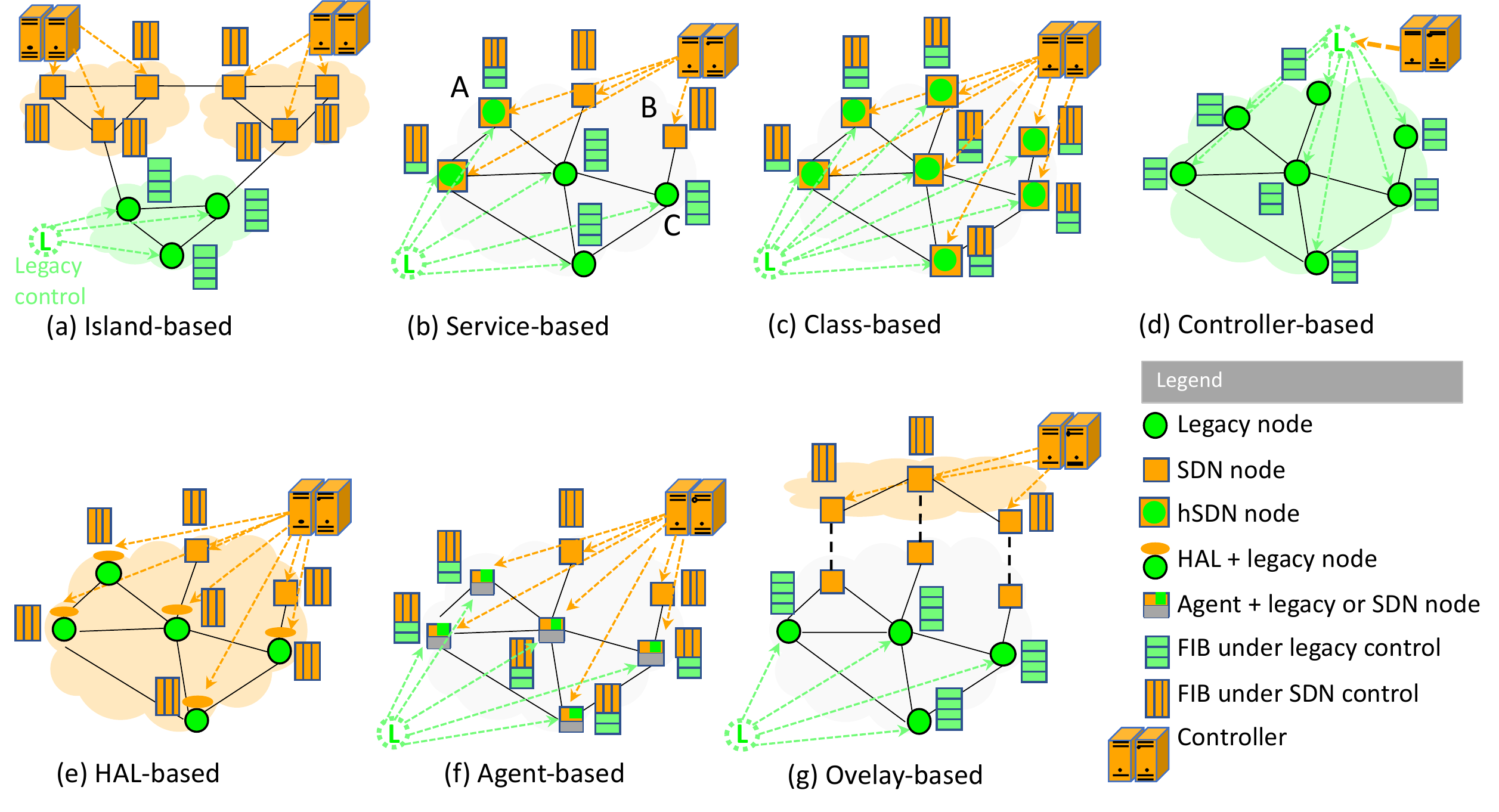}
\centering
\caption{\acs{hsdn} models}
\label{fig:arch_models}
\end{figure}

The issue related to the capacity of \ac{sdn} forwarding tables has been already investigated and tackled by leveraging the properties of the \ac{sdn} paradigm \cite{cacheflow,mms1,mms2}. In this regard, \ac{hsdn} can be seen not only as a transitional/testing deployment strategy towards a full adoption of the \ac{sdn} technologies, but also as an opportunity to maximise the potential of the \ac{sdn} technologies, thus overcoming the memory limitations through a coordinated mixture of \ac{sdn} and traditional networking. All these aspects are highlighted in previous works \cite{amin, SANDHYA201735, Huang19survey, Vissicchio14Opportunities} where a variety of \ac{hsdn} deployment strategies are discussed:

\begin{enumerate}[label=(\alph*)]

\item	\textbf{Island-based.} The network is topologically divided in areas, also called islands, consisting of either legacy devices or \ac{sdn} nodes (see Figure \ref{fig:arch_models}(a)). 
This deployment model is often adopted to gradually introduce and test the \ac{sdn} technology in confined regions, with limited impact on existing network operations. An example of the island-based model is B4 \cite{Branch:B4}, a \ac{sdwan} implemented for connecting Google sites (see Section \ref{subsec:sd-branch} for a detailed view). Like in other deployments, the backbone network is managed with \ac{sdn}, where coarse-grained network policies are required. Whereas the traffic in the \acp{dc} is routed using legacy protocols and long fine-grained forwarding tables. Figure \ref{fig:arch_models}(a) shows an example of the island-base model composed of three islands. In the figure, the orange color is used to represent \ac{sdn} elements (nodes, \acs{fib}, controller and networks), while the green color represents legacy elements (nodes, \acs{fib} and networks).

\item \textbf{Service-Based.} This deployment model follows the requirements of the services to be provisioned onto the network. While advanced services are usually implemented using \ac{sdn} technologies (e.g., \ac{lb}, \acp{acll}, etc.), others are typically managed through traditional techniques (\ac{vpn}, \ac{mpls}, IPSec, etc.). As depicted in Figure \ref{fig:arch_models}(b), the service-based strategy might use hybrid nodes, where both distributed and the \ac{sdn} control planes are supported. This type of nodes are obtained by adding \ac{sdn} capabilities to legacy routers or switches through executing software agents, upgrading the hardware components or 
installing a software \ac{sdn} switch implementation (such as \ac{ovs} \cite{openvswitch}). In hybrid nodes, the forwarding tables are managed by both legacy and \ac{sdn} paradigms and can combine \ac{tcam} with standard memory (SRAM/DRAM).   

\item \textbf{Class-based.} This model is based on partitioning the network traffic into different classes, some controlled with \ac{sdn}, others with traditional mechanisms. A common way to define the classes is using the \ac{vlan} tag. Other options include separating the traffic based on the transport protocol (TCP or UDP), application protocol (HTTP, FTP, etc.) or service (e.g., \ac{vpn}). Like in the service-based model, the network devices (at least a subset of them) are required to support both traditional and \ac{sdn} networking (see Figure \ref{fig:arch_models}(c)). Hence, the traffic-class strategy can be adopted to gradually and safely transition from traditional networking to \ac{sdn}, as one of the two paradigms might also serve as a backup solution. 

\item \textbf{Controller-based.} Advanced \ac{sdn} controllers such as ONOS and OpenDaylight, implement a range of \acp{sbi}, some of which allow them to interact with legacy devices. This is very convenient in deployments where a centralized control plane is the main requirement. In this scenario, the \ac{sdn} controller interacts with the control plane of the devices to manipulate the forwarding tables or to tune the routing protocol settings. The controller-based strategy, illustrated in Figure \ref{fig:arch_models}(d), often represents the first step of the migration towards a fully-fledged \ac{sdn} network environment.   

\item \textbf{\acs{hal}-based.} The main objective of a \ac{hal} is to make a legacy network device \ac{sdn}-compatible through a set of abstractions. Specifically, the \ac{hal} hides the technology and hardware-specific features of the network device, with the aim of presenting an abstracted \ac{sdn}-compatible device to the controller. This approach can be also employed to extend the capabilities of \ac{sdn}-enabled nodes, e.g. network virtualization support, compatibility with multiple versions of the OpenFlow protocol \cite{14-Parniewicz}. This invasive approach, requires the installation of hardware-specific software components onto the device. The \ac{hal} takes the full control of the \ac{fib}, as represented in Figure\ref{fig:arch_models}(e).

\item \textbf{Agent-based.} A software module, called agent, is executed on the network device to enable traffic control and forwarding through \ac{sdn}. The opposite is also possible, though less common, where a software agent is installed on \ac{sdn}-enabled devices to support the traditional routing/switching protocols (Figure\ref{fig:arch_models}(f)). Unlike the \ac{hal}, the agent allows the coexistence of both legacy and the \ac{sdn} control planes, hence sharing the forwarding tables of the device. This is also called dual-stack mode~\cite{Kandoi2015DeployingSN}.  

\item \textbf{Overlay-based.} As the name says, this model is based on building an overlay \ac{sdn} network on the top of a mixed legacy/\ac{sdn} network (see Figure \ref{fig:arch_models}(g)). The overlay network consists of virtual \ac{sdn} devices that are mapped onto the \ac{sdn} portion of the physical network, and are interconnected through virtual links. A virtual link can be mapped onto one or more physical links in the physical substrate. The \acp{fib} of the \ac{sdn} devices includes traffic forwarding rules, plus a set of entries that are required to build the virtual view of the network for the \ac{sdn} controller (e.g., the Big Virtual Switch \cite{bigswitch}).   

\end{enumerate}

The aforementioned models have been deeply discussed in previous surveys \cite{SANDHYA201735,Huang19survey}. We refer the reader to them for more details on the benefits and drawbacks of the different  deployment strategies. Instead, in the rest of this chapter, we review the state-of-the-art in \ac{hsdn} with particular focus on the strategies introduced above.

\subsection{\texorpdfstring{The \ac{hsdn}}{hSDN} Data Plane}
\label{sec:data_plane}
The data plane executes traffic processing (e.g., CRC checking, \ac{qos} queueing, TTL update, etc.) and traffic forwarding based on the decisions of the control plane. In \ac{hsdn} networks, the data plane combines traditional devices (typically \ac{l2} Ethernet switches, \ac{l3} routers, and other hardware appliances) with \ac{sdn}-enabled  devices, based on the adopted deployment strategy (cf. Section \ref{sec:deployment_strategies}). 
The \ac{sdn} portion of the data plane is composed of simple forwarding elements with no embedded intelligence. The decisions on how to forward or process the network traffic are taken by a remote and logically centralized entity called controller, which interacts with the \ac{sdn}-enables devices through its \ac{sbi}. Modern \ac{sdn} controllers support multiple \acp{sbi} such as \acf{of} \cite{ONF:TR-502},  ForCES \cite{yang2004forwarding} and some others \cite{RFC7426}. Other \acp{sbi} such as \ac{netconf} \cite{RFC6241} and \ac{snmp} can be also used to this purpose, even though they might be not suitable for addressing low-latency or high-bandwidth requirements \cite{RFC7426}.  
It is also worth mentioning that there exist control protocols that natively support hybrid data planes. One is \ac{of} starting from version v.1.3, whose specification defines the \ac{of}-hybrid mode switch~\cite{openflow13}. In hybrid mode, the "NORMAL port" action programs the switch to forward packets based on the legacy protocols stack. Also, the interface to the Routing System I2RS \cite{13-Hares} defines a \ac{sbi} that allows the  manipulation of the routing control on a modified node, using an agent to control legacy routing tables. \ac{sdn} controls some specific flows and leaves legacy routing protocols the management of other flows.  

In this section we review the scientific works available in the literature that have tackled the issues related to the integration of traditional and \ac{sdn} data planes.

\subsubsection{\texorpdfstring{\ac{hal}}{HAL}-based data planes}

\begin{figure}[ht]
\includegraphics[scale=0.6]{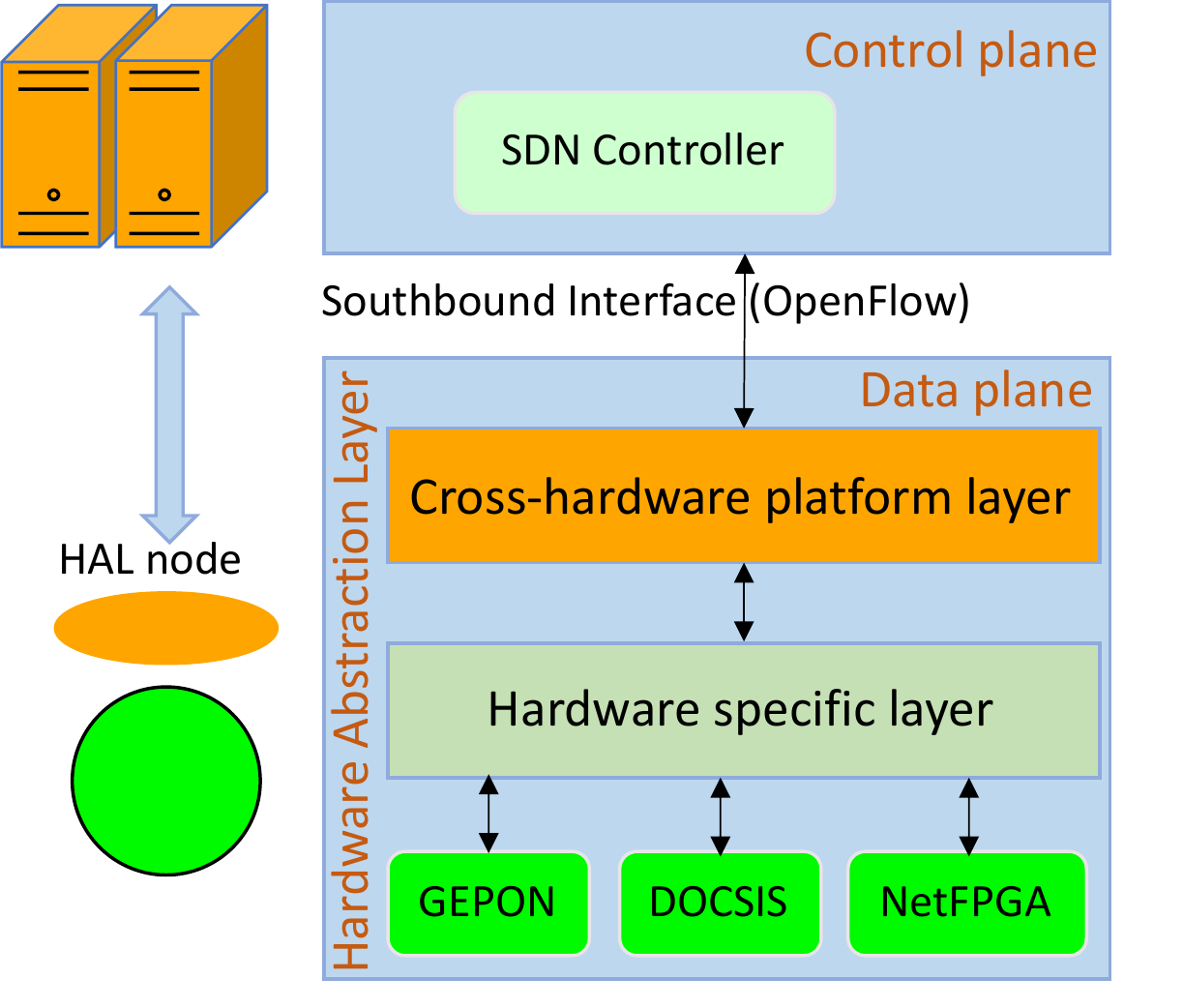}
\centering
\caption{HAL-based model}
\label{fig:hal}
\end{figure}

As introduced in Section \ref{sec:deployment_strategies}, \ac{hal} is a software layer that turns a legacy device into an \ac{sdn}-compatible switch.  
In \cite{14-Parniewicz}, a \ac{hal} has been designed to provide OpenFlow support to traditional devices. The proposed \ac{hal} is divided in two layers: the first is a common platform independent layer, called Cross-hardware platform layer, which is in charge of the node abstraction and configuration. The second layer is platform dependant and overlooks the translation and execution of commands in the node (see Figure \ref{fig:hal}).
This \ac{hal} has been implemented for various \ac{hw} platforms, such as Programmable network processors, light-path devices, Point to multi-point access networks e.g. Gigabit Ethernet Passive Optical Network (GEPON), Data Over Cable Service Interface Specification (DOCSIS)), Net Field Programmable Gate Array (\acs{fpga}).

Szalay et al. \cite{szalay2017harmless,csikor2020transition} propose HARMLESS, a framework that relies on a server to implement an \ac{sdn} control plane legacy L2 switches. Every port has to be configured with a unique \ac{vlan} identification on a legacy switch and this configuration can be repeated in the rest of the  switches. Commodity switches are connected to the server through the trunk ports and received frames are sent to the server through these ports with its input port \ac{vlan} ID. Afterwards, when the server receives the frame with the input port \ac{vlan}, in a special \acs{of} translator switch, it converts the \ac{vlan} ID. into a virtual port number that connects this switch with a second standard \ac{sdn}-enabled switch. Once the packet has being processed by the \ac{sdn}-enabled switch, that packet is returned to the translator switch through a different virtual port, tagged with a different \ac{vlan} ID and finally sent back to a legacy switch. 

Similar \ac{hal}-based approaches are presented in other recent surveys \cite{Huang19survey}.
 
 \subsubsection{P4-based data plane}\label{sec:p4}
P4 is a domain-specific programming language for expressing how the network traffic is processed and forwarded in the data plane, including P4-enabled forwarding device such as switches, routers, network cards, etc. ~\cite{bosshart2014p4}. The workflow of P4 programming model comprises three main elements: (i) \textit{the P4 architecture}, which identifies the P4-programmable data plane blocks and interfaces, (ii) \textit{the P4 program}, written in P4 language and compiled for the target hardware architecture or software device, and (iii) \textit{the target}, either a software switch or a hardware element, including programmable network cards~\cite{p4spec}. P4 enables programmers to define a variety of protocols and data plane behaviours to be implemented in nodes without depending on manufacturers to implement new features. However, despite its great potential, the adoption of P4 in hybrid environments is still very limited.       
 
Feng et al. \cite{19-Feng} propose Cl\'{e}, a solution for improving the security of legacy networks with P4-based programmable data planes.  
Cl\'{e} finds the minimum set of critical legacy switches that should be upgraded to maximise the security of the network through \ac{sdn} functionalities. In the \ac{sdn} switches, forwarding and security functions (basic firewalling and \ac{ids}) are implemented using the P4 language~\cite{14-Bosshart}.

Martinez et al. \cite{martinez2019arp} propose ARP-P4, a P4-Runtime implementation of the data plane and a new L2 legacy protocol combined with \ac{sdn}. ARP-P4 defines a data plane that forwards any ingress packet locally via ARP-Path or remotely via P4 Runtime installed rules through an \ac{sdn} controller.

Recently, various solutions to translate P4 programs for traffic processing into \acp{fpga} code have been proposed on the market and in the scientific literature. Among others, we mention Netcope P4~\cite{netcope}, P4FPGA \cite{P4FPGA}, P4$\rightarrow$FPGA ~\cite{10.1145/3289602.3293924} and P4toFPGA~\cite{8976091}. The main idea behind these initiatives is to combine the high-level P4 programming language with the performance of \ac{fpga}-based packet processing. 
 
\subsubsection{\texorpdfstring{\acf{sr}}{Segment Routing}} 
\acl{sr} is a data plane technology that can be used with the \ac{hsdn} paradigm, as it allows a source node to specify a path as an ordered list of segments (instructions that a node executes on the incoming packet) in order to guide a packet across the network. A segment scope can be local to a \ac{sr} node or global within an \ac{sr} domain. In \ac{sr}, the header has sufficient information to route the packets from the ingress to the egress node; besides, it can be applied to a technology with the source routing capability such as \ac{mpls} and IPv6. \\
  \ac{sdn} and \ac{sr} technologies can be deployed together to enable key benefits such as scalability (\ac{sr} minimizes state changes by maintaining the state at the source node only), agility (rapid response to topology changing in the source node only),  and centralized manageability (\ac{sdn} provides a global view of the network topology across multiple domains) \cite{abdullah2018segment}. One of the latest proposals, Serement et al. \cite{vseremet2020advancing}, implements and compares two scenarios with \acs{sr}  combined with \ac{mpls}-\ac{te} in one scenario and \ac{sdn} in the second. They conclude that \ac{sr} jointly with \ac{sdn} presents two outstanding characteristics that the \ac{mpls}-\ac{te} scenario does not:  
(i) Global view of the network, in \ac{mpls}-\ac{te}, when a tunnel is going to be opened, the source router calculates the best path that meets the resource requirements without taking into account other network nodes' requirements. As a result, this  could lead to an inefficient usage of network resources. Conversely, \ac{sdn} controller has the global view of the network which results in optimal usage of network resources.
(ii) Continuous link monitoring, in \ac{mpls}-\ac{te}, the nodes react by rerouting tunnels only when there is a link failure. The \ac{sdn} controller can be programmed to monitor congestion with certain threshold. If the defined threshold is exceeded, the \ac{sdn} controller can establish a new segment path and load balance traffic between the old and the new path. 

Zhang et al. \cite{zhang2020multipath} have recently proposed a mechanism for \ac{rt} multimedia services based on multipath transport, \ac{sdn}  and \ac{sr}.  
They take advantage of the global view of \ac{sdn}, the path allocation algorithm in the controller, considers future traffic distribution and current links load in order to reduce network congestion. In addition, \ac{sr} is used to eliminate the scalability problem of \ac{sdn} with multipath transmission. Their proposal balances network traffic as well as it improves the quality of end-user’s service experience.

 \subsubsection{\texorpdfstring{\ac{hsdn}}{hSDN} in mobile networks}
 
Poularakis et al. \cite{TestBeds:manet} propose a \ac{hsdn} architecture for mobile networks in order to alleviate the reliability limitations and risks of service interruption inherent in these kind of networks. 
 The data plane nodes use a solution based on \ac{sr} \cite{abdullah2018segment}. In order to deploy \ac{sr}, the network is divided into segments and the \ac{sdn} controller is in charge of the inter-\acs{sr} and it periodically broadcasts the list of segment labels (endpoint nodes of the segments) to the nodes instead of flow rules.
The routing inside a segment is made by a legacy routing protocol such as OLSR (Optimized Link State Routing Protocol). 
The control plane node forwards packets depending on the previously distributed \ac{sdn} segment label. The packets are forwarded through the segment nodes domain to the endpoint node of the segment. Then, this last node, which is also connected to next segment, forwards the packet through the next segment.  The process is repeated until the packets reach their destination. 

\subsubsection{\acfp{fpga}}
Kaljic et al. \cite{kaljic2019implementation} present a hybrid \acs{fpga}/CPU node architecture with a dedicated central processing unit that aims at overcoming some \ac{of} limitations. The \ac{of} switch programmability is limited to the level of the flow table configuration. Thus, an \ac{sdn} node can not implement new or legacy (e.g. ARP) protocols or advanced packet processing functionalities such as encryption, transcoding, \ac{dpi}, etc. The hybrid architecture is chosen because the \acs{fpga} technology achieves the \ac{sdn} high-speed packet processing while the CPU allows the implementation of new protocols and advanced packet processing functionalities.

\subsection{\texorpdfstring{\ac{hsdn}}{hSDN} Control Plane}

The control plane is responsible for instructing the data plane in accordance with the application plane requirements. Due to the fact that a hybrid control plane has to manage both legacy and SDN forwarding devices, there are  many legacy protocols that can be part of the controller in addition to the OpenFlow (\ac{of}) protocol. For example, there are some legacy \ac{l2} protocols such as \ac{stp}, \ac{vlan}, \ac{arp}, \ac{lldp}, \ac{snmp} to control connected \ac{l2} data planes. Similarly, protocols like \ac{ospf}, \ac{bgp}, \ac{isis}, control the \ac{l3} data plane.

One of the limitations of the \ac{sdn} control plane in hybrid networks is that the de-facto  protocol for discovery topology, the \ac{lldp} can not discovers legacy nodes. This can be addressed by a recent proposal, the Hybrid Domain Discovery Protocol \cite{alvarez2020hddp}, which discovers all nodes in a wired network.  

The remainder of this section elaborates on the \ac{hsdn} control plane. For the sake of coherence, we follow the classification of \ac{hsdn} models presented in Section~\ref{sec:deployment_strategies}.

\subsubsection{Island-based model proposals}
\label{Arch:Island_Based}

\begin{figure}[ht]
\includegraphics[width=0.9\columnwidth]{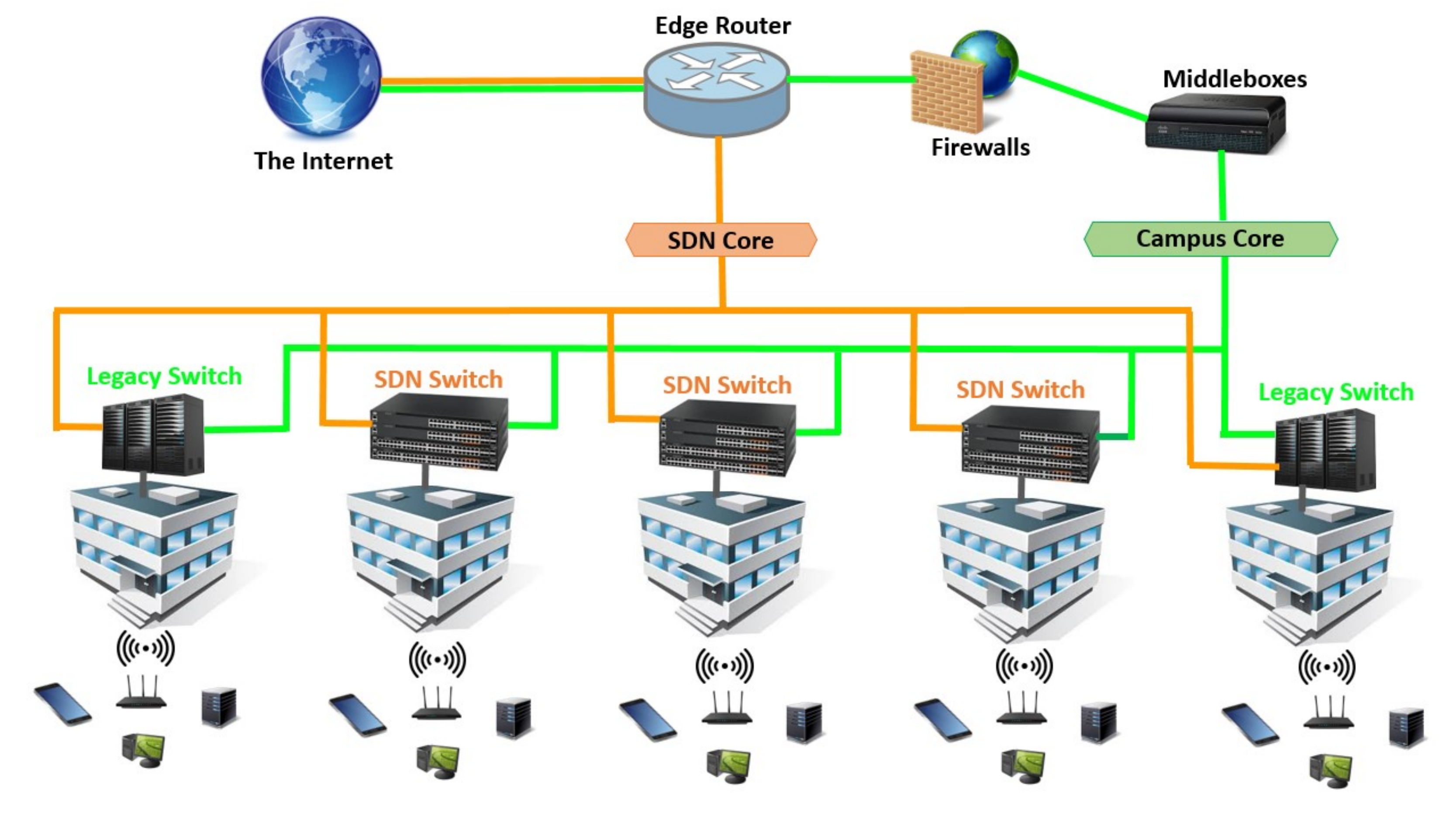}
\centering
\caption{Island-based model at the Kentucky University campus}
\label{fig:island-based}
\end{figure}

In an island-based model, the network is topologically split into areas consisting of only legacy nodes, whereas the other areas of the topology have just \ac{sdn} nodes.
Shi et al. \cite{TestBeds:Kentucky} propose an island-based  approach to deploy a new high-speed connection service (100 times faster than usual) for authenticated users. This model is implemented at the Kentucky University campus, represented in Figure \ref{fig:island-based}, where a portion of the buildings adopt \ac{sdn}-based networking, while other use Cisco legacy devices. To route the traffic from legacy networks to the SDN network, the authors developed a new controller module designed to communicate with legacy switches using \ac{pbr}.

Another example of the island-based model is the application of \ac{sdn} in the \ac{bgp} inter-domains communications to ease and improve its control. Lin et al. \cite{lin2016btsdn} propose a practical transition to \ac{sdn}, called BTSDN. The key idea in this approach is to integrate the border \ac{bgp} routers, which have high price and performance, with \ac{sdn}. It is noteworthy that in BTSDN, \ac{bgp} still works in the same way as it does in the Internet. \ac{bgp} routers in inter-domain routing run \ac{ebgp} while they run \ac{ibgp} in intra-domain routing tasks. Inside the BTSDN domain there is a controller which also runs the \ac{ibgp} protocol and acts as an \ac{ibgp} router to learn the Internet routing table. The controller is also responsible for configuring intra-domain and inter-domain flow for the hosts communications.

\subsubsection{Agent-based model proposals}

Feng and Bi \cite{feng2015openrouteflow} propose an architecture called “OpenRouteFlow”. OpenRouteFlow provides a centralized control over legacy nodes by upgrading the legacy node software with an agent called “OpenRouteFlow”. This agent communicates the distributed routing information to the controller. Moreover, the OpenRouteFlow agent receives, in parallel, the application oriented control instructions from the controller, via \ac{of} messages, which are then map into \ac{acll} and \ac{qos} operations of the router.

The complementary case is also possible, where an \ac{sdn} node is upgraded with a legacy agent. Tilmans and Vissicchio \cite{tilmans2014igp} propose an “\acs{igp}-as-a-Backup \ac{sdn}” architecture. On one hand, the controller configures the route using the \ac{of} protocol and takes advantage of the optimal network operation in the long term. On the other hand, the legacy agent builds backup paths using the \ac{ospf} protocol. Every time a network failure occurs, the \ac{igp} agent quickly restores the connectivity using the local routing information. In this way, network failures are repaired faster than in a pure \ac{sdn} network.  

Alvarez-Horcajo et al. \cite{alvarez2017new} propose a hybrid switch with partial delegation of basic bridging and new cooperative mechanisms between controller and these switches. The flow processing mode by default is done using a \ac{l2} legacy protocol. The \ac{sdn} controller can also installs forwarding rules on the switches. Finally, there is a "cooperative" flow processing mode based on the fact that the controller has knowledge of the operation of the switches, it could help then with some specific tasks because of its global knowledge of the underlying network, for instance to compute restoration paths or to help in recovery mechanisms.

\subsubsection{Class-based model proposals}

\begin{figure}[ht]
\includegraphics[scale=0.6]{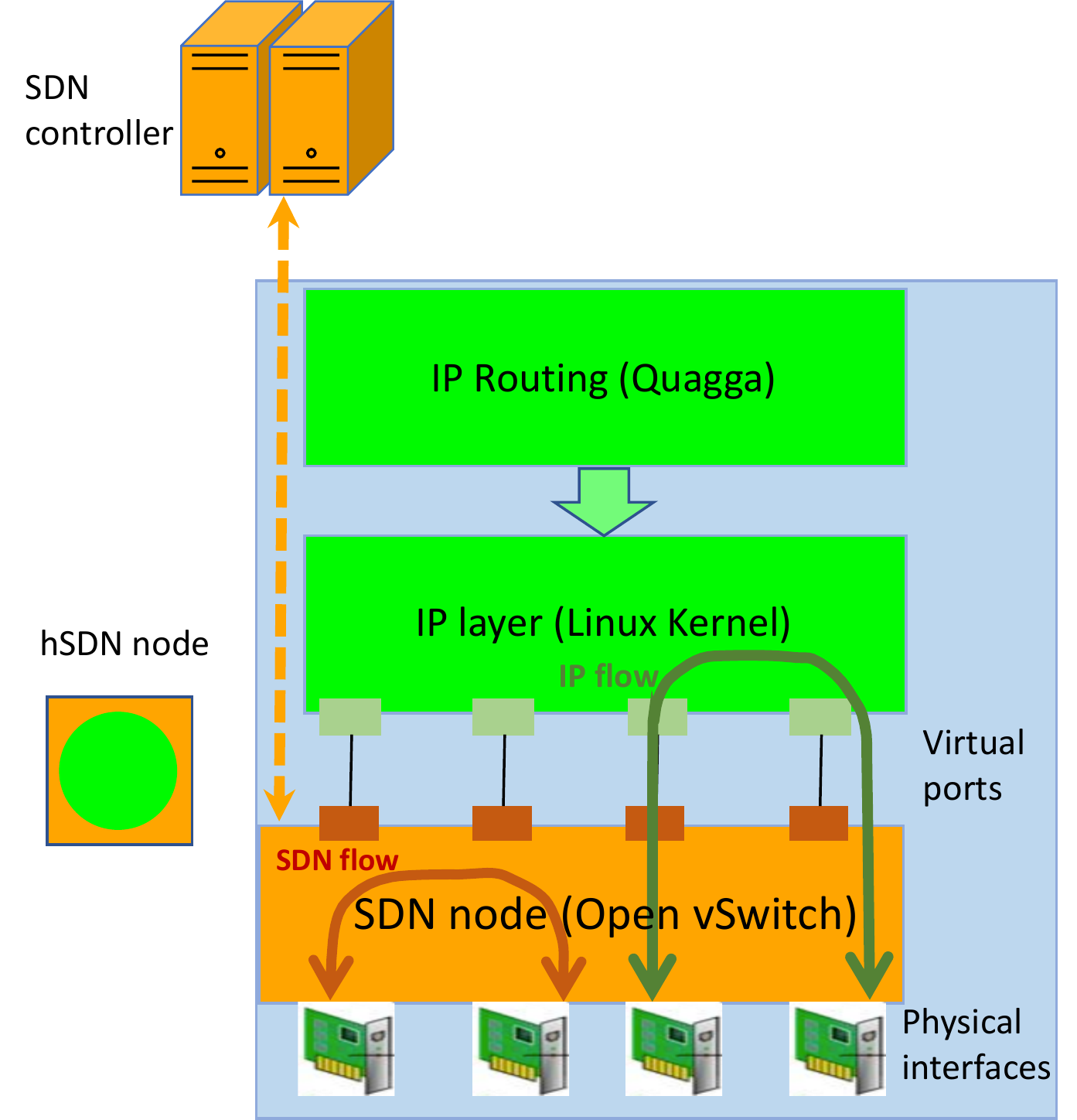}
\centering
\caption{OSHI Hybrid IP/SDN node architecture}
\label{fig:class-based}
\end{figure}

Open Source Hybrid IP/\ac{sdn} (OSHI) nodes \cite{salsano2014} combine \ac{sdn} with IP Linux kernel routing to enable research in hybrid IP/SDN carrier network scenarios. The traffic traversing OSHI nodes is divided into classes, some of them are forwarded to the next hop based on the decisions of an external \ac{sdn} controller, others using the Linux routing tables (Figure \ref{fig:class-based}). The \ac{sdn}-enabled forwarding plane of a OSHI node is implemented with \ac{ovs}~\cite{openvswitch}, a software framework that implements a fast forwarding path in the Linux kernel. The legacy IP routing is implemented with a combination of the Linux kernel IP networking and Quagga~\cite{Quagga}, which acts as a routing daemon. 

\subsubsection{Controller-based model proposals}

Kuliesius et al. present \cite{kuliesius2019sdn}, a practical deployment of \ac{hsdn}, moving two classic functions of the \ac{sdn} control plane, namely \textit{network discovery and control}, to the management plane. They develop each function on a module that interacts with the control plane via \ac{rest} \acsp{api} so that they can fit any controller that supports \ac{rest} interfaces. Topology discovery for \ac{sdn} nodes is based on the Broadcast Domain Discovery Protocol (BDDP). This protocol has the same structure as the \ac{lldp}/\ac{ofdp} except that the destination multicast address is changed to a broadcast address and the protocol field is also changed. Furthermore, to reveal information about legacy devices, the topology discovery module generates \ac{snmp} requests to register devices and/or gets \ac{snmp} trap messages. To register switches, they must be pre-configured with an \ac{snmp} IP address and authentication data. The module also uses the port traffic load to detect devices. If the port statistics show activity but the \ac{bddp} discovery does not find the neighbor on this active port, the module should turn the port down. The \ac{sdn} controller then converts \ac{of} rules into \ac{snmp} in order to to communicate and configure the legacy device.

Caesar et al. \cite{caesar2005design} develop an example of how to apply  the controller-based model to manage external traffic in a \ac{bgp} domain. They propose a similar device to a central controller called “routing control platform” (RCP). It establishes an \ac{ibgp} session with all the routers in the domain. RCP collects information about external destinations and internal topology and select the \ac{ibgp} routes for each router in the domain and sends them using \ac{ibgp}. RCP reaction is fast to the link failures in the network and it also provides scalability. 

Vissicchio et al. \cite{Vissicchio14Opportunities} propose a controller to manage a legacy network through fake nodes. In their next work, Vissicchio et al. \cite{vissicchio2014sweet} propose a central controller called \textit{“Fibbing”}. It manipulates the input of traditional routing protocol by introducing fake nodes in the network through the injection of fake \acp{lsa}. Based on the path requirements, it injects fake \acp{lsa} in the network to introduce fake nodes in the network topology. \textit{Fibbing} provides \ac{lb}, traffic steering, and providing a backup path as well.

\subsubsection{Service-based model proposals}

\ac{mpls}-\ac{te} is efficient in achieving a control plane based on both legacy and \ac{sdn}, protocols. Legacy \ac{mpls}-\ac{te} control allows the manipulation of the routing and bandwidth allocation of each tunnel to avoid congestion and meet the bandwidth and latency requirements \cite{birk2016evolving}. When a tunnel is heavily congested, an efficient way to alleviate it is to open new tunnels and split traffic among them. If congestion decreases, tunnel merging is performed by concentrating the traffic back to a single tunnel and removing the other tunnels. Legacy \ac{mpls}-\ac{te} does not have built-in splitting and merging capability and the \ac{sdn} control plane takes over these limitations.

\subsubsection{Overlay-based model proposals}

In this model an \ac{sdn} is built as an overlay on the top of the legacy network. The logical overlay is composed of the \ac{sdn} nodes and of the logical links that connect them. These logical links are implemented over one or more legacy nodes and their links. 
Levin et at. \cite{levin2014panopticon} propose an implementation of the overlay-model called \textit{"Panopticon"} where \ac{sdn} nodes are connected with legacy nodes. Pure \ac{sdn} applications can run over the Panopticon network. The key idea of this implementation is the application of traffic at the reference point which forces every packet, going from source to destination, to traverse an \ac{sdn} switch. Upon reaching the \ac{sdn} switch, the \ac{sdn} controller handles the packet. In this way each packet is processed as it would be in an \ac{sdn} network.

\begin{figure}[h!]
\includegraphics[scale=1]{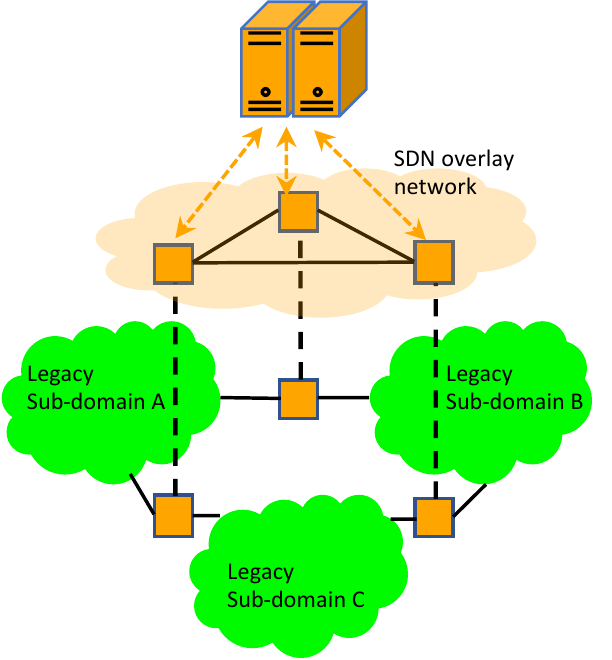}
\centering
\caption{Overlay-based model}
\label{fig:overlay-based}
\end{figure}

Caria et al. \cite{caria2016sdn} propose to divide  an \ac{ospf} network into separate sub-domains that are connected with \ac{sdn} nodes, see Figure \ref{fig:overlay-based}. Changes in the network trigger broadcasting of \ac{lsa} and re-computation of the routing table in legacy routers. These \acp{lsa} are also received by \ac{sdn} nodes and passed to the \ac{sdn} controller, where the hybrid network manager processes them. The manager optimizes the sub-domain routing configurations for \ac{lb} by computing the \ac{ospf} link costs based on the network division. Then, through the \ac{sdn} nodes in the corresponding boundary of sub-domains, these configurations are applied to the network via \acp{lsa}.

\subsubsection{Combined proposals}

Poularakis et al. \cite{17-Poularakis} propose three methods (region-based, overlay-based and agent-based) to combine distributed and \ac{sdn} control planes in mobile networks:

\begin{enumerate}
	
	\item	\textbf{ Dynamic migration of \ac{sdn} control protocol:} Part of the nodes in the \ac{sdn} network are enabled to dynamically swap to an IP routing protocol (e.g. Optimized Link State Routing Protocol (OLSR)) and to ignore the \ac{sdn} forwarding rules. As a legacy protocol can adapt to network changes faster than a remote \ac{sdn} controller, dynamic migration can be adopted in regions of the network where topology changes are frequent.  
	
	\item	\textbf{Cluster-based hierarchical control:} This method divides the mobile network into several zones and runs an IP routing  protocol for each zone independently from the rest. The centralized \ac{sdn} controller will be responsible for the traffic routing in the overlay network formed by the zones. 
	
	\item	\textbf{The distribution of backup \ac{sdn} rules:} In this third method, \ac{sdn} proactively distributes backup \ac{sdn} rules in legacy nodes that can be taken to repair network failures. As a proof-of-concept for this third method, they implement a prototype with a local agent on a smartphone. They also install \ac{ovs} so the smartphone becomes an \ac{sdn} switch. The agent periodically sends "heartbeat" messages to neighbor nodes to decide whether the link is up or down. If it is down the agent takes the backup \ac{sdn} rule. 
	
\end{enumerate}

In Table \ref{Table:test}, we provide a summary of the most relevant above-mentioned deployment methods and their specifications, proposed by the research community, over \ac{hsdn} networks for the data and control planes.

\begin{table}[h!]
\scriptsize
\centering
\caption{Comparison of \acs{hsdn} Main Proposals}
\begin{tabularx}{\textwidth}{|>{\hsize=1.0\hsize}X|>{\hsize=0.65\hsize}X|>{\hsize=0.4\hsize}X|>{\hsize=0.45\hsize}X|>{\hsize=1.5\hsize}X|>{\hsize=1.7\hsize}X|>{\hsize=0.9\hsize}X|>{\hsize=1.4\hsize}X|}

\hline
\textbf{Proposal}  & \textbf{Model} & 
\textbf{Plane} & \textbf{Node} & \textbf{Use Case} & \textbf{Key idea} & 
\textbf{Protocols} & \textbf{Evaluation} \\ 

\hline
\textbf{\textit{Parniewicz et al.} \cite{14-Parniewicz}} & \ac{hal} & DP & L & \acs{te} in an European network & Install \ac{hal} in legacy nodes & \ac{of} & Commercial implementation \\ 

\hline
\textbf{\textit{Szalay et al.} \cite{szalay2017harmless}, \cite{csikor2020transition}} & \ac{hal} & DP & L & \ac{acll}, \ac{lb}, Access gateway & Traffic tags with \ac{vlan} & \ac{vlan}, \ac{of} & Testbed \\

\hline
\textbf{\textit{Feng et al.} \cite{19-Feng}} & Service & DP & L & Avoid security attacks & Attract traffic to \ac{sdn} nodes & \ac{arp}, \ac{of} & - \\

\hline
\textbf{\textit{Kaljic et al.} \cite{kaljic2019implementation}} & Service & DP & L \& \ac{sdn} & Implementation of new protocols & \acs{fpga} run \ac{sdn}, CPU run new protocolos & \ac{of} & Small testbed \\

\hline
\textbf{\textit{Poularakis et al.} \cite{TestBeds:manet}} & Service & DP & L \& \ac{sdn} & \ac{hsdn} in mobile network & \ac{sdn} Distributes \acs{sr}  & \ac{of}, OLSR & Small testbed \\

\hline
\textbf{\textit{Shi et al.} \cite{TestBeds:Kentucky}} & Island & CP & L & High speed connections & Control legacy nodes with commands \& \ac{snmp}  & \ac{of}\& \ac{snmp} & Campus network \\

\hline
\textbf{\textit{Lin et al.} \cite{lin2016btsdn}} & Island & CP & L \& \ac{sdn} & Integration with \ac{bgp} domains & Integrate \ac{bgp} routers with \ac{sdn} & \ac{of} \& \ac{bgp} & Quagga Testbed \\

\hline
\textbf{\textit{Feng and Bi} \cite{feng2015openrouteflow}} & Agent & CP & L & Efficent routing & Integrate an agent in legacy node & \ac{of} \& \ac{igp} & Click Router testbed \\

\hline
\textbf{\textit{Poularakis et al.} \cite{17-Poularakis}} & Class & CP & L \& \ac{sdn} & Avoiding service interruption & An agent follows backup \ac{sdn} routes  & \ac{of} & Proof-of-concept \\

\hline
\textbf{\textit{Salsano et al.} \cite{salsano2014}} & Class & DP & L & \acs{te}, Backup & Quagga \& \ac{ovs} integration & IP, \ac{ospf}, \ac{bgp}, \ac{of} & Simulation \& Implementation \\

\hline
\textbf{\textit{Kuliesius et at.} \cite{kuliesius2019sdn}} & Controller & CP   & L \& \ac{sdn} & Legacy and \ac{sdn} integration & Control legacy nodes with \ac{snmp}  & \ac{of}, \ac{snmp}, \ac{netconf} & Testbed \\

\hline
\textbf{\textit{Levin et at.} \cite{levin2014panopticon}} & Overlay & CP  & L \& \ac{sdn}  &  Deploy of \ac{sdn} in operating networks & Waypoint enforcement of traffic  & \ac{of}, \acs{vlan} & Hardware testbed \& Mininet \\

\hline
\textbf{\textit{Caria et at.} \cite{caria2016sdn}} & Overlay & CP  & \ac{sdn}  &  Traffic Engineering & Sub-domain controlled by \ac{sdn}   & \ac{of}, \ac{ospf} & Numerical analysis \\

\hline \hline
\multicolumn{8}{|c|}{\bf Plane: DP (Data Plane) or CP (Control Plane)}  \\
\hline
\multicolumn{8}{|c|}{\bf Modified node: L (Legacy) or \ac{sdn})}  \\
\hline

\hline
\end{tabularx}

\label{Table:test}
\vspace{-1mm}
\end{table}

\subsubsection{Controller placement}
\label{subsec:Controller_placement}

Implementing an \ac{sdn} controller in a single server leads to scalability and robustness issues. A physically distributed but logically centralized \ac{sdn} controller architecture resolves these issues. Nonetheless, the distributed \ac{sdn} controller implies the problem of the location and number of controller instances and the node-to-controller mapping. This problem is referred to as the \ac{cpp} \cite{lu2019survey}. A good controller placement policy is vital to guarantee a reliable and efficient network management. Nevertheless, previous surveys dedicated to \ac{hsdn} networks~\cite{sinha2017survey,Surveys:HSDN_approaches,saraswat2019challenges} do not cover the problem of the controller placement in depth. In this regard, we analyze recent \ac{cpp} surveys in pure \ac{sdn} networks. 

Prakasa et at.\cite{Prakasa19controller} review the existing literature on the controller placement. They propose a complete taxonomy of controller placement that is divided based on : (i) Network type (wired or wireless), (ii) Traffic characteristics (static or dynamic), (iii) Controller characteristics (if the capacity of controllers is limited or not), (iv) Solution approach (optimization formulation, heuristic approaches, network partitioning, game theory), (v) Reliability of network elements (controller, links, nodes) and (vi) optimization objective(s) (latency, connectivity, cost, etc.).

Another survey \cite{kumari2019survey} presents a deep study on different solutions, which are classified based on the following metrics: latency, flow setup time, availability, capacity, energy, cost and some combinations of them.
Lu et al. \cite{lu2019survey} survey \ac{cpp} from the perspective of optimization as the main objective. They classify \ac{cpp} into four aspects: latency, reliability, cost and multi-objective. In the rest of this section, we discuss some of the most relevant works and solutions about \acsp{cpp} in pure \acsp{sdn}.

The algorithm proposed by Mohanty et al. \cite{mohanty2020simulated} aims to reduce the routing cost of the network considering the demands of nodes and controllers capacity. 
Schutz and Martins \cite{schutz2020comprehensive} propose a  mathematical formalization of the \ac{cpp}, which takes into consideration constrains of propagation latency and controller capacity, and determines the minimum number of controllers, their location and the mapping of nodes to each, while keeping a balanced load distribution among controllers. Simulations on 16 real network scenarios show that the proposed approach produces balanced and resilient solutions, which are proven to be optimal for most of the evaluated networks.

Two interesting works have been recently published on the \ac{cpp} in \ac{hsdn}. Guo et al. \cite{guo2019joint} study the joint optimization of upgrading the numbers of \ac{sdn} switches and controllers in \ac{hsdn} networks under a budget constraint. Deploying multiple controllers should take the following two factors into account: First, controllers should be able to process a number of flow requests in a timely manner, otherwise it will impact in the network performance. Second, the distance is a significant parameter in \acp{wan} because the propagation delay is part of the total flow request delay. They show that the problem is NP-hard and propose heuristic for small and large networks.\\
Yuan et al. \cite{yuan2019latency} focus on controller placement in \acs{hsdn} \acp{wan}. Having only one controller may not be convenient in large  \acp{wan} for reasons such as latency (caused by large distances), load balancing, scalability and reliability. The authors argue that the latency of the control messages might be affected by the legacy routers they traverse. 
In this regard, they propose a dynamic controller assignment approach that considers the impact of legacy nodes on the latency on the control channel. \\

\subsubsection{The control plane scalability } \label{subsec:cp-scalab}

\begin{figure}[ht]
\includegraphics[scale=1.1]{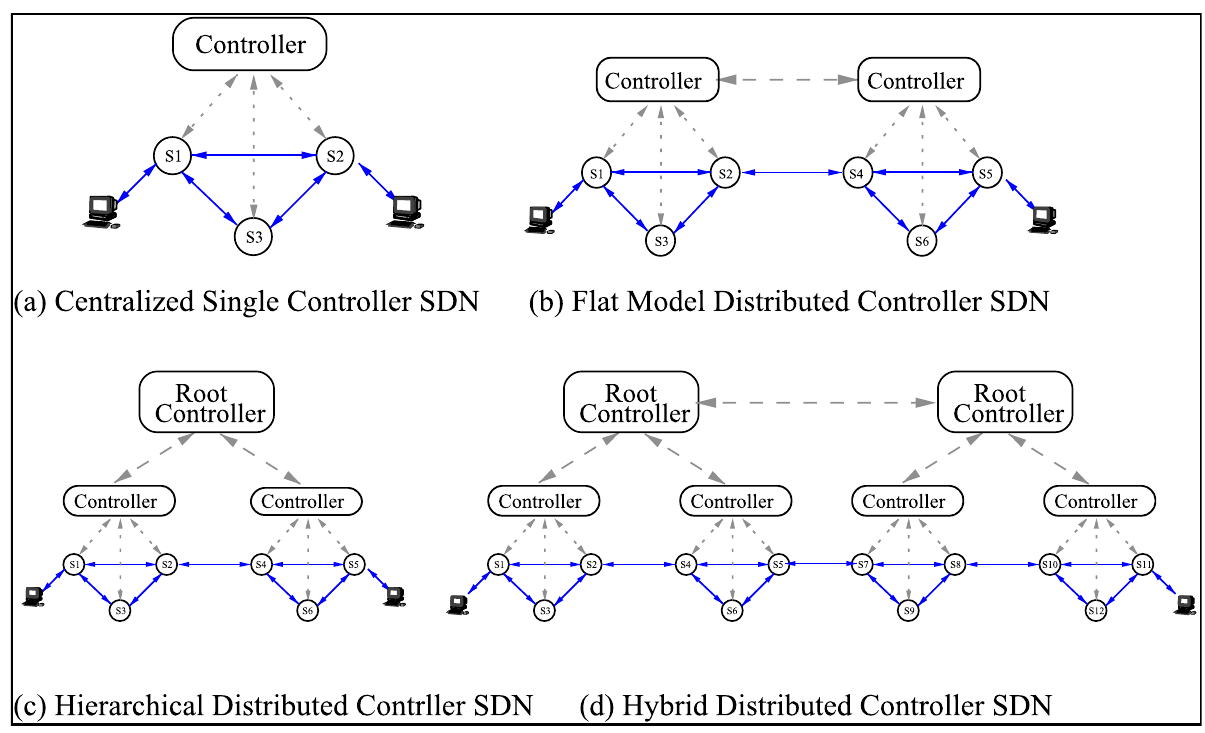}
\centering
\caption{\ac{sdn} network architectures}
\label{fig:sdn-architectures}
\end{figure}

One of the main problems with \acs{cpp} is the control plane scalability due to different types of existing \ac{sdn} network architectures, as depicted in Figure \ref{fig:sdn-architectures}  \cite{saraswat2019challenges}.

\renewcommand{\labelenumi}{(\alph{enumi})}
\renewcommand{\labelenumii}{\arabic{enumii})}

\begin{enumerate}
    \item A \textbf{single controller} based \ac{sdn} architecture consists of only one controller to maintain the global network state and control of the entire data plane of the network (Figure \ref{fig:sdn-architectures} (a)). In this scenario, a single controller can become a bottleneck for the network. To deal with scalability issues in single-controller deployments, a range of different approaches are envisioned in \cite{abuarqoub2020review}. They involves multi-threading strategies, optimization of the flow tables in the switches and machine learning techniques for traffic classification and prediction. 
\item In the \textbf{flat topology}, the network is divided into multiple domains, each of these domains is managed by a controller (Figure \ref{fig:sdn-architectures} (b)). The controllers exchange information about their domains between each other. 
\item In the \textbf{hierarchical}\textbf{ topology}, local controllers run \ac{sdn} applications which do not require a global view of the network. A global controller coordinates the local ones (\ref{fig:sdn-architectures} (c)).

\item In the \textbf{hybrid} distributed controllers architecture, flat and hierarchical architectures are used to organize the controllers in the network (Figure \ref{fig:sdn-architectures} (d)). In this scenario, the local controllers are connected with root controllers, and root controllers are directly connected with each other.
Multiple domain controllers, i.e. root controllers in the network, maintain global network information by coordinating among themselves at regular intervals. Each root controller is responsible for handling multiple local controllers. Network discovery, routing, and other \ac{sdn} services are run both at the local and the root controllers.


\end{enumerate} 

\section{Security and privacy in Hybrid SDN} \label{sec:HSDN_Sec_and_Privacy}

The adoption of \ac{sdn} technologies in legacy networks has fostered the design of novel security solutions, enabling the implementation of security services tailored to the specific needs of the users and their applications, as well as allowing a faster and coordinated reaction to zero-day network threats.   

Despite these benefits, \ac{sdn} also might bring an increase in the attack surface. The \ac{sdn} controller, the \ac{sdn} control channel and the size of the \ac{tcam} of \ac{sdn} switches are specific features of the \ac{sdn} architecture that may represent additional targets for network intrusions or other malicious cyber-attacks.
In a hybrid environment, they can be exploited by cyber-criminals to undermine the whole network, legacy devices included. In this section, we analyse the security risks in hybrid \ac{sdn} environments and the solutions proposed by the research community.

\subsection{Vulnerabilities and Threats}\label{sec:HSDN_Vulnerabilities}
As analysed in recent surveys \cite{Huang19survey,SANDHYA201735,Surveys:HSDN_approaches}, the coexistence of different control planes can be source of security issues in hybrid \ac{sdn} networks. For example, updating a hybrid network might lead to forwarding inconsistencies \cite{Huang19survey}. Moreover, security loopholes can be caused by a poorly secured interaction between legacy and \ac{sdn} devices, like in the scenario where a legacy switch reports statistics to the \ac{sdn} controller over an unsecured connection \cite{SANDHYA201735}. It is also worth mentioning the vulnerability to \ac{arp} spoofing attacks that might be determined by autoconfiguration mechanisms enabled in \ac{hsdn} networks to adapt the traffic routes to topology changes \cite{Surveys:HSDN_approaches}. 

On the other hand, hybrid \ac{sdn} is also open to a wide range of security issues inherited from traditional and software-defined networking. In this regard, the remainder of this section focuses on classifying and discussing the specific vulnerabilities of legacy and \ac{sdn}-enabled networks with the aim to provide a comprehensive overview of the attack surface of hybrid deployments.

\subsubsection{Unauthorized access}
Both traditional and \ac{sdn} network devices expose monitoring, management and control interfaces. An attacker can either use the default credentials, or obtain weak credentials, to log into the devices through such interfaces. This scenario opens the door to a number of malicious activities ranging from unauthorized re-configuration of network devices to the manipulation of traffic routing policies~\cite{10.1007/978-3-319-11599-3_14,7116184}.   

Specific to \ac{sdn}, the control channel interfaces the controller with the \ac{sdn}-enabled switches. As per \ac{sdn} functional specification, multiple controllers can access the data plane for load balancing and redundancy. This opens the opportunity for an attacker to impersonate an \ac{sdn} controller, hence to control how the traffic is routed in the network by injecting custom rules in the flow tables of the switches, potentially disrupting the normal operations through configuring black-holes, loops in the networks, or simply for re-directing part of the traffic towards a remote host for inspection~\cite{securing_sdn}. As modern \ac{sdn} controllers allow the dynamic activation/deactivation of the \ac{sdn} applications, with direct access to the \ac{sdn} controller an attacker could disable a firewall or other security-specific \ac{sdn} applications that are protecting the network~\cite{6980437,7929660}. 

\subsubsection{Vulnerabilities from applications}
Besides the vulnerabilities and threats introduced in the network by malicious software (known as \textit{malware}) executed on end hosts, network security can be also compromised by malicious/malfunctioning \ac{sdn} applications. Indeed, as \ac{sdn} applications can operate network control through the northbound interface of the controller, they can potentially harm the normal network operations and management. In the case of in-house applications, the risks for the operator/network administrator are mainly related to software bugs. On the other hand, \ac{sdn} applications coming from third-party repositories (as envisaged, for instance, by authors of \cite{netide_netsoft}) might embed malware code intentionally designed to damage the network.  

In addition, advanced \ac{sdn} controllers (e.g., ONOS \cite{Ctrl:ONOS} and OpenDaylight \cite{medved2014opendaylight}) allow multiple applications to be executed in parallel to perform different tasks on the same network traffic (e.g., monitoring, load balancing, packet filtering, network address translation, etc.). If not properly coordinated, such applications might introduce inconsistencies in the forwarding rules, which can cause unexpected traffic routing or security breaches. In \cite{netide}, the authors propose a so-called \textit{Network Engine} to mitigate these issues.       

\subsubsection{Man-in-the-Middle attacks}\label{sec:mitm}
The vulnerability of user data to threats such as eavesdropping and tampering is a general concern in network security. In this regard, the objective of \ac{mitm} attacks is to steal sensitive information by monitoring the communication between endpoints, or by impersonating one of them.

A \ac{mitm} attack can target the user traffic in the data plane as well as the \ac{sdn} control channel between the \ac{sdn} controller and the data plane. In a hybrid scenario, an attacker can build a \ac{mitm} attack to impersonate the \ac{sdn} controller, hence to inject custom rules in the network, or to mimic the behaviour of the \ac{sdn} switches with the aim to send crafted messages to the controller (e.g., false topology updates). This vulnerability can be avoided by implementing a strong authentication/encryption mechanism between switches and controllers. For instance, the OpenFlow specification defines \ac{tls} as the mechanism (although optional) for mutual authentication between the OpenFlow controller and the switches. However, due to its non-trivial configuration procedure~\cite{10.1145/2491185.2491222}, \ac{tls} is not always enabled by operators and network administrators, making the control channel vulnerable to \ac{mitm} attacks. 

In virtualized \ac{sdn} environments, network hypervisors (FlowVisor \cite{flowvisor}, OpenVirteX \cite{openvirtex} etc.) logically operate between control and forwarding planes to assign ``slices'' of the network traffic to multiple controllers. Unauthorized access to the hypervisors allows a hacker to build a \ac{mitm} attack on the control channels of all the virtual networks. This issue can be partially mitigated by implementing the virtualization mechanism directly in the data plane \cite{datapath-centric}. A similar level of vulnerability to \ac{mitm} attacks is enabled by composition frameworks, such as CoVisor \cite{covisor}, NetIDE \cite{netide}, which can be placed between the \ac{sdn} controller and switches to coordinate the execution of \ac{sdn} applications written for different controllers.    

\subsubsection{Misconfigurations}

Communication networks are complex environments, whose management procedures are often based on manual (hence error-prone) intervention of expert personnel. In recent years, network administrators have started adopting \ac{sdn} technologies to implement more advanced network automation methodologies.
Nevertheless, the deployment of the \ac{sdn} components and their correct configuration pose several technical challenges. On the one hand, as discussed in Section \ref{sec:mitm}, the configuration of the \ac{tls} mechanism between \ac{sdn} controller and switches is a complex task. A misconfigured \ac{tls} channel, or the adoption of weak authentication mechanisms and certificates can expose the network to several security threats \cite{kreutz-survey}.        
In addition, as multiple \ac{sdn} applications can access the forwarding plane through the northbound API of the controller and multiple controllers can control the same network, a lack of coordination between the different forwarding logics might lead to inconsistencies  in the network. These aspects have been tackled by a number of scientific works \cite{flowbricks,corybantic,covisor,netide}, where the authors propose methods for the composition of forwarding rules and the resolution of conflicts between different \ac{sdn} applications.      

\subsubsection{\texorpdfstring{\ac{dos}}{Denial of Service (DoS)} attacks} 
\ac{dos} attacks are common and harmful types network intrusions in which the cyber-criminals generate malicious traffic from compromised devices with the aim of disrupting the normal functions of the target networks or computing systems.  In hybrid environments, the \ac{sdn} controller can bring clear advantages in the mitigation of this type of the attacks. Indeed, thanks to its privileged position in the network, the controller can programmatically insert traffic filtering rules in strategic \ac{sdn}-enabled devices to block \ac{dos} attacks efficiently~\cite{sahay2017aroma, harikrishna2020sdn,hameed2017leveraging}. However, both \ac{sdn} controller and switches can be subject to a variety of \ac{dos} attacks: 
\begin{enumerate}
    \item \textit{Volumetric attacks}, which try to overwhelm the resources of the target system by flooding the victim with large volumes of traffic. In \ac{sdn}-enabled networks, an attacker may send a high rate of unique packets to force a continuous communication between the switches and the controller, saturating the \ac{sdn} controller with control messages. Depending on the control logic implemented in the controller, this might also lead to a saturation of the (limited) capacity of the \ac{tcam} memory of the switches~\cite{10.1007/978-3-319-58469-0_2}.
    \item \textit{Protocol attacks}, which  exploit the weaknesses or a specific behaviour of protocols. The objective of these attacks is to consume the computational resources of end hosts and network devices such as firewalls. One example in this category is represented by Syn Flood attacks, which exploit the TCP three-way handshake to consume the resources of the victim server and making it unavailable to the legitimate users. The \ac{sdn} controller can be vulnerable to this type of \ac{dos} attacks.
    \item  \textit{Application layer} attacks, where the attacker exploits the specific logic of an \ac{sdn} application or service to crash the controller, making it inaccessible to the \ac{sdn}-enabled switches. These attacks usually produce lower volumes of traffic compared to volumetric and protocol attacks. This aspect makes them harder to detect.
\end{enumerate}

Of course, also the legacy portion of the \ac{hsdn} network is vulnerable to \ac{dos} attacks. Legacy network devices can be attacked through the management interface \cite{juniper-ddos} or can enable hackers to build \ac{dos} attacks against the network infrastructure \cite{cisco-ddos}.  

\subsection{Threat detection and mitigation}\label{sec:treat_detection_mitigation}

The \ac{hsdn} paradigm can be exploited to implement advanced network security solutions by combining state-of-the-art technologies available for traditional and softwarized networks. Nevertheless, the integration of traditional hardware appliances and software-defined functions is not trivial and can be influenced by multiple factors, such as: (i) the \ac{hsdn} deployment model (cf. Section \ref{sec:Hybrid_Net_Arch}), which might determine how the traffic is routed within the network and where the security functions can be executed \cite{7962178,9098994}, (ii) the specific security policies and best-practices of the operator/enterprise, which define how each class of network traffic should be processed~\cite{shameli,pess2}, and (iii) the user requirements in terms of \ac{qos} (usually maximum end-to-end latency and minimum bandwidth), as busy links and devices might create bottlenecks in the network. 

In terms of \ac{qos}, specialized hardware appliances, also called \textit{middleboxes}, are widely used in traditional networks to achieve line-rate traffic monitoring, shaping, filtering and other network functions. However, they often run proprietary functions and operating systems, and are fixed in their position in the network. \ac{hsdn} can partially overcome this limitation by complementing traditional middleboxes with advanced security functions implemented in software and executed either in the control plane, the data plane, or both. This includes \ac{sdn} controller and switches and \ac{nfv}-enabled servers/end-hosts.
In this regard, network programmability enabled by \ac{sdn} allows the implementation of efficient network monitoring solutions entirely executed in the data plane \cite{INT_specs,ding2}, as well as the design of advanced threat detection methods that can exploit the global view of the network state from the logically centralized control plane \cite{sandra,7091959}.
In addition, \ac{sdn} and \ac{nfv} enable the provisioning of complex network security services in the form of sequences (often called chains) of \acp{vsnf}, i.e., security-specific \acp{vnf} such as firewalls and \acp{ips}, running on physical or virtual machines~\cite{shameli,pess2,8847058}. This technique is called \acf{sfc}~\cite{sfc}. The dynamic network management enabled by \ac{sfc} allows the implementation of highly-customisable security services and a faster adaptation of the system to unknown threats. On the other hand, compared to specialized hardware appliances, \acp{vsnf} may have a negative impact on the performance of the network and on the \ac{qos} level experienced by the users. The virtualization overhead, the utilization level of the servers and the techniques adopted to implement the \acp{vsnf} are the most significant contributors to the performance degradation. To mitigate the problem, recent approaches propose to either adopt kernel bypass technologies such as the \ac{dpdk} \cite{dpdk:webpage}, Netmap \cite{rizzo2012netmap} or \ac{vpp} \cite{vpp}, or to move the software packet processing in kernel space using eBPF/XDP \cite{miano}.  

An overview of existing security solutions for \ac{hsdn} deployments is presented in the next section.

\subsection{Frameworks or Existing Security Modules}\label{sec:security_frameworks}

There is small body of work in the scientific literature proposing methods, algorithms or architectures for the security of \ac{hsdn} networks. The most relevant papers are categorized in this section. 

\begin{table}[hbt!]
	\centering
	\caption{Security frameworks for \ac{hsdn}}
	\label{tab:security_frameworks}
	\begin{tabular}{|p{3.5cm}|l|l|} \hline
		{\bf Article} & {\bf Category} & {\bf Technique} \\ \hline \hline
		Heng et al.~\cite{10.1145/2491185.2491209}& Firewall & \begin{tabular}{@{}l@{}}Adaptation of traffic filtering rules to different types\\ of switches \end{tabular}\\ \hline
		Gamayunov et al.~\cite{gamayunov}& Firewall & \begin{tabular}{@{}l@{}}Translation of firewall \acp{acll}\\for legacy networks into flow rules for \ac{sdn} devices \end{tabular}\\ \hline
		Fiessler et al.~\cite{fireflow}& Firewall & \begin{tabular}{@{}l@{}}Combination of \ac{sdn} techniques with traditional\\firewalls \end{tabular}\\ \hline
		Feng et al.~\cite{19-Feng}& \begin{tabular}{@{}l@{}}Partial deployment\end{tabular} & \begin{tabular}{@{}l@{}} Framework for efficient upgrade of legacy devices\end{tabular}\\ \hline
		Ding et al.~\cite{ding1}& \begin{tabular}{@{}l@{}}Partial deployment\end{tabular} & \begin{tabular}{@{}l@{}} Greedy algorithm for incremental deployment of\\ \ac{sdn} devices\end{tabular}\\ \hline
		Cheng et al.~\cite{bgp-security-2019}& \begin{tabular}{@{}l@{}}Threat prevention\\and mitigation\end{tabular} & \begin{tabular}{@{}l@{}} A solution for \ac{sdn} controllers to protect the \ac{bgp}\\peering in hybrid networks\end{tabular}\\ \hline
		Ubaid et al.~\cite{arp_spoofing}& \begin{tabular}{@{}l@{}}Threat prevention\\and mitigation\end{tabular} & \begin{tabular}{@{}l@{}} Mechanism for the localization of the attackers in\\ \ac{arp} spoofing attacks\end{tabular}\\ \hline
		Sebbar et al.~\cite{sebbar}& \begin{tabular}{@{}l@{}}Threat prevention\\and mitigation\end{tabular} & \begin{tabular}{@{}l@{}} Algorithm for the mitigation of a variety of security\\threats \end{tabular}\\ \hline
		Zkik et al.~\cite{amsecp}& \begin{tabular}{@{}l@{}}Distributed hybrid\\ \ac{sdn} networks\end{tabular} & \begin{tabular}{@{}l@{}} Security layer for hybrid distributed \ac{sdn} networks\\designed to detect and mitigate network intrusions\end{tabular}\\ \hline
		Zkik et al.~\cite{zkik}& \begin{tabular}{@{}l@{}}Distributed hybrid\\ \ac{sdn} networks\end{tabular} & \begin{tabular}{@{}l@{}} Modular security plane composed of firewall and\\anomaly detection modules \end{tabular}\\ \hline
		Amin et al.~\cite{amin,kpartite}& \begin{tabular}{@{}l@{}}Network monitoring\end{tabular} & \begin{tabular}{@{}l@{}} Automated re-configuration of \ac{acll} policies upon\\changes in the topology\end{tabular}\\ \hline
		Lorenz et al.~\cite{lorenz}& \begin{tabular}{@{}l@{}}NFV-based solutions\end{tabular} & \begin{tabular}{@{}l@{}} Patterns for the integration of \ac{sdn}/\ac{nfv}-based\\security solutions into traditional enterprise networks\end{tabular}\\ \hline
		Doriguzzi-Corin et al. \cite{pess2}& \begin{tabular}{@{}l@{}}NFV-based solutions\end{tabular} & \begin{tabular}{@{}l@{}} Algorithms for the progressive and application-aware\\placement of security functions in telecom networks\end{tabular}\\ \hline
		H. Jmila et al. \cite{8847058}& \begin{tabular}{@{}l@{}}NFV-based solutions\end{tabular} & \begin{tabular}{@{}l@{}} Methods for the provisioning of security-aware\\service requests\end{tabular}\\ \hline
		Shameli et al.~\cite{shameli}& \begin{tabular}{@{}l@{}}NFV-based solutions\end{tabular} & \begin{tabular}{@{}l@{}} Algorithms for the placement of security functions\\in data centers\end{tabular}\\ \hline
	\end{tabular}
	\vspace{-2mm}
\end{table}

\subsubsection{Firewall for \texorpdfstring{\ac{hsdn}}{hSDN} networks}

FlowAdapter \cite{10.1145/2491185.2491209} is a firewall designed to work in environments with mixed \ac{sdn} and legacy devices. FlowAdapter translates the filtering rules from the controller flow table pipeline to switch hardware flow table pipeline, so that the same rules can be fitted into different types of hardware. In the opposite direction, Gamayunov et al. \cite{gamayunov} propose a solution for translating the firewall access lists for legacy networks into flow rules for \ac{sdn} devices. 

The authors of \cite{fireflow} propose a hybrid firewall solution called FireFlow, which combines the standard controller-switch \ac{sdn} architecture with a software-based firewall in the data plane. FireFlow offloads the complex decisions to the firewall, such as those based on string matching rules.

\subsubsection{Partial deployment of \texorpdfstring{\ac{sdn}}{hSDN}-based security functions}

Authors of \cite{19-Feng} argue that upgrading only a few legacy switches is sufficient to achieve the desired level of security for all the network infrastructure. In this regard, the authors propose \textsc{Cl\'e}, a system composed of three main modules: the \textit{Device Upgrader} determines which legacy devices must be upgraded to \ac{sdn}, the \textit{Unified Controller} re-directs the traffic towards the \ac{sdn} switches present in the network for threat analysis, and \textsc{Cl\'e} \textit{Data plane} encompasses legacy and P4 switches. Security functions such as firewall and \ac{ips} are implemented in P4 language and installed onto the  P4 switches.

In \cite{ding1}, the authors propose a greedy algorithm for the incremental deployment of \ac{sdn} programmable switches in legacy infrastructures that aims at maximising the number of monitored network flows.

\subsubsection{Threat prevention and mitigation}

In \cite{bgp-security-2019}, the authors propose S-SDBGP (Secure SDBGP), a solution for \ac{sdn} controllers to protect the \ac{bgp} peering in \ac{hsdn} networks. S-SDBGP includes three security-related software modules. A TCP authentication module that generates an authentication code for each \ac{bgp} packet. The objective is to avoid unauthorized modifications of the \ac{bgp} data. Another module is in charge of verifying the legitimacy of the prefixes announced by the \ac{as}. Finally, the \ac{as} path validation module leverages BGPsec \cite{bgpsec} to prevent attackers inserting malicious \acp{as} in the \ac{as} path. Interestingly, it is not clarified in the paper the reason why BGPsec is not enough to provide the same level of security as the proposed framework.

In \cite{arp_spoofing}, the authors propose a graph-based traversal mechanism for the localization of the attackers in the case of \ac{arp} spoofing attacks. The solution is studied for \ac{hsdn} networks, where \ac{arp} spoofing can affect the applications running on the controller. The approach is based on configuring both legacy and \ac{sdn} switches to send the \ac{arp} packets to a server for analysis. The experimental results show that the proposed solution can detect and mitigate \ac{arp} spoofing attacks.

KPG-MT \cite{sebbar}, is an algorithm for the mitigation of a variety of security threats such as \ac{mitm}, \ac{dos} and malware. KPG-MT has been implemented at different levels of the \ac{sdn} architecture. Evaluation results demonstrate the effectiveness of the proposed algorithm in different \ac{sdn}-enabled environments, \ac{hsdn} networks included.

\subsubsection{Distributed \texorpdfstring{\ac{hsdn}}{hSDN} networks}

AMSecP \cite{amsecp} is a security layer for distributed \ac{hsdn} networks designed to detect and mitigate network intrusions such as identity usurpation, \ac{mitm} and \ac{dos} attacks. Despite the complex software architecture, the threat detection logic relies on a simple algorithm that monitors the flow rates and takes decisions based on pre-defined thresholds for benign and malicious traffic. Indeed, the evaluation is focused on the execution time of the software components rather than the detection accuracy.

In \cite{zkik}, the authors present an architecture for securing logical distributed \ac{hsdn} networks. The main contribution is a centralized modular security plane for network threat detection and mitigation. The security plane includes an anomaly detection module, a network \ac{ips} and a stateful firewall.

\subsubsection{Network monitoring}

The authors of \cite{amin} argue that network topology changes might affect network policies (e.g., \acp{acll}) configured at the interfaces of forwarding devices. They propose a solution based on a graph difference technique to auto-detect the affected interfaces and network policies and to re-configure the policies based on the new topology state. An evolution of this work is presented in \cite{kpartite}, where the authors propose an approach to optimize the implementation of the \ac{acll} policies in \ac{hsdn} networks. The proposed algorithm is based on constructing a K-partite graph to search for the possible placement of the \ac{acll} policies. The set of candidate nodes for \ac{acll} policy implementation is determined by computing the shortest path algorithm on the graph.
        
\subsubsection{NFV-based solutions}

Authors of \cite{lorenz} cover different architectural design patterns for the integration of \ac{sdn}/\ac{nfv}-based security solutions with traditional enterprise networking. By analysing a stateful firewall use-case, the authors conclude that the \ac{hsdn} approach is the best choice due to good scalability with low latency, although it may be complex to implement and monitor. Other works \cite{shameli,pess2} propose strategies for provisioning privacy/security services in \ac{sdn}/\ac{nfv}-enabled networks by means of chains of software functions executed on \ac{sdn}/\ac{nfv}-enabled devices in telecom networks or \acp{dc}. 

\section{Network Management in Hybrid SDN} \label{sec:HSDN_Net_Mgmt} 
Network management is a challenging task, especially in large networks. Through efficient network management processes, it should be guaranteed that a network is highly available, scalable, and reliably performant. Besides, network management sometimes entails more sophisticated tasks such as network isolation
across complex network boundaries and decoupling logical
and physical infrastructure through network virtualization solutions.
The diverse requirements of managing a hybrid heterogeneous network can be intrinsically challenging. Accordingly, existing literature tend to use a hybrid network controller which is responsible to map legacy networking state and commands to \ac{sdn} and vice versa. Besides, the controller should be able to communicate with both legacy and \ac{sdn} switches while it needs to expose well defined \acp{api} to communicate with existing \ac{sdn} controllers.\\
For example, in a hybrid network management solution proposed by Arora et al. \cite{arora2016hybrid}, a hybrid network controller controls a network consists of \ac{sdn} switches and legacy Switches. It includes a user interface configured to display network information and to receive high-level network management commands and an \ac{sdn} controller configured to communicate with the \ac{sdn} Switches to update and manage the \ac{sdn} switches. There is also a \ac{rpc} manager configured to communicate with the legacy switches to update configuration and rules in them along with a hybrid manager configured to a route between two communicating nodes in the network based on a network topology, to calculate legacy switches and \ac{sdn} switches in the route to update, and to create a connection and network isolation using the \ac{rpc} manager and \ac{sdn} switch configuration.

\begin{figure}[ht]
\includegraphics[width=0.6\columnwidth]{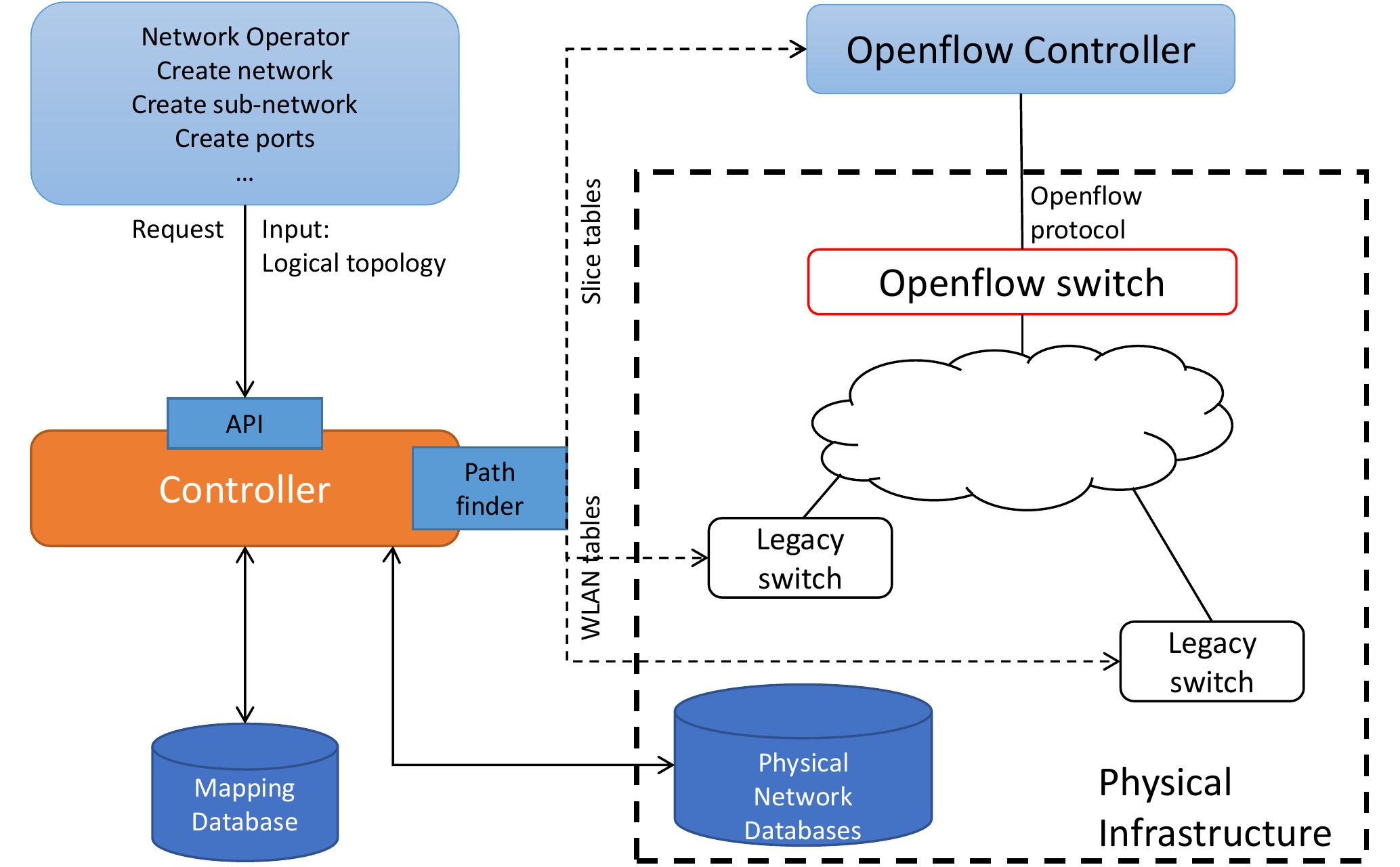}
\centering
\caption{Hybrid-controller architecture in HybNET \cite{TestBeds:HybNET}}
\label{fig:HynNET}
\end{figure}

Another implementation is HybNET \cite{TestBeds:HybNET}, a framework for automated network management of hybrid network infrastructures. As Figure \ref{fig:HynNET} depicts, HybNET consists primarily of three
components including a physical infrastructure, a path finder, and a controller. The physical infrastructure describes the basic mechanism that HybNET connects to the network infrastructure, and the path-finder provides a basic algorithms to find viable paths in the network for
connectivity. The main component and workflow is a part of the controller which manages and orchestrates the entire management framework. Each network operator request 
is handled as an atomic transaction. HybNET checks if the rules have been updated successfully, and reports any errors while maintaining state consistency at all times. Additionally, HybNET supplies a common \ac{api} for network operators to use to process transactions. For each transaction, configurations are prepared and sent to legacy switches or \ac{sdn} controllers through remote communication mechanism, such as \ac{rpc} and \ac{rest} calls, and a persistent state view is kept in the mapping database.

\subsection{Network Update}

Today's network deployments have to be highly available, scalable, manageable, and respond to different changes in an agile manner. Therefore, network-update is a frequent operation to tackle with possible network failures, or respond to the fast policy changes \cite{reitblatt2012abstractions, ghorbani2012walk}. Nonetheless, any network update should be safe meaning that making any alteration to the network device states needs to be done with minimum or no service disruption. Moreover, any network update should lead to an incremental transition from an initial correct network state to a final correct state with reasonable computation and network bandwidth overhead. In other words, a successful network update should guarantee strong consistency in the \ac{fib} of different network nodes \cite{vissicchio2017safe}. In an \ac{hsdn} environment where legacy and \ac{sdn} network devices/protocols coexist, consistency models can be applied to various network properties such as forwarding rules (for both \ac{sdn} and traditional portions of the network), network policy and even performance (mostly for the \ac{sdn} portion of the hybrid environment). Besides, network-update consistency can also be enforced in per-packet and per-flow levels \cite{reitblatt2012abstractions}.

\subsubsection{Network updates within a traditional network Domain}\label{subsubsection:Network updates within a Traditional Network Domain}

In legacy networks where network updates are mainly pertinent to forwarding paths, distributed routing protocols play an important role in network updates. Within a network with a uniform administration policy, link state \ac{igp}s are typically used to compute forwarding paths within a network. This can be adjusted by network administrators form a network logical
view in terms of weighted graph which. Such a view is then being shared among all routers in a network domain \cite{foerster2018survey}. When a router's logical network view changes for any reason, the router tries to rebuilt a new logical network view while it starts disseminating messages to its neighbors. Each neighbor also follows the same procedure until all the nodes in the network build their own new and consistent logical network view. During this period of time, which is referred to as routing convergence time \cite{francois2005achieving}, there is no guarantee on when and in what order the routers receive messages about new network states. This introduces transitory network state inconsistencies which potentially leads to usually short-lived forwarding disruptions between some of the network nodes. \\
Accordingly, both industry and academia made effort to propose feasible solutions for such a problem. Defining \ac{igp} extensions has been discussed extensively in  \cite{shaikh2006avoiding, francois2007avoiding, moy2003graceful,ammireddy2012expedited, shand2008restart}. While such solutions are ideal in theory, they have some serious limitations in practice. For example, extending routing protocols does not support all network update scenarios. It sometimes can be difficult to implement such extensions and it also takes time for them to be accepted by the community. As a consequence, other solutions such as network update optimization and scheduling \cite{keralapura2006optimal, raza2009graceful, raza2011graceful} and "Ships in the Night" (\ac{sitn}) strategy \cite{vanbever2011seamless, vanbever2012lossless} emerged.

\subsubsection{Network updates between autonomous systems}\label{subsubsec: Network updates between autonomous systems}

Network update within an \ac{as} is associated with state distribution among network nodes sharing similar network administration policy. However, the core of network updates among \acp{as} is based on information propagation in terms of route distribution. Therefore, any routing convergence problem, network state inconsistencies, or anomalies in the network behavior  potentially originate from route distribution issues \cite{le2008shedding}. Since \ac{mpls} traffic between \acp{as} is managed by routing protocols such as \ac{bgp}, applying solutions discussed in section \ref{subsubsection:Network updates within a Traditional Network Domain} may result in undesirable problems \cite{vanbever2013cure} (e.g. loop forwarding). Appropriately, authors in \cite{francois2007avoiding} proposed a solution by completing \ac{bgp} alternate route distribution and forward traffic through them before the routing session gets expired. Enhancing virtual router migration \cite{wang2008virtual}, \ac{bgp} configuration migration \cite{keller2010seamless}, and applying \ac{sitn} to \ac{bgp} \cite{vissicchio2012improving} are other examples of proposed solutions in this context. While all the above-mentioned solutions are protocol-dependant, \cite{alimi2008shadow} allows routers to run several configurations in parallel by proposing a generalization in the \ac{sitn} strategy.

\subsubsection{Network updates in \texorpdfstring{\ac{sdn}}{Software Defined Networking (SDN)} environments}

In \ac{sdn} deployments, data plane nodes are dummy devices which rely on the control plane for decision making on packet processing tasks. Due to the logically centralized control plane which maintains a holistic view of the network at any time, issues such as rerouting traffic in case of path failure, route convergence, and reacting to the network state changes can be addressed almost immediately in \ac{sdn} \cite{rexford2016purpose}. In \ac{sdn} environments, the controller frequently checks the network state while it also enforces security policies and tune network performance. Consequently, the consistent network update that the controller performs not only has an influence on network forwarding behavior, but also affects other network properties such as security and performance. \\
Network update operations in an \ac{sdn} can be extended to network characteristics other than forwarding states. This is in contrast to the updates in legacy networks \cite{reitblatt2011consistent,reitblatt2012abstractions}.
Through a labeling technique, the authors enforce update consistency in packet granularity. Many research works extend what is proposed in \cite{reitblatt2011consistent,reitblatt2012abstractions}. For instance, \cite{brandt2016consistent, forster2016power,zheng2015minimizing} propose more guarantees during network updates such as loop-free and lossless update processes with minimum congestion. Another examples are \cite{dudycz2016can,forster2016consistent,ludwig2016transiently, foerster2017loop} where a customized update algorithm has been devised. These algorithms can selectively relax some of the extra consistency requirements in \ac{sdn} deployments. 

\subsection{Network Automation in \texorpdfstring{\ac{hsdn}}{Hybrid SDN (hSDN)}}
Network automation helps network administrators and network providers configure large network deployments with complex topologies in a fast systematic way. Network automation is based on collecting data from network devices, process them taking into account network administration policies and applying necessary changes to the network devices under control \cite{edelman2018network}. With the advent of \ac{ai} and \ac{ml}, network automation gradually gains more popularity in network control and management solutions \cite{rafique2018machine}.\\
In a hybrid environment, network device configuration files are created consistently through \emph{network configuration templating} technique. These templates then reliably determine which part of the configuration file needs to be static or dynamic. Decoupling templates from configuration data and leveraging configuration
management systems such as Ansible and Salt facilitate automated management of a hybrid network.

\subsubsection{Data collection for management plane}

\begin{figure}[ht]
\includegraphics[width=0.8\columnwidth]{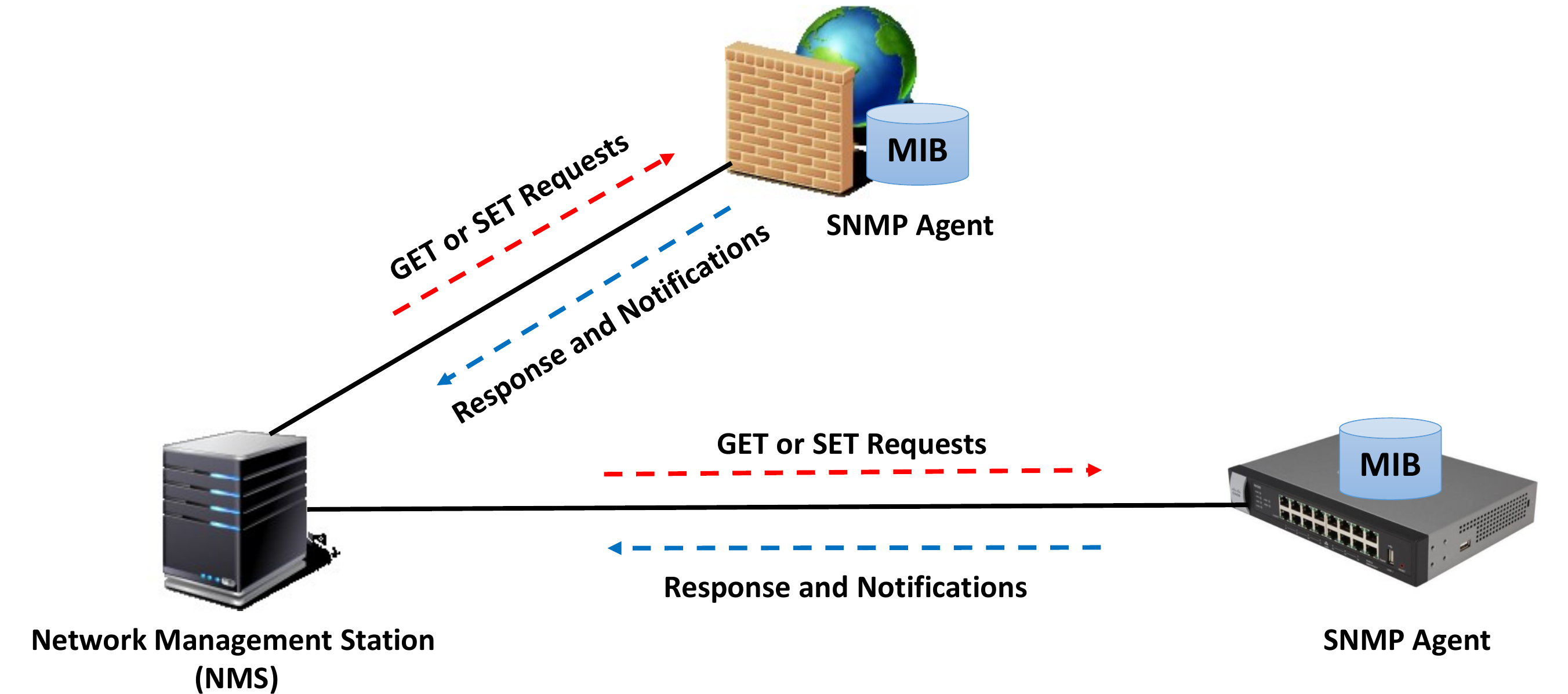}
\centering
\caption{Simple Network Management Protocol (SNMP)}
\label{fig:snmp}
\end{figure}

Network data collection is another domain in which network automation plays an important role. Data can be pulled from the network nodes (i.e. pull model) or the network nodes can stream data to the data collection agents (i.e. streaming telemetry \cite{paolucci2018network} or push model). Regardless of which data collection model is being used, the interface used to connect to the management plane of the network nodes is of essence. \ac{snmp} is a widely deployed protocol which can be used to collect data from network nodes based on the pull model. As Figure \ref{fig:snmp} illustrates, a network node plays the role of a Network Management Station (NMS) which is an \ac{snmp} server. The server collects exposed data from managed network nodes (aka \ac{snmp} agents) using Management Information Bases (MIBs). \ac{snmp} can fetch MIB information through \emph{Get Requests} while it also supports updating MIB information using \emph{Set Requests}. Although \ac{snmp} has been around for a few decades, it was not designed as a real-time \ac{api} to network devices. Therefore, it is gradually being replaced with next generation network data collection tools and management protocols \cite{zhang2015sdnmp}.

\begin{figure}[ht]
\includegraphics[width=0.6\columnwidth]{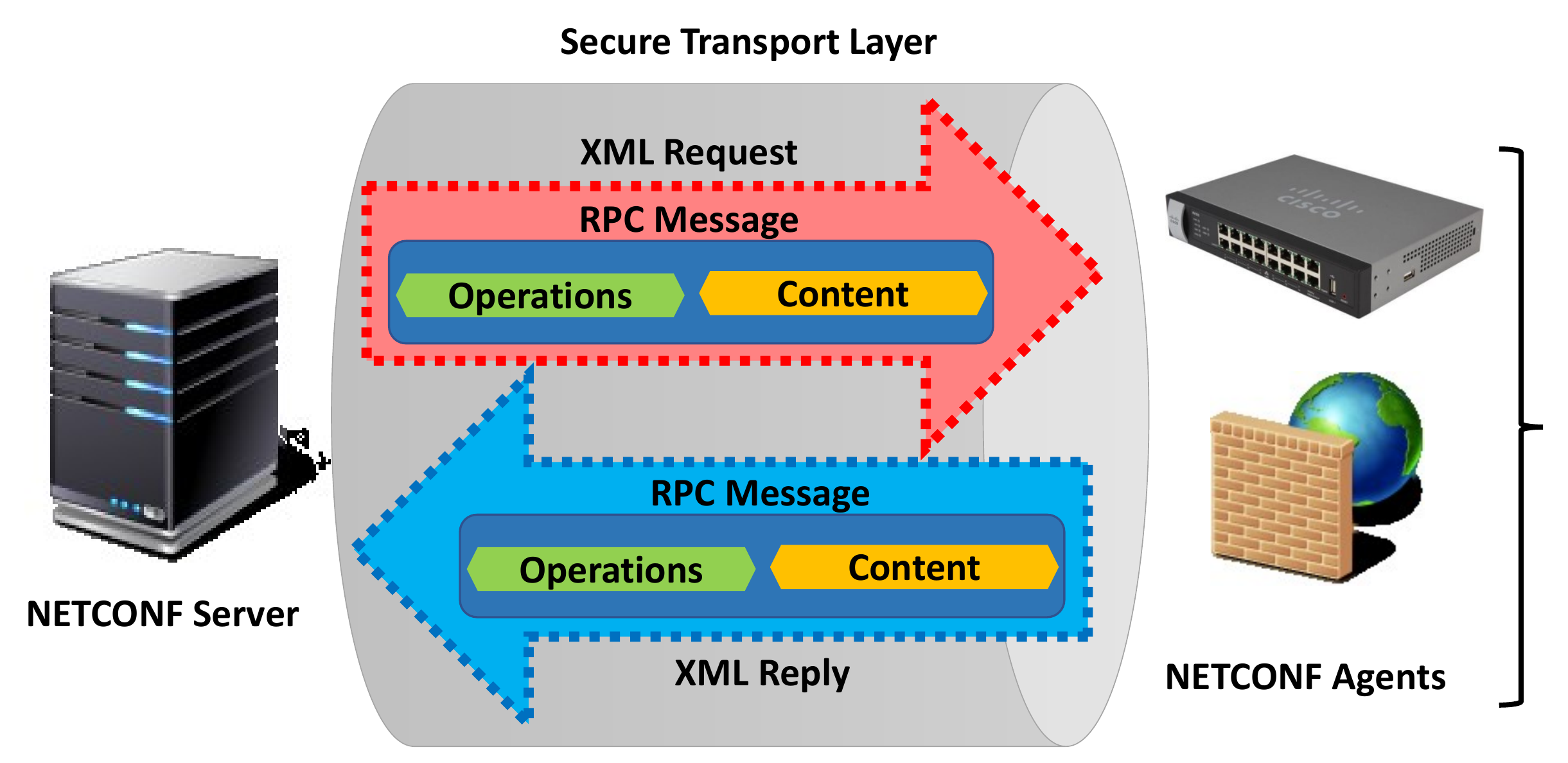}
\centering
\caption{Network Configuration protocol (NETCONF)}
\label{fig:netconf}
\end{figure}

Another network management protocol through which management plane can collect data is \ac{netconf}. It is a connection-oriented protocol standardized by \ac{ietf}. As Figure \ref{fig:netconf} shows, RFC6241 conceptually partitioned the \ac{netconf} protocol into four layers namely the content layer, the operations layer, the messages layer, and the secure transport layer.\ac{netconf} leverages \ac{xml} to encode data and send it securely between the \ac{netconf} client and the \ac{netconf} server. It uses \ac{xml} to encode a request in the form of \ac{rpc} which triggers the execution of prearranged operations in a transaction-based manner \cite{katiyar2015auto}. 

\subsubsection{Network auto configuration}

Upon successful data collection form different network nodes, various networking tasks can be performed by automation. This brings considerable flexibility and agility to network management solutions and let the solutions keep up with dynamic nature of hybrid network topologies.

Auto configuration of \acf{acll} policies after network topology changes in \ac{hsdn} are discussed in \cite{amin2016auto}. In a hybrid network, frequent topology changes may violate network policies in terms of \ac{acll} at the
interfaces of forwarding devices. This may result in security vulnerabilities and network performance degradation. Consequently, the authors propose a new approach for \ac{hsdn} that auto-detects the interfaces of forwarding devices in addition to the network policies that are affected due to changes in the network topology. Afterwards, a graph which represents the topology of the network is created and a graph difference technique is applied to detect changes in the topology. The network management system then uses the graph and its analysis results to verify policies for updated topology. If there is any violation in policy implementation, affected interfaces along with associated policies are detected and necessary revisions are being applied.

Auto configuration of \ac{sdn} data plane in a \ac{hsdn} deployment proposed by Katiyar et al. \cite{katiyar2015auto}. When a
new \ac{sdn} switch attaches to a \ac{hsdn} network, automation of the switch initial configuration reduces the installation costs and maintains the consistency of the updated hybrid network. However, this is a challenging task due to the heterogeneity of a hybrid network. Katiyar et al. achieved this goal by detecting a new \ac{sdn} switch in a hybrid network. It then provides the switch with
appropriate configuration parameters and ensures seamless service among
\ac{sdn} and non-\ac{sdn} network segments. This solution also
introduced \acs{dhcp}-\ac{sdn} as an extension of \ac{dhcp} to support \ac{sdn} compatible nodes. It enables a new \ac{sdn} switch to
automatically start operating even though the intermediate
switches are not configured with the \ac{sdn}-specific \ac{vlan}
setup.


\subsection{Reliability, Resiliency, Fault-tolerance, and Load balancing}

A \ac{hsdn} can be seen as a distributed system. Hence, various distributed systems issues including reliability, resilience, \ac{lb} and fault-tolerance need to be managed for seemless operation of the network as a whole. In this section, we briefly explore existing research on these topics.

Chu et al.~\cite{7218482} propose an approach to guarantee traffic reachability in the presence of any single link failure. By redirecting traffic on the failed link to \ac{sdn} switches through pre-configured IP tunnels, the proposed approach is able to react to the failures quickly. With the help of coordination among \ac{sdn} switches, it is also able to explore multiple backup paths for failure recovery. This allows the proposed approach to avoid potential congestion in the recovered network by choosing proper backup paths. The proposed scheme requires a small number of \ac{sdn} switches in the \ac{hsdn} network to achieve fast recovery and guarantee 100\% reachability from any single link failure. The proposed approach is able to better load-balance the post-recovery network compared to IP Fast Reroute and shortest path re-calculation.

Yasunaga et al.~\cite{7412516} propose a load balancing (\ac{lb}) method for symmetrically routed \ac{hsdn} networks that can handle existing distributed routing parameters, link cost and path selection. The proposed method optimizes the link cost and the path selection simultaneously under a symmetrically routed condition, while traditional methods optimize them individually. The \ac{lb} performance of the proposed method is better and more stable than any of the traditional methods. Furthermore, the proposed method has high versatility and provides a good \ac{lb} performance even in networks without distributed routing protocols.

Vissicchio et al.~\cite{vissicchio2017safe} study the problem of computing operational sequences to safely and quickly update arbitrary networks. They characterize cases, for which this computation is easy and propose a generic sequence-computation approach, based on two new algorithms that they combine to overcome limitations of prior proposals. The proposed approach always finds an operational sequence that guarantees strong consistency throughout the update with very limited overhead. Moreover, it can be applied to update networks running any combination of centralized and distributed control-planes, including different families of \acp{igp}, OpenFlow or other \ac{sdn} protocols, and \ac{hsdn} networks. The 
proposed approach therefore supports a large set of use cases, ranging from traffic engineering in \ac{igp}-only or \ac{sdn}-only networks to incremental \ac{sdn} roll-out and advanced requirements in partial \ac{sdn} deployments.

Shozi et al.~\cite{8851009} study multi-path \ac{lb} in a \ac{hsdn} data plane. The load balancer’s performance is tested using four different \ac{hsdn} topologies,  namely, Fat Tree, Ring, Mesh and Torus topology with the OpenDayLight controller. The main contribution of this work is balancing the load in congested data plane links (either before or after network failure) in \ac{hsdn} environments. The obtained results display an improved overall  network performance and reduced network delay.

Link fault protection becomes a key problem in hybrid networks with both \ac{sdn} and non-\ac{sdn} switches. Existing solutions need large computation time and high configuration overhead (e.g., tunnels and flow table entries for each switch). To address the above limitation, Jia et al.~\cite{8406823} propose a solution named Hybrid-Hie to achieve fast reroute and load balancing in \ac{hsdn} networks. The proposed solution configures the split ratio on each backup path in advance by predicting the link utilization. It can achieve efficient fault recovery for inter-domain links, and achieve better load-balance and recovery path stretch.

Yang et al.~\cite{8967024} aims to minimize the repair path length to reduce the delay experienced by the rerouted traffic and saving network bandwidth. The key enabler is a tunneling mechanism for constructing multi-tunnel repair paths by leveraging multiple \ac{sdn} switches. Multi-tunnel repair paths provide additional flexibility in rerouting and a new \ac{sdn} candidate selection algorithm is then designed to take advantage of them. To further reduce the repair path length, they propose to replace the link-based tunneling adopted by fault-detection routers by destination-based tunneling. If the network traffic distribution is available, they further show that destination-based tunneling can be used to avoid network congestion. A brief summary of related issues including reliability, resilience, fault-tolerance, and load balancing is given in Table~\ref{te3}.

\begin{table}[hbt!]
\centering
\caption{Reliability, resilience, fault-tolerance, and load balancing in \ac{hsdn}}
\label{te3}
{\small
\begin{tabular}{|l|l|l|} \hline
{\bf Article} & {\bf Objective} & {\bf Technique} \\ \hline \hline
Chu et al.~\cite{7218482} & Single Link Failure Recovery & Pre-configured IP tunnels  \\ \hline
Yasunaga et al.~\cite{7412516} & Load Balancing  &  Joint optimization of link cost and path selection \\ \hline
Vissicchio et al.~\cite{vissicchio2017safe}& Safe Update  &  Compute operational sequences for safe update \\ \hline
Shozi et al.~\cite{8851009}& Load Balancing  & Multi-path load balancing  \\ \hline
Jia et al.~\cite{8406823}& Link Fault Protection  &  Fast reroute and load balancing \\ \hline
Yang et al.~\cite{8967024}& Link Failure Protection  &  Minimize repair path length \\ \hline
\end{tabular}
}
\end{table}
\section{Traffic Engineering} 
\label{sec:Traffic_Eng}


\acf{te} aims to measure and analyze \ac{rt} network traffic, and design routing
mechanisms to improve utilization of network resources. In \ac{te}, a wide range
of optimization techniques is applied to achieve maximum network performance parameters. In
traditional networks, most common traffic engineering mechanisms are \ac{mpls} \cite{10.17487/RFC3031} which uses short path labels and GMPLS\cite{mannie2004generalized} which extends \ac{mpls} to manage further class of interfaces and switching technologies. In \ac{sdn}, a centralized controller
communicates with the forwarding elements and maintains global network views, network topology,
traffic demand and link state information to determine routing paths. Capability of fine-granularity
network control over flows renders \ac{sdn} a leading candidate for many \ac{te} solutions
and several mechanisms have been proposed for \ac{sdn} networks \cite{abbasi2016traffic,akyildiz2016research}. In this section, we provide
an overview of \ac{te} mechanisms proposed for \ac{hsdn} networks.

\subsection{Traffic Measurement}

Traffic measurement includes monitoring, measuring and collecting network status information. Traffic measurement is
challenging in \ac{hsdn} since only a subset of routers are
\ac{sdn} capable and legacy routers do not have the flexibility in traffic statistics collection. In this section, we focus on traffic measurement schemes proposed for \ac{hsdn}.

Measurement of all the flows in \ac{hsdn} is too costly and not really necessary. Cheng et al.~\cite{8468206} propose a method
to collect load of a subset of significant links and estimate the rest with estimation error of 5\%.
Polverini et al.~\cite{7460177} provides definition and assessment of an effective criterion, based on the flow spread parameter,
to identify the flows to be measured that minimise the estimation error. Experimental results show that a small percentage of flows are enough
to drive the estimation error an order of magnitude lower than that obtained with the classical solution solely based on link
load measurements. The proposed algorithm is also able to distribute measurement tasks fairly among network nodes, taking into account the available forwarding tables space. 

Another challenge in \ac{hsdn} is measurement of \ac{rt} link-loads. The latency for collecting the global link-load information can be prohibitively long in a \acf{wan} due to the long routing protocol convergence time. Cheng et al.~\cite{8468206} propose a comprehensive traffic monitoring method for collecting \ac{rt} load information of all links. In this method, the controller only needs to collect the load of a small subset of significant links and then, it estimates the link load of the rest. Minimum number of \ac{sdn} routers is placed to cover these links. Experiment results show that the proposed method can quickly estimate the global link load at an estimation error rate of 5\% within sub-seconds. Compared with state-of-the-art methods, the proposed method adapts better to dynamic traffic changes and can reduce the maximal link usage by up to 40\%.

Recently, new paradigms are investigated to select critical
flows for the traffic matrix. Zhang et al.~\cite{zhang2020cfr} propose a reinforcement learning based scheme for this purpose.
Reinforcement learning is dynamically learning by adjusting actions based on continuous feedback to maximize a reward.
To mitigate the impact of network disturbance, one interesting \ac{te} solution is to forward the majority of the traffic flows using Equal-Cost Multi-Path and to selectively reroute a few critical flows using \ac{sdn} to balance link utilization of the network. However, critical flow rerouting is not trivial due to the vast solution space for critical flow selection. Moreover, it is impossible to design a heuristic algorithm for this problem based on fixed and simple rules, since rule-based heuristics are unable to adapt to the changes of the traffic matrix and network dynamics. Zhang et al.~\cite{zhang2020cfr} propose a reinforcement Learning-based scheme that learns a policy to select critical flows for each given traffic matrix automatically. The proposed scheme then reroutes these selected critical flows to balance link utilization of the network by formulating and solving a simple Linear Programming problem. Extensive evaluations show that the proposed method achieves near-optimal performance by rerouting only 10\%-21.3\% of the total traffic.

Monitor placement for network measurement is vital for effective
measurement of link latencies. Tian et al.~\cite{9046249} investigate link latency measurement problem in scenarios where conventional routers can only support the shortest path routing protocol, and where conventional routers can support source routing protocol. For both scenarios they show how to deploy a minimum number of monitors and how to construct measurement paths between monitors to measure all the link latencies. Several algorithms are presented to solve these problems and the evaluations on different topologies show the performance benefits of the proposed methods.

A brief summary of traffic measurement schemes for \ac{hsdn} is given in Table~\ref{te1}.
\begin{table}[hbt!]
\centering
\caption{Summary of Traffic Measurement and Traffic Management}
\label{te1}
{\small
\begin{tabular}{|l|l|l|} \hline
{\bf Article} & {\bf Objective} & {\bf Technique} \\ \hline \hline
\multicolumn{3}{|c|}{\bf Traffic measurement} \\ \hline
Polverini et al.~\cite{7460177}& Identify flows that reduce & Flow spread based algorithm\\
                               & estimation error most    & \\ \hline
Cheng et al.~\cite{8468206}& Compressive traffic monitoring  & Collect load of small subset of links,\\
                           &                                 & estimate the link load of the rest \\ \hline
Zhang et al.~\cite{zhang2020cfr}& Selection of critical flows & Linear programming approach\\
                                & for traffic matrix using ML    & \\ \hline 
Tian et al.~\cite{9046249} & Link latency measurement & Minimize number of monitors  \\ \hline \hline
\multicolumn{3}{|c|}{\bf Traffic management} \\ \hline
Casado et al.~\cite{10.1145/2342441.2342459} & Vendor-neutral network model & Use \ac{mpls} for \ac{hsdn} \\ \hline
Tu et al.~\cite{6903516} & Mitigate congestion & Tunnel splicing between \ac{mpls} and \ac{sdn}\\ \hline
Sugam et al.~\cite{6567024}& Improve network utilization & Fully Polynomial time approximation schemes \\ \hline
He et al.~\cite{7145321} & Optimize TE performance & Provable approximation guarantees \\ \hline
Ren et al.~\cite{8985283} & Distributed algorithm for TE &
Integer Linear Programming formulation \\ \hline
Guo et al.~\cite{6980429}& Minimize maximum link utilization & Change weights and flow-splitting ratio\\ \hline
Nakahado et al.~\cite{7000328}& Decrease network congestion & Edge nodes replaced with \ac{sdn} switches\\ \hline
Sharma et. al.~\cite{6573054}& Integrated network management & Integration of dynamic control for flows \\
                             & and control system            & \\ \hline
Galan-Jimenez et al.~\cite{8826373}& Traffic matrix assessment & Mixed measurement and estimation\\ \hline
Ren et al.~\cite{7841819}& Minimize maximum link utilization & Jointly determine appropriate switch \\
                         &                                   &             and splitting fractions \\ \hline
Guo et al.~\cite{7410320}& Optimize migration sequence       & Genetic algoritm based heuristic                     \\
                         & of legacy routers                 & \\ \hline
Wang et al.~\cite{7996839}& Forwarding graph construction    & Forwarding graphs with high throughput   \\ \hline
Guo et al.~\cite{8038397}& Optimize routing over multiple    & Problem NP-hard. Heuristic algorithm \\
                         & traffic matrices                  &                                      \\ \hline
Davoli et al.~\cite{7313628}& \ac{te} in IP carrier network & Segment routing \ac{te} model\\ \hline
Seremet et al.~\cite{9066291}& Investigate Benefits of & Comparison of scenarios with/out \ac{sdn} \\
                             & Segment routing for TE & \\ \hline
Wang et al.~\cite{7980165}& Heterogeneity in forwarding      & \ac{sdn} placement to enhance controllability\\
                          & characteristics in \ac{hsdn}         & \\ \hline
\end{tabular}
}

\end{table}

\subsection{Traffic Management}

Traffic management investigates how to manage and schedule network traffic based on the network status
information provided by traffic measurement. In this section, we focus on traffic management schemes proposed for \ac{hsdn}.

\noindent Casado et al.~\cite{10.1145/2342441.2342459} argue that an ideal network design would involve \ac{hw} that
is simple, vendor-neutral and future-proof and the ideal \ac{sw} control plane should be flexible. They indicate
that \ac{sdn} falls short of these goals and propose Fabric which is a better form of \ac{sdn} by retrospectively
leveraging the insights underlying \ac{sdn}.

\noindent Tu et al.~\cite{6903516} propose a tunnel splicing mechanism for heterogeneous networks with \ac{mpls} and \ac{of} routers.
Two key mechanisms are suggested. The first abstracts the underlying network devices into uniformed nodes in order
to shield the details of various equipments. The second strips the manipulation of flow table and \ac{mpls} label switch table
from controller and fulfills it in an independent module. The proposed paradigm has been developed on a Linux system and tested in experiment networks. Emulation results are also provided to show its feasibility and efficiency.

\noindent A hybrid network model which encompasses both \ac{sdn} and \ac{mpls}-based functionality is presented and experimentally
evaluated with the help of a prototype implementation in~\cite{7939157}. This model
gives the \ac{poc} for the co-existence of \ac{mpls} and \ac{sdn}-based network elements in the network. The
switches can be configured using a centralized controller to monitor and push configurations.
The model is able to accomplish conflict-free separation of centralized and decentralized controls.
Also, a centralized controller is able to provision traffic flows better as compared to
decentralized based approaches using \ac{mpls}. Therefore, this model provides a better alternative for the smooth
transition of a legacy network to an \ac{sdn}.

\noindent Sugam et al.~\cite{6567024} investigate how to leverage the centralized controller to get significant
improvements in network utilization as well as to reduce packet losses and delays. They formulate the
\ac{sdn} controller's optimization problem for \ac{te} with partial deployment and develop 
fast fully polynomial time approximation schemes (FPTAS) for the problem resolution. The reason for
using FPTAS is that it is simpler to implement and runs significantly faster than a general linear programming solver for medium and large size problems. They show that deploying even a few
strategically placed \ac{sdn} forwarding elements in the network can lead to improved network performance.

\noindent He et al.~\cite{7145321} investigate near-optimal \ac{te} to optimize the \ac{te} performance over
all network links shared with uncontrollable conventional traffic. Two modes are investigated. In \textit{barrier
mode} two forms of traffic (conventional and \ac{sdn}) are routed in separated capacity spaces. In \textit{hybrid mode}
each link can be fully occupied by either form of traffic. Fast algorithms are proposed for both scenarios with
provable approximation guarantees.

\noindent Ren et al.~\cite{8985283} propose a distributed algorithm for
near optimal \ac{te}. In \ac{hsdn}, redirecting flows from every source-destination pair through at least one \ac{sdn} node,
can improve \ac{te} performance and obtain flow manageability, while on the other hand leading to increasing
demands of \ac{tcam} resources in \ac{sdn} nodes. They formulate the \ac{te} problem as an \ac{ilp} model to minimize \ac{mlu} and solve it in a centralized manner, where \ac{sdn} waypoint selection and splitting fractions for each flow are jointly determined. They develop a
distributed algorithm deriving from Lagrangian decomposition theory to effectively solve the \ac{te} problem.
The simulation results indicate that, when 30\% of the \ac{sdn} nodes are deployed, the proposed \ac{te}-aware distributed routing algorithm obtains \ac{mlu} comparable to that of full \ac{sdn}, and has a limited influence on the routing efficiency.

\noindent Guo et al.~\cite{6980429} propose the SOTE algorithm in which \ac{ospf} weights and flow-splitting ratio of the \ac{sdn} nodes can both be changed. The controller
can arbitrarily split the flows coming into the \ac{sdn} nodes. SOTE can obtain a lower \ac{mlu} and performs better compared with legacy network and the hybrid network with fixed weight setting. SOTE shows that near optimal performance can be achieved when only 30\% of the \ac{sdn} nodes are deployed.

\noindent Nakahado et al~\cite{7000328} propose a \ac{hsdn} strategy, where only edge nodes need to be replaced by \ac{sdn} switches and other nodes obey conventional \ac{ospf} routing.  Only edge nodes distribute the traffic incoming to the \ac{ospf} network from other networks,
and other nodes operate \ac{ospf} routing. This technique can decrease network congestion ratio.

\noindent Sharma et al.~\cite{6573054} argue that in future networks, traditional network management techniques and
controller-based mechanisms need to co-exist and work in an integrated manner to enable incremental deployment
of \ac{sdn} capabilities in legacy networks. They propose an integrated network management and control system framework that combines legacy network management functions such as discovery, fault detection with the end-to-end flow provisioning and control enabled by \ac{sdn}.

\noindent Galan-Jimenez et al.~\cite{8826373} focus on \ac{tm} Assessment problem from the perspective of an Internet Service Provider. Since the migration from legacy IP to fully-deployed \ac{sdn} networks needs to be incremental due to budget and technical constraints, they propose a mixed measurement and estimation scalable solution for
hybrid IP/\ac{sdn} networks to accurately solve the \ac{tm} Assessment problem by exploiting the availability of flow rule counters in \ac{sdn} switches.
The performance evaluation shows that the proposed error-tolerant solution is able to assess the \ac{tm} with a negligible estimation error by only
measuring a small percentage of traffic flows, overcoming other state-of-the-art algorithms proposed in the literature. The
performance analysis of the proposed implementation using an emulated network environment shows that a trade-off between the quality of the assessed \ac{tm} and its impact on the network in terms of control messages can be found by properly tuning the number of measured flows.

\noindent Ren et al.~\cite{7841819} propose a flow routing and splitting algorithm that jointly determines an appropriate \ac{sdn} switch for every flow as the waypoint, and optimize the traffic splitting fractions for every
\ac{sdn} switch among its outgoing links to minimize the \ac{mlu}. Simulation results with various \ac{sdn} deployment
rates indicate that, when 20\% of the \ac{sdn} switches are deployed, the proposed algorithm can obtain a lower \ac{mlu} compared with other state-of-art works. The proposed algorithm can generate paths that are about 50\% longer for
every flow on the average. However, it has a limited influence on the routing efficiency.

\noindent Guo et al.~\cite{7410320} search for an optimal migration sequence of legacy routers to \ac{sdn}-enabled routers to determine the order the
routers should be migrated. They propose a genetic-heuristic algorithm to find a migration sequence of the routers that maximizes the \ac{te} benefit. They evaluate the algorithm by conducting simulation experiments, making comparison to the greedy migration algorithm and static migration algorithms. The experiments show that the genetic algorithm outperforms the other migration algorithms in searching for a migration sequence. When properly deployed, about a migration of 40\% of routers reaps most of the benefit.

\noindent Wang et al.~\cite{7996839} argue that the effectiveness of 
\ac{te} in \ac{hsdn} strongly depends on both the structures of forwarding graphs and traffic distribution, while existing approaches mainly focus on the latter. They define the consistent forwarding graph, and then construct forwarding graphs with potential high throughput for effective \ac{te}, while maintaining forwarding consistency. The evaluation results show that the proposed forwarding graph construction approach improves network throughput and achieves better load balancing compared with existing simple forwarding path constructing approaches, especially with more fraction of \ac{sdn} deployment.

\noindent Guo et al.~\cite{8038397} formulate the problem of optimizing routing for both the average case and worst case over multiple traffic matrices. They prove the problem is NP-hard and propose a heuristic algorithm and evaluate the algorithm with real traffic datasets. Through extensive experiments, they observe that the worst case performance of  routing can be dramatically improved by about 30\% with a little sacrifice of the average case performance by about 2\% and demonstrate the effectiveness of their algorithm in optimizing both the average case and worst case performance of routing.

\noindent \acf{sr} is also proposed for \ac{te} in \ac{sdn}.
Davoli et al.~\cite{7313628} have examined the \ac{te} problem in the context of an IP carrier network and proposed an \ac{sdn}-based
segment routing \ac{te} model. \ac{sdn} controller allocates network traffic to the links according to link capacity. For flow
assignment, a heuristic based approach minimizing network traversal time is used. Seremet et al.~\cite{9066291} shows the benefits of \ac{te} in \ac{sdn} by comparing two scenarios: Segment routing in classical IP/\ac{mpls} networks and segment routing in IP/\ac{mpls} networks with \ac{sdn} controller.

\noindent Wang et al.~\cite{7980165} propose a solution to handle the heterogeneity caused by distinct forwarding characteristics of \ac{sdn} and legacy switches. They plan \ac{sdn} placement to enhance the \ac{sdn} controllability over the hybrid network, and conduct \ac{te} considering both the forwarding characteristics of \ac{sdn} and legacy switches. The experiments with various topologies show that the \ac{sdn} placement planning and hybrid forwarding yield better network performance especially in the early 70\% \ac{sdn} deployment.

A brief summary of traffic management schemes for \ac{hsdn} is given in Table~\ref{te1}.

\subsection{Quality-of-Service Routing}

\ac{sdn} provides an open control interface to support flexible network traffic scheduling strategies rendering it the ideal framework for satisfying \ac{qos} requirements of many applications such as voice and video.

\noindent Egilmez et al.~\cite{6411795} propose OpenQoS, a design based on an \ac{of} controller for multimedia delivery with end-to-end \ac{qos} support. 
It is based on \ac{qos} routing where the routes of multimedia traffic are optimized dynamically to fulfill the required \ac{qos}.
They measure the performance of OpenQoS over a real test network and compare it with the performance of the current state-of-the-art, HTTP-based multi-bitrate adaptive streaming. Experimental results show that OpenQoS can guarantee seamless video delivery with little or no video artifacts experienced by the end-users. Moreover, unlike current \ac{qos} architectures, in OpenQoS the guaranteed service is handled without having adverse effects on other types of traffic in the network.

\noindent Ongaro et al.~\cite{7069395} exploit the use of \ac{sdn} in conjunction with the \ac{of} protocol to differentiate network services with quality level assurance and to respect agreed \acp{sla}. They define a \ac{mano} architecture that allows to manage the network in a modular way. They provide a seamless integration of the proposed architecture and the \ac{sdn} following the separation between the control and data planes. They provide an \ac{ilp} formulation of the problem of enhancing \ac{qos} and \ac{qoe} in \ac{sdn} networks in terms of packet loss and delay, taking into account  network constraints and the requirements of real-time applications, i.e., maximum acceptable packet loss and delay rates. 

\noindent HiQoS~\cite{7112035} provides \ac{qos} guarantees using \ac{sdn}. Moreover, HiQoS uses multiple paths between source and destination and queuing mechanisms to guarantee \ac{qos} for different types of traffic. Experimental results show that the proposed HiQoS scheme can reduce delay and increase throughput to guarantee \ac{qos}. In addition, it recovers from link failure very quickly by rerouting traffic from failed path to other available path.

\noindent Sieber et al.~\cite{7140446} demonstrate an implementation of a Network Services Abstraction Layer (NSAL) on top of the network control and management plane. They introduce a unified data model for both \ac{sdn} and legacy devices that allows managing and configuring both networks in a unified way in order to achieve \ac{qos} for time-critical tasks (e.g. VoIP). Due to the unified data model, network operators are able to manage their network through one interface. They demonstrate a use case by implementing a VoIP scheduling application on top of the NSAL and evaluate VoIP call quality in a distributed heterogeneous network.

\noindent Lin et al.~\cite{7557432} propose a \ac{qos}-aware adaptive routing (QAR) in multi-layer hierarchical \ac{sdn} networks. Specifically, the distributed hierarchical control plane architecture is used to minimize signaling delay in large \ac{sdn} networks via three-levels design of controllers, i.e., the super, domain (or master), and slave controllers. The QAR algorithm is proposed with the aid of reinforcement learning and a \ac{qos}-aware reward function, achieving a time-efficient, adaptive, \ac{qos}-provisioning packet forwarding. Simulation results confirm that QAR outperforms the conventional Q-learning solution and provides fast convergence with \ac{qos} provisioning, facilitating the practical implementations in large-scale software service-defined networks.

\noindent Bi et al.\cite{8863947} consider a hybrid Industrial network consisting of conventional routers (e.g., running \ac{ospf}) and \ac{sdn}-enabled switches (e.g., running \ac{of}), and propose an intelligent \ac{qos}-aware forwarding strategy to improve \ac{qos} in industrial applications, by utilizing a single path minimum cost forwarding scheme and a K-path partition algorithm for multipath forwarding. Simulation results demonstrate that the proposed scheme guarantees the \ac{qos} requirements of industrial services and efficiently utilizes bandwidth resources by balancing traffic load in the \ac{sdn}/\ac{ospf} hybrid industrial network. 

\noindent Mondal et al.~\cite{9107218} study the problem of \ac{qos}-aware data flow management in the presence of heterogeneous flows. The optimal data flow management in this is NP-hard and they propose a game theory-based heterogeneous data flow management scheme, that is capable of reducing network delay by 77–98\%, while ensuring 24–47\% increase in network throughput. 

\noindent Chienhung et al.~\cite{LIN2017242} propose an \ac{sdn} hybrid network architecture which can discover existence of legacy switches using the Spanning Tree Protocol to have global view of the \ac{sdn} hybrid network. Authors enable OpenFlow switches to cooperate with legacy switches by using the Learning Bridge Protocol without requiring any modification on legacy switches. By utilizing the characteristics of \ac{sdn}, \ac{sdn} applications can dynamically find routing paths according to pre-defined \ac{qos} requirements and current network status.

\noindent Alouache et al.~\cite{doi:10.1002/dac.4521} 
propose a \ac{hsdn} based geographical routing protocol with a clustering approach for vehicular networks. It takes into account three different criteria to select the best relay to send data: (1) the contact duration between vehicles, (2) the available load of each vehicle, (3) and the log of encountered communication errors embedded in each cluster head. The multi‐criteria strategy allows the selection of the most reliable vehicles by avoiding communication problems and ensuring connection availability. Once the hybrid control plane has finds the next eligible neighbor, the data plane is in charge of dividing and sending data. Simulation results show that proposed scheme A achieves good performance with respect to the average routing overhead, the packet drop rate, and the throughput.

\begin{table}[hbt]
\centering
\caption{Summary of \ac{qos} solutions in \ac{hsdn}}
\label{te2}
{\small
\begin{tabular}{|l|l|l|} \hline
{\bf Article} & {\bf Objective} & {\bf Technique} \\ \hline \hline

Egilmez et al.~\cite{6411795} & Multimedia delivery & Dynamic optimization\\ \hline
Ongaro et al.~\cite{7069395}& Meet service level agreements & ILP formulation\\ \hline
HiQoS~\cite{7112035}& Differentiated services for & Multi-path and queuing \\
                    & bandwidth guarantee         &                        \\ \hline
Sieber et al.~\cite{7140446}& Simpler configuration of \ac{hsdn} & Unified data model\\ \hline
Lin et al.~\cite{7557432}& Multi-layer hierarchical \ac{sdn} &  Minimize signaling delay\\ \hline
Bi et al.\cite{8863947}& Hybrid industrial network & Multipath forwarding\\ \hline
Mondal et al.~\cite{9107218}& Data Flow Management  &  Game theory based scheme \\ \hline
Chienhung et al.~\cite{LIN2017242}        &    Discover legacy switches  &  Spanning tree protocol  \\  \hline
Alouache et al.~\cite{doi:10.1002/dac.4521}        & Geographical routing     &  multi-criteria strategy  \\   \hline
Hin et al.~\cite{8254456}       &  Energy-aware routing    &  Safely turn off equipment  \\   \hline

\end{tabular}
}
\end{table}

\noindent Hin et al.~\cite{8254456} propose an algorithm to enable energy aware routing in \ac{hsdn}. Since in real life, turning off network equipment is a delicate task as it can lead to packet losses, proposed work provides several features to safely enable energy saving services: tunneling for fast rerouting, smooth node disabling and detection of both traffic spikes and link failures. Proposed work is validated by extensive simulations and by experimentation.

\noindent A brief summary of \ac{qos} schemes for \ac{hsdn} is given in Table \ref{te2}.

\section{Hybrid SDN in Emergent Networks: Implementation and Deployment} \label{sec:HSDN_Emergent_Networks} 


During the 2010s the scientific and academic community has experienced a period of development of new and innovative network technologies with evolving complexity and requirements e.g., Next Generation Mobile Communication Networks (5G), \ac{ns} \cite{NGMN:Network_Slice}, \ac{nfv} \cite{ETSI:NFV_Framework}, Cloud Computing \cite{Cloud_Comp}, \ac{iot} \cite{IoT:Definition}, Blockchain \cite{Digital_Transformations}, etc. This section analyzes the impact and use of the \ac{hsdn} concept over various of these emergent technologies.

\subsection{5G Mobile Communications}

5G next generation mobile communication technology has been raised as the response to the increasing volume of traffic and number of connections generated by mobile communication networks and also as a need to fulfill the requirements that enable the deployment and implementation of new use cases for mobile communications and even new applications for Industry 4.0. 

\begin{figure}[ht]
\includegraphics[scale=0.24]{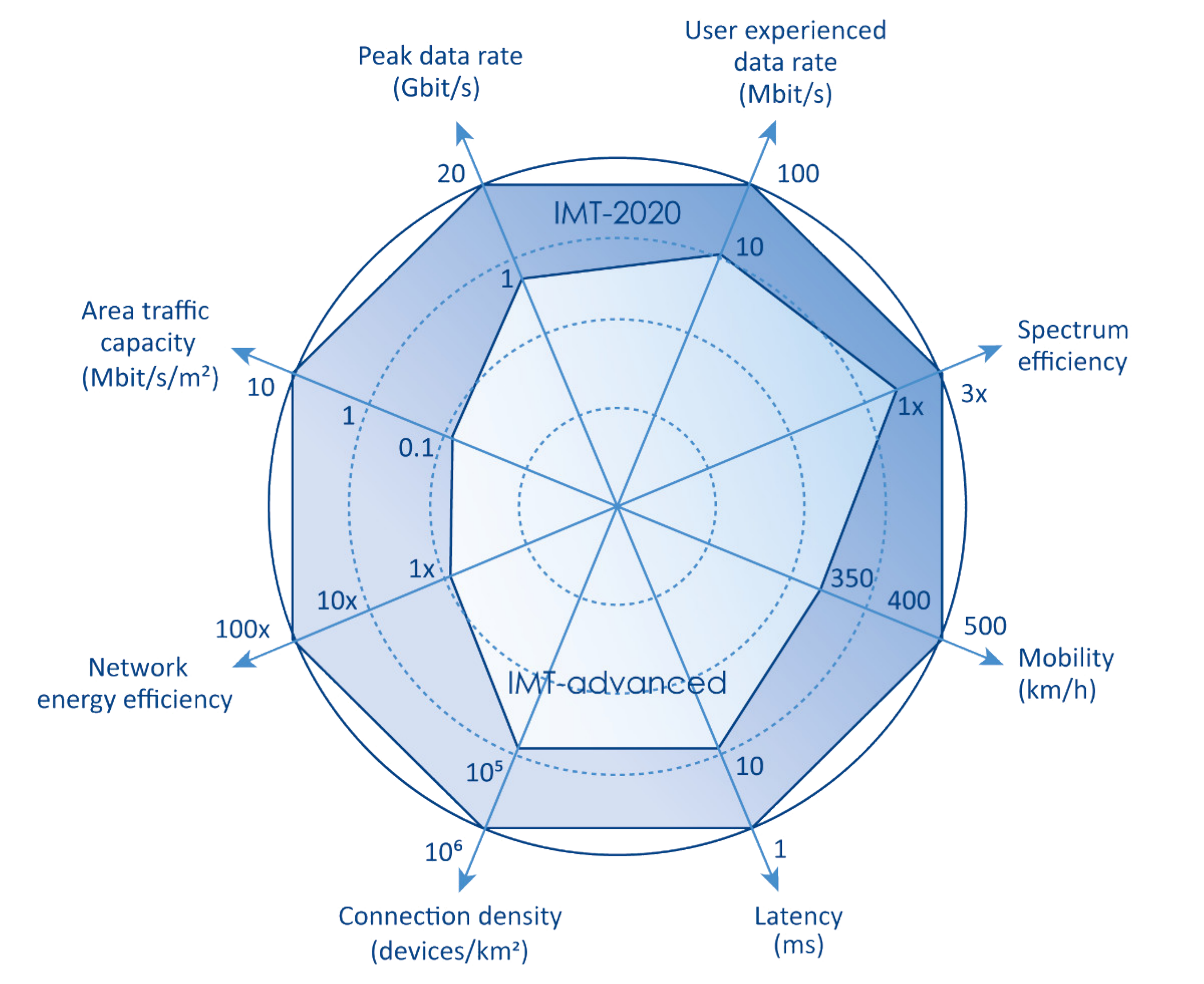}
\centering
\caption{IMT-Advanced (4th Generation) VS. IMT 2020 (5th Generation) key capabilities according to ITU-R M.2083}
\label{fig:4gvs5g}
\end{figure}

The \ac{itur} M.2083 \cite{ITU:M.2083-0} distinguishes three main usage scenarios for this technology: \ac{embb} \cite{5G:uRLLC_eMBB}, \ac{mmtc} \cite{5G:mMTC} and \ac{urllc} \cite{5G:uRLLC_eMBB_2}; each of them with different requirements ranging from high data rates (+50Mbps) and extensive cell coverage (+10 Tbps/Km\textsuperscript{2}), to low delays (\textless5ms) or low data transmission rates. 5G is a complex technology that requires a flexible architecture that provides the possibility to implement all the previously-mentioned requirements through the usage of novel technologies e.g., \ac{mimo} \cite{5G:MIMO}, \ac{ns}, \ac{nfv}, \ac{sdn} and, potentially, \ac{hsdn}. This subsection will survey the  use of the \ac{hsdn} paradigm over these architectures and study the impact on its key-technology enablers (\ac{nfv}, \ac{ns} and \ac{mec} \cite{ETSI:MEC}).

\subsubsection{Architectures and technologies}

\begin{figure}[ht]
\includegraphics[scale=0.75]{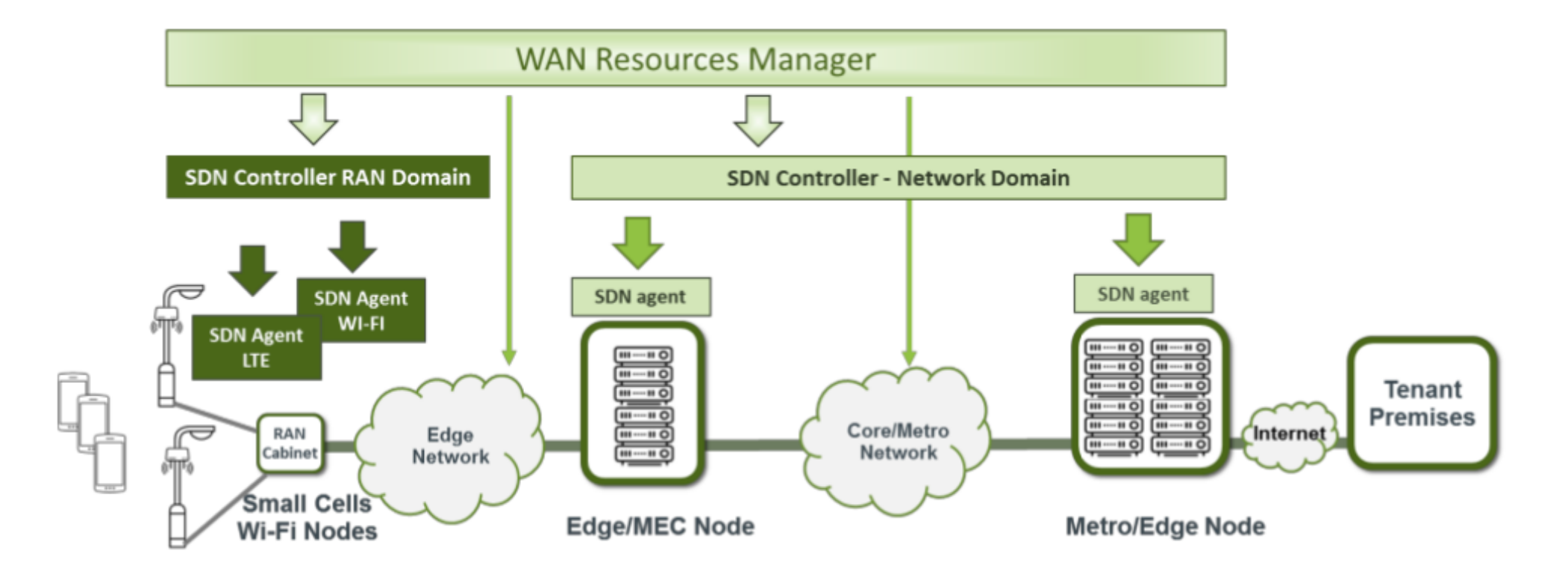}
\centering
\caption{\ac{5gppp} \ac{sdn} Architecture for the data plane Programmability in the 5G Architecture (Source \cite{5G:5gppp_Arch}).}
\label{fig:5gppp_sdn_arch}
\end{figure}

As mentioned in previous sections, \ac{sdn} proposes the separation of the control and data planes enabling a centralized control of all the data flows and paths inside a network domain, that is, the fundamental role of the CUPS concept introduced by the \ac{3gpp} TS 129.244 \cite{5G:3gpp_CUPS} for \ac{lte}. This straight relation between the \ac{sdn} technology basis and the CUPS concept was one of the main triggers in order to stablish \ac{sdn} as a key-enabling technology for 5G. The standardization of the \ac{sdn} architecture was raised by the \ac{onf} firstly in June 2014 in \ac{tr}-502 \cite{ONF:TR-502}. It has evolved throughout different \acp{tr} until the current definition has been achieved, which is divided in various \acp{tr}: \ac{tr}-518 \cite{ONF:TR-518}, \ac{tr}-521 \cite{ONF:TR-521} and \ac{tr}-522 \cite{ONF:TR-522}. The \ac{onf} released the aforementioned architecture and several 5G-related \acp{sdo} took it as an input for the construction of the 5G architecture. In fact, it is worth to mention that the overall 5G technology evolution is flourishing around the definitions and standards of multiple \acp{sdo} e.g., \ac{3gpp}, \ac{5gppp}, \ac{etsi}, \ac{ietf}, \ac{ngmn}, etc. This combination is generating several benefits, since never before in the history of mobile communications have so many \acp{sdo} get involved around the same technology; however, there are also some disadvantages derived from this approach which lead to collisions between the definitions of 5G concepts and to a slower developing road-map of the overall technology.\\
As a result of this conjunction of standards and definitions sources, there are numerous proposed 5G architecture recommendations with a strong inter-dependence of the involved technologies. The leading 5G architecture is the one defined by the \ac{5gppp} in \cite{5G:5gppp_Arch}. It is based on the 5G architecture defined by the \ac{3gpp} in the TS 23.501 \cite{5G:3gpp_Arch} but enhancing it with the addition of key-enabler technologies: \ac{sdn}, \ac{nfv} and \acf{ns}. It explores three levels of network programmability through the usage of the \ac{sdn} technology: the data plane, Transport Network and Network Function in \ac{ran}. The data plane programmability is the most dependent on \ac{sdn} features as it requires to be fully integrated with the \ac{mano} plane and agnostic to the underlying \ac{hw} infrastructure. Figure \ref{fig:5gppp_sdn_arch} shows the \ac{5gppp} proposed \ac{sdn} architecture for the data plane programmability. This architecture is divided in three layers: \textit{(1)} \ac{wan} Resource Manager, the \ac{sdn} Application in charge of triggering  the \ac{sdn} control plane operations and translating the orchestrator external link information to network domain-specific paths; \textit{(2)} \ac{sdn} Controllers, allocated in the network and \ac{ran} domains (each of them supported by \ac{sdn} agents on the respective domain); and \textit{(3)} the data plane, comprised of \ac{lte} small cells, Core and Edge \ac{nfvi} and \ac{wlan} Access Points. Moreover, the advantages of introducing the \ac{sdn} solution in this technology helps in the interconnection of \acp{vnf} and Network Slice instances. However, this architecture was one of the first steps towards the integration of all these technologies inside a single architecture and is not fully \ac{sdn}-enabled (current \ac{sdn}/\ac{nfv} architectures integrate the \ac{sdn} controller in the \ac{vim} and derive the \ac{sdn} operation management to the \ac{nfvo}).

\acp{sdo} have developed several 5G architecture recommendations which allow the integration of these technologies e.g., \ac{onf}’s \ac{sdn} Architecture for \ac{ns} TR-526 \cite{ONF:TR-526}, \ac{etsi}-\ac{nfv} architecture for \ac{sdn} integration with \ac{nfv} \cite{ETSI:SDN_Revision} or \ac{etsi}-\ac{mec} modified architecture for \ac{nfv}/\ac{sdn} integration with \ac{mec} \cite{ETSI:MEC_NFV}. These architectures do not assume the \ac{hsdn} paradigm as a default but leave room for implementing the \ac{hsdn} approach instead of the pure \ac{sdn} vision.\\
The research community has made great advancements in this topic also, the first approaches were towards unifying \ac{sdn} and \ac{nfv} technologies in order to evolve the \ac{sdn}-agnostic architectures to fully \ac{sdn}-enabled architectures \cite{wood2015toward,hakiri2015leveraging,moyano2017user}. Another intermediate step before reaching 5G \ac{hsdn} architectures was adding to the previous bundle of technologies the \ac{ns} technology as explored by \cite{ordonez2017network} and, finally, integrating all those technologies with the current 5G mobile communications architecture proposed by the \ac{3gpp} (see Table \ref{Table:5G_SDN-Archs}):

\begin{table}[h]
\small
\caption{5G SDN/NFV Academic Architecture Proposals}
\centering
\begin{tabularx}{\textwidth}{| l | X |}
\hline
\textbf{Article}                                                 & \textbf{Brief Description}\\ 

\hline
\textit{Ordonez et al. \cite{ordonez2017network}}                           & Presents the key-enablers to allow the integration of \ac{ns} in the \ac{sdn} architecture proposed by the \ac{onf} \cite{ONF:TR-502} and theoretically demonstrate the what is needed and what may be supplied by \ac{nfv} in order to implement it.\\ 

\hline
\textit{Yousaf et al. \cite{yousaf2017nfv}}                            & Surveys the challenges and issues of the current 5G network requirements. Presents each technology advantages/disadvantages and explores several \ac{sw} solutions for the different elements that comprise the 5G network e.g., \ac{mano}, \ac{vim}, \ac{vnfm}, etc.\\ 

\hline
\textit{Sun et al. \cite{sun2015integrating}}                                & Articulates the challenges around novel technologies such as \ac{mimo}, mmWave, HetNet… and their impact in the overall 5G architecture. Besides, they develop a fully functional architecture that combines the \ac{nfv}, \ac{sdn}, 4G LTE and \ac{sdr} technologies.\\ 

\hline
\textit{Trivisonno et al. \cite{trivisonno2015sdn}}                             & Proposes a flexible 5G architecture focused on the challenges presented by \ac{v2v} and \ac{iot} scenarios that combines \ac{sdn} and \ac{nfv} with a \ac{mec} module. The proposed 4G data plane instantiation within the 5G architecture requires that the \ac{sdn} data plane acknowledges how to interact with legacy protocols; therefore, this is the first partial (lacks the \ac{ns} technology) 5G academic architecture towards the \ac{hsdn} paradigm.\\

\hline
\textit{Barakabitze et al. \cite{barakabitze2019novel}} & Introduces the concept of \ac{qoe}\textit{-Softwarization} through \ac{qoe}-aware \ac{sdn}-\ac{nfv} architecture for delivering multimedia services over 5G networks.\\ 

\hline
\textit{Bouras et al. \cite{bouras2017sdn}}      & Small survey about the combination of the \ac{nfv} and \ac{sdn} technologies in 5G environments.\\

\hline
\end{tabularx}
\label{Table:5G_SDN-Archs}
\end{table}

The final step is the constitution of fully \ac{hsdn} 5G architectures that integrate all the aforementioned technologies. Irshad et al. \cite{irshad2019hybrid} propose a 5G hybrid architecture based on optical wireless network units that aims to easily unify legacy technologies with 5G-enabled technologies all-at-once. To demonstrate the capabilities of the proposed architecture, they conduct several simulations using VMware, MiniNet \cite{Sim:mininet} and the FloodLight \cite{Ctrl:floodlight} \ac{sdn} controller. These demonstrations show that this theoretical architecture outperforms, at a data rate level, those architectures based on a multi-\ac{ran} approach. \\
A similar example can be found in \cite{tawbeh2017hybrid}. In this case, the main objective is to virtualize the \ac{lte} \ac{epc} using \ac{sdn} and \ac{nfv}. In this architecture, the combination of both technologies is applied on each gateway in order to find the optimal path towards different sets of bearers, with the same \ac{qos} class identifier without any consequence over the \ac{qos}. Finally, two instances of a direct application of the \ac{hsdn} paradigm in 5G architectures are \cite{zhang2018enabling} and \cite{nunez2015service}. The first of them surveys the current challenges of applying this combination of technologies to \ac{sfc} (another key-enabler in 5G fully-virtualized networks) and, the latter, presents a service-based \ac{hsdn} model that allows to efficiently manage wireless mesh back-hauls  and achieves higher performance if compared with canonical \ac{sdn} models (rates of x6 less delay).\\
In summary, these last architectures offer potential solutions to face 5G \acp{kpi} such as the interfacing between the \ac{3gpp} slicing management system and other management systems, such as the Transport Network, by introducing domain-specific \ac{sdn} controllers (wireless, optical, satellite...) which will  allocate  transport  resources  properly according  to  network  slices’ requirements (see \ref{fig:5gppp_sdn_arch}).

It is worth mentioning that there is a set of projects in the \ac{5gppp} H2020, Phase-2 and Phase-3, \ac{ict}-17 and \ac{ict}-19 EU programmes that aim at creating advanced 5G infrastructures and testbeds, running verticals experimentation trials and testing 5G \acp{kpi} in Europe. From these projects, two of them are in charge of generating the 5G infrastructure that will be used by \ac{5gppp} Phase-3:Part 3 projects, such as 5G-Growth \cite{5G:Growth}, 5G-Solutions \cite{5G:Solutions}, 5G-Heart \cite{5G:Heart}. To validate their specific  vertical use cases, those are: 5G-VINNI \cite{5G:VINNI} and 5G-EVE \cite{5G:EVE}. The first one proposes several architectural principles for 5G architectures that are implemented on each of the main facility sites \cite{5G:VINNI_d1.1}, which may be migrated to the \ac{hsdn} paradigm. On the other hand, 5G-EVE also proposes multiple architectures \cite{5G:EVE_d2.1} for the main facility sites in order to be \ac{3gpp} Release-16 compliant, which implies the usage of OpenDayligh \cite{medved2014opendaylight} as the \ac{sdn} controller.

\subsection{Cloud and Data Center Networking}
\label{subsec:Emergent_CC_DCN}

Before the 2010s, the provisioning of computing and networking resources was based on the acquisition of large amounts of \ac{hw} by application providers in order to create their own \ac{dc}. The main drawbacks of this approach are the huge costs of buying and maintaining the whole infrastructure and that it is not an easily-scalable or flexible method. The appearance of cloud computing has introduced a shift to a new paradigm, the cloud utility model \cite{Cloud_Comp}, where application providers do not need to buy and maintain these expensive infrastructures but, instead, are able to lease computing resources with just a couple of clicks and a lesser upfront investment. This section analyzes the current status and impact of the \ac{sdn} and \ac{hsdn} paradigms on cloud computing environments. 

\noindent Currently, cloud computing supports three service delivery models which are usually categorized as follows: 
\begin{enumerate}
	\item \textbf{\ac{saas}}: Provides any kind of \ac{sw} as a cloud solution to cloud customers e.g., e-mail services, Virtual Private Network(\acs{vpn}) services, schedulers, etc. 
	\item \textbf{\ac{paas}}: Provides a specific platform so that cloud customers are able to develop multiple operations and manage applications without the underlying complexity of maintaining the required infrastructure for developing an application.
	\item \textbf{\ac{iaas}}: Provides access to a completely virtualized infrastructure such as \acp{vm}, \acp{sdr} and any kind of computing or networking resource. 
\end{enumerate}

To provide these services, cloud providers must operate several \acp{dc} among multiple locations around the globe. These \acp{dc} are comprised of a humongous quantity of switches and servers providing response to application requests through the division of their computing resources in \acp{vm}. Therefore, the \ac{dcn} is usually designed is such a way that the interchange of these amounts of traffic between the \acp{vm} and containers is done in the most efficient and optimal possible way; however, legacy networks offer a poor solution to these highly complex scenarios as they are not easy to manage and scale. The \ac{sdn} technology provides a clear separation of the control and data planes with a centralized control logic and, thus, with a global vision of the overall status of the network that enables a flexible, programmable and dynamic solution to the management and scalability issue of the \acp{dcn}. There are relevant cloud providers that have already adopted this approach, for instance Google and Microsoft (see \textit{Section \ref{subsec:sd-branch} \ac{sdwan} and Branch}).

\begin{figure}[ht]
\includegraphics[scale=0.71]{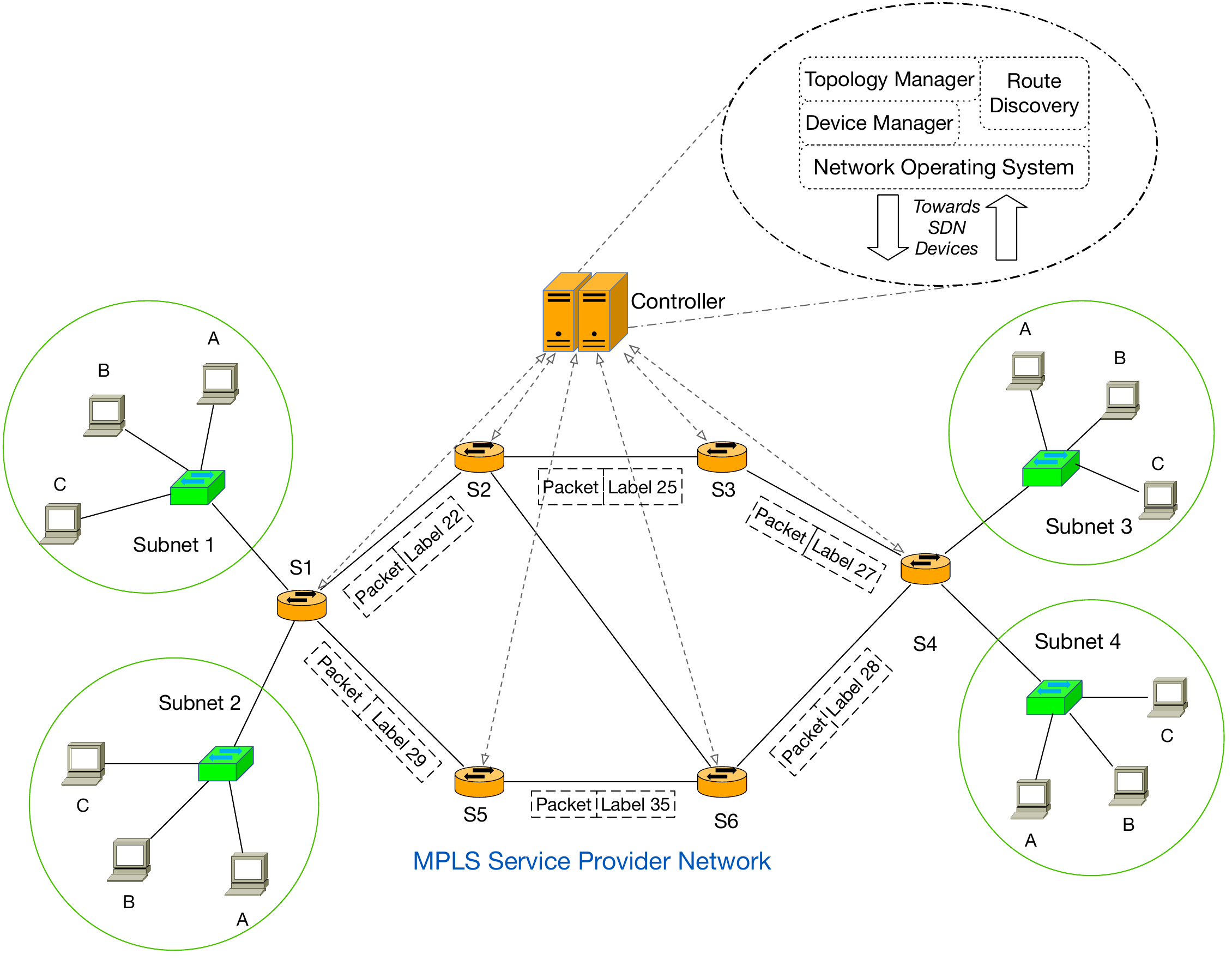}
\centering
\caption{Energy efficient \ac{hsdn} architecture for \acp{dcn}. Edited from source: \cite{paliwal2019effective}.}
\label{fig:HSDN_MPLS}
\end{figure}

Whereas Cloud Computing, \acp{dcn} and \ac{sdn} are highly researched topics,  (see Table \ref{Table:CC_Surveys}) there are still few published articles that make a specific mention or usage of the \ac{hsdn} paradigm. Chekired et al. propose ScalCon \cite{chekired2018hybrid}, an \ac{hsdn} architecture with a decentralized hierarchical control plane focused at efficiently managing \acp{dcn} with multiple \ac{sdn} controllers. This architecture is similar to ParaCon \cite{qiu2017paracon}, as it is shown bellow, but with some remarkable differences: ScalCon contemplates three different levels for the control plane which may be governed by one or more \ac{sdn} controllers (Fog layer, Edge layer and Cloud layer); in ParaCon, every controller has links associations with all the switches within the network, ScalCon only allocates links associations with the switches within the corresponding cluster of a layer; and, finally, ScalCon utilizes two asynchronous algorithms with a novel multi-priority queuing model, instead of the one used in ParaCon, in order to improve the performance for path computation. Besides, a comparison between both architectures and \ac{sdn} controllers ,ONOS \cite{Ctrl:ONOS} and POX \cite{kaur2014network}, is driven at the end of the article (Section IV) which demonstrates that ScalCon outperforms all the previously mentioned schemes in terms of path computation latency, transmission overhead, and end-to-end delay.\\
Paliwal et al. \cite{paliwal2019effective} have created an energy efficient resource management technique intended for those \acp{dc} where the traffic patterns are highly correlated to the traffic demand over sets of operational hours. This approach aims at enhancing the advantages of combining \ac{hsdn} with \ac{mpls} to conform real-time dynamic provision of network elements and energy saving in the \ac{dcn} (see Figure \ref{fig:HSDN_MPLS} for the architectural view). The core network of the proposed architecture is composed of hybrid switches that are able to interpret both \ac{mpls} tags and \ac{of} protocol rules; besides, the Network administrator module interacts with the \ac{sdn} controller through a Northbound interface developed with \ac{rest} \acp{api}. The method is based on switching specific network elements with small loads off and re-allocating the traffic to an alternate active network element taking into account the traffic demand. Simulations show that energy consumption is reduced approximately a 77\% in \textit{Fat-Tree and BCube-like} \acp{dcn}.

\begin{table}[h]
\small
\caption{SDN \& Cloud Computing surveys }
\centering
\begin{tabularx}{\textwidth}{| l | X |}
\hline
\textbf{Survey}                                                 & \textbf{Brief Description}\\ 

\hline
\textit{Wang et al. \cite{wang2015survey}}                           & Performs an exhaustive exploration of the current impact of different cloud computing architectures depending on the physical \ac{dcn} topology, the virtualized infrastructure management and monitoring, and the \ac{dcn} routing. Moreover, a profound analysis is driven towards the capabilities of different kind of \acp{dc}: optical, electrical, wireless, \ac{sdn}-based… It mentions the unique opportunity that the \ac{hsdn} paradigm brings to \ac{sdn}-based \acp{dc}: reducing deployment costs and overcoming the \ac{sdn} techniques issues to provide optimized \ac{qos} on demand in the cloud.\\ 

\hline
\textit{Son  et al. \cite{son2018taxonomy}}                            & Provides a detailed taxonomy of the state of art of both \ac{sdn} and cloud computing. The authors address a set of topics that range from \ac{sdn}-enabled cloud computing architectures to energy-efficient \acp{dcn} designs and \ac{sdn}-based \ac{qos} network management methods and policies.\\ 

\hline
\textit{Jararweh  et al. \cite{jararweh2016software}}                                & Presents a comprehensive survey that aims at exploring the existing Software-Define Clouds and its comprising subsystems (security, storage, networking…). Besides, the authors develop a \ac{poc} towards a fully automated software-based control framework for cloud computing environments. \\ 

\hline
\textit{Toosi et al. \cite{toosi2014interconnected}}                             & Mainly focused on the issues that arise, from a cloud-provider perspective, when interconnecting multiple \acp{dc}. It describes and provides several solutions for the interoperability scenarios that are generated under the above-mentioned circumstances multiple \acp{dc}. \\

\hline
\textit{Jain et al. \cite{jain2013network}} & Brief survey that analyses the benefits and drawbacks of \ac{sdn}/\ac{nfv}-based \acp{dc}. Besides, it presents the initial development and capabilities of OpenADN,  an \ac{sdn}-based network application focused on application partitioning and delivery in  multi-cloud scenarios. \\ 

\hline
\end{tabularx}
\label{Table:CC_Surveys}
\end{table}

There are some other minor examples of the \ac{hsdn} applied to Cloud Computing environments .The authors of \cite{hong2016incremental, khorsandroo2017experimental} present a model for the incremental deployment of \ac{sdn} in ISP and \ac{dc} networks. This model combines \ac{sdn}-enabled switches with upgraded legacy switches that are able to interact with the OpenFlow protocol in such a manner that it effectively reduces \ac{mlu} for congestion mitigation with only 20\% deployment of \ac{sdn} switches. \cite{guo2014traffic} and \cite{salsano2014} introduce, respectively, a \ac{te} view for \ac{sdn}/\ac{ospf} hybrid network architectures where some \ac{ospf} flows and weights are modified from the centralized \ac{sdn} controller. Besides, a hybrid IP/\ac{sdn} architecture known as \textit{“Source Hybrid IP/SDN (OSHI)” \ref{fig:class-based}} is also described. This hybrid network architecture implements IP routing with the \ac{sdn} forwarding control in the core of the network.

In conclusion, Cloud Computing and \acp{dcn} are crucial pillars of the current technological evolution towards \textit{"Everything as a Service"} and the necessity of developing networks that are highly flexible and scalable. In order to achieve these goals but without the burden of huge investments, there is an existing dependence between these technologies and the inclusion of the \ac{hsdn} paradigm in them. \ac{hsdn} applied in \acp{dcn} academic and research projects are starting to emerge as it is a major research area that is even being exploited by the industry (see section \ref{subsec:sd-branch}).

\subsection{Internet of Things}
\label{subsec:IoT}

The concept of \ac{iot} was born in 1999 by Kevin Ashton \cite{IoT:Ashton} in the context of supply chain management. Since then, as other technologies evolved, so did the concept of \ac{iot} and, currently, it refers to a network of interconnected devices that gathers environmental information through the interaction with its environment  and even uses the existing Internet Services to generate novel communication and application services. \\
According to the study carried by \textit{IoT Analytics} \cite{IoT:Analytics}, in 2011 the number of interconnected devices was greater than the number human beings in the planet. As of writing, in 2020, over \(21.2*10^9\) devices exist (46\% of them are IoT devices – non mobiles phones, laptops, servers, etc.) coexist with the human race and it is expected that more \(32.4*10^9\) devices will be used by 2025.. As it can be seen, the expected expansion and evolution of this technology has surpassed its initial goals, gathering multiple benefits and advantages that come attached within this technology. In order to achieve those advantages, \ac{iot} devices interact with back-end systems (cloud, edge and fog servers…) that process the volumes of data generated by the sensing activities. A critical aspect upon this technology is the implementation of new architectures that are able to adapt the overall network capabilities to the requirements of its inherent extremely dynamic environments e.g.,  highly scalable, avoidance of new potential attack surfaces and vectors, network programmability, high security for e-Health and Industry 4.0 devices due to potential data flaws, etc. \\
Several works have been created around \ac{iot} architectures: \cite{IoT:Survey_SDN_Sec,IoT:Sample_Arch_1,IoT:Sample_Arch_2,IoT:Sample_Arch_3,IoT:Sample_Arch_4}. These architectures are characterized by the softwarization and virtualization of the network and, consequently, \ac{sdn} and \ac{nfv} technologies are becoming key enablers for the generation of this second phase of evolved \ac{iot} architectures. This section gives an \textit{“au courant”} status of the \ac{iot} deployments that implement the \ac{hsdn} paradigm.

Pure \ac{sdn} integration and deployments are being vastly researched and implemented within \ac{iot} scenarios to achieve the above-mentioned upgrades over common \ac{iot} networks and other applications. Fichera et al. \cite{IoT:5G_1} and Tello-Oquendo et al. \cite{IoT:5G_2} are two examples of the usability of \ac{sdn} in 5G \ac{iot} networks. Aside from these theoretical architectures, there are lots of experimental implementations which try to solve \ac{iot} security flaws. \cite{IoT:Sec_1,IoT:Sec_2,IoT:Sec_3} are clear examples of frameworks that exploit \ac{sdn} capabilities to achieve  traffic anomalies detection and mitigation in \ac{iot} cloud networks, secure cluster formation for large \ac{iot} grids, \ac{mitm} attack prevention in fog-based \ac{iot} scenarios and security improvement through a flow-based SD-GW method, respectively.

Notwithstanding, some researches are leading the approach to the \ac{hsdn} paradigm in the \ac{iot} realm. On first place, Bendouda et al.\cite{IoT:HSDN_1} explore a partial approach towards a complete \ac{hsdn} \ac{iot} deployment, proposing a \ac{hsdn} architecture that integrates both, a centralized and a semi-distributed approach for the control plane. The “hybrid” concept of this proposal is directly related with the division of the pure \ac{sdn} centralized control plane into three layers: \textit{(1)} Main Controller, has a view of the global network status and coordinates lower-level controllers; : \textit{(2)} Secondary Controller, acts as an entry point to address network constraints related to wireless technologies and routing processes; \textit{(3)} Local Controller, lower layer controller, its main role is to manage and interchange control packages with the \ac{iot} island gateway (these nodes are selected by the \textit{Connected Dominating Set} algorithm presented in the same article). The resulting architecture is capable of auto-selecting the level-3 local controllers depending on multiple network parameters (distance from gateway, average link quality, etc.) and therefore obtains remarkable performance results if compared with traditional one-layer controller-plane \ac{sdn} approaches.\\
Lee et al. \cite{IoT:HSDN_2} discover that the \ac{hsdn} paradigm may be a good model for large \ac{iot} architectures (more than 34 connected devices) if compared with an efficient set of \ac{te}, load-balancing and flow management policies in order to reduce the delay and upgrade the throughput of the overall network. The implementation of a \ac{hsdn} scenario was developed due to the traditional throwbacks of pure \ac{sdn} scenarios e.g scalability, availability, reliability and cost. This architecture is comprised of two kind of networks: Distributed Networks, composed of legacy devices that support \ac{ospf} and \ac{sdn}-enabled networks, composed by hybrid or \ac{sdn} devices. Additionally, those networks are populated with three different types of devices: legacy, hybrid (run IP routing-based protocol stack in the Distributed Network and \ac{of} forwarding-based stack in \ac{sdn}) and \ac{sdn}. They modify the idea of a basic forwarding table from \cite{IoT:6900880} and apply it to the hybrid nodes using \ac{tcam} tables for \ac{of}-based routing entries and Static Random-Access Memory (RAM) tables for IP-based routing entries and emulate the architecture using MiniNet \cite{Sim:mininet} with the Floodlight controller \cite{Ctrl:floodlight}.\\
Saadeh et al. \cite{IoT:HSDN_3}, on the other hand, propose a \ac{hsdn} architecture that combines \ac{icn}  and \ac{sdn}. The proposed architecture aims to expand PPUSTMAN architecture \cite{saadeh2018ppustman}  (see Zhang et al. survey \cite{IoT:ICN_Survey} for detailed \ac{icn}-\ac{sdn} architectures information) by separating the controlling function into three controlling planes: Operational, Tactical, and Strategic planes. Additionally, each of those planes is composed of three controlling units: Control, View, and Model. This project delves into the advantages of joining both technologies in \ac{iot} scenarios:

\begin{enumerate}
	\item \textit{Enhancing the performance of the controllers} 
	\item \textit{Generation of more abstraction levels for node interactions} 
	\item \textit{\acp{icn} easier deployments} 
	\item \textit{Better inter-paradigm operability} 
	\item \textit{Decrease in traffic volume using in-network caches for SDN actions publication} 
\end{enumerate}

Lastly, technologies similar to \ac{iot} such as \ac{iov} \cite{IoT:IoV_Def} are starting to implement the \ac{hsdn} paradigm. \cite{IoT:IoV_1,IoT:IoV_2} present a clear demonstration of this path towards the research of \ac{hsdn} in \ac{iov} environments.\\
Once again, as with previous technologies, the \ac{iot} realm is not an exception and there is still a lot of work to be done in order to have production-level \ac{iot} networks with a full \ac{hsdn} implementation. However, the first projects combining both technologies are starting to arise and its advantages being discovered.

\subsection{Others}
\subsubsection{Blockchain}

Since the implementation of the Blockchain Technology in 2009 by a person (or group of people) with the pseudonym of “Satoshi Nakamoto” \cite{nakamoto2019bitcoin} for the renamed cryptocurrency Bitcoin, this technology has been applied to numerous applications other than pure cryptocurrencies \cite{BC:Kudos,BC:Health,BC:Health,BC:IoT} with great benefits to the scientific community due to its four main characteristics: decentralization, persistency, anonymity and auditability. In the field of legacy and \ac{sdn} networks, the most important of these characteristics might be the decentralization as it avoids traffic and security bottlenecks in high density networks such as \ac{iot} environments, \ac{mmtc} scenarios, \acp{dcn}. But, at the same time, this technology has a major throwback which is crucial for all of these networks, \textit{Scalability}. Public Blockchain may be addressed as a public distributed ledger or database where event records (transactions) are stored in a chain of blocks that continuously grows as soon as new block are added. Therefore, the blockchain tends to be very heavy if the amount of transactions per day is high which, as a direct consequence, makes this technology not that suitable for large networks.\\
Fortunately, at this point is where the \ac{sdn} and the \ac{hsdn} technologies appear as a pillar that helps to manage this issue in these kind of networks as, combined with Blockchain, they are capable of providing the flexibility, programmability and scalability required by them. In fact, almost all of the papers reviewed in this survey about this topic are related to \ac{iot} networks, vehicular networks and vehicle to grid scenarios. This section surveys the current research status and impact of this technology in \ac{hsdn} scenarios.

In \cite{BC:Survey_IoT}, T.M. Fernández-Carames and P. Fraga-Lamas carry-out a research on the applications of Blockchain in \ac{iot} architectures (BIoT) and they reference two articles where Blockchain capabilities are upgraded using \ac{sdn} to control fog nodes. These novel architectures are able to deploy distributed fog nodes with a central \ac{sdn} controller that enables a huge reduction of the overall delay of the network, greatly increases the throughput and balances the load when flooding attacks are launched against them. As mentioned before, the combination of these innovative technologies leads to impressive architectures that overcome the disadvantages of each of these technologies, if implemented independently e.g., Kumar Sharma et al. \cite{BC:IoT_Arch_1} propose a distributed \ac{sdn} fog node-based cloud architecture for \ac{iot}, comprised of three layers (device, fog and cloud) that achieves high resiliency, scalability and availability  with full decentralization and true redundancy if compared with traditional \ac{iot} networks. The key enablers of that architecture are each of the fog nodes which, basically, are small \ac{sdn}/\ac{hsdn} controlled blockchain-based distributed networks that are responsible of service delivery, data analytics of each of the device-layer isles and sending the already-processed data back to the device-layer or to the cloud layer if needed.\newline
A similar work called DistBlockNet is proposed by almost the same authors in \cite{BC:IoT_Arch_2}, the main difference between these two projects is that DistBlockNet is conceived as a large horizontal single-layered distributed mesh of \ac{sdn} controllers, that are able to communicate with each other thanks to the Blockchain technology, instead of a three-layered vertical architecture. Moreover, this architecture is more focused on solving legacy \ac{iot} network security issues (reducing the attack window, threat prevention, data protection, mitigation of cache poisoning/\ac{arp} spoofing, \ac{ddos}/\ac{dos} attacks) than on optimizing the overall performance of the network. In any case, although both architectures are presented as pure \ac{sdn}-controlled architectures, due to their characteristics, a \ac{hsdn} approach could be implemented as far as the deployed legacy devices do not collide with the generated Blockchain algorithms.\\
More recent proposals have driven to the application of this novel combination of technologies into the field of energy trading in vehicular networks. \cite{BC:BEST} presents BEST which is a dedicated Blockchain-based secure energy trading scheme for electric vehicles (EVs) where the blockchain algorithm is applied to validate EVs' requests in a distributed manner and where \ac{sdn} is utilized to pass the verified EVs' requests to a global \ac{sdn} controller in a flexible, scalable and programmable way. This proposal includes some miner nodes that are the ones in charge of carrying out the validation operation based on several factors (dynamic pricing, connectivity records, etc.).  Furthermore, Jindal et al. present SURVIVOR \cite{BC:SURVIVOR}, a blockchain based edge as a service framework for secure energy trading in \ac{sdn}-enabled Vehicle to Grid environments, that merges \ac{sdn} and Blockchain to achieve a framework with a low-latency network backbone and secure energy transactions based on the proof-of-work method. Once again both frameworks do not collide with the \ac{hsdn} paradigm. Additionally, \cite{BC:VANETS} and \cite{BC:Fog_5G} present two projects where these technologies are combined with further technologies such as 5G and vehicular ad-hoc Networks (VANETS) but, in this case, it is not that clear if an \ac{hsdn} approach can be added.

In summary, Blockchain is a promising technology that all-along has lots of advantages but that, if combined with \ac{sdn} and \ac{hsdn} technologies, it enables great advantages if applied in large-scale and dense networks. However, it is still a novel technology if applied within the \ac{hsdn} paradigm and, therefore, there are not many finished studies that address this topic.

\subsubsection{Software-Defined WAN and Branch}\label{subsec:sd-branch}

\begin{figure}[h]
\includegraphics[scale=0.72]{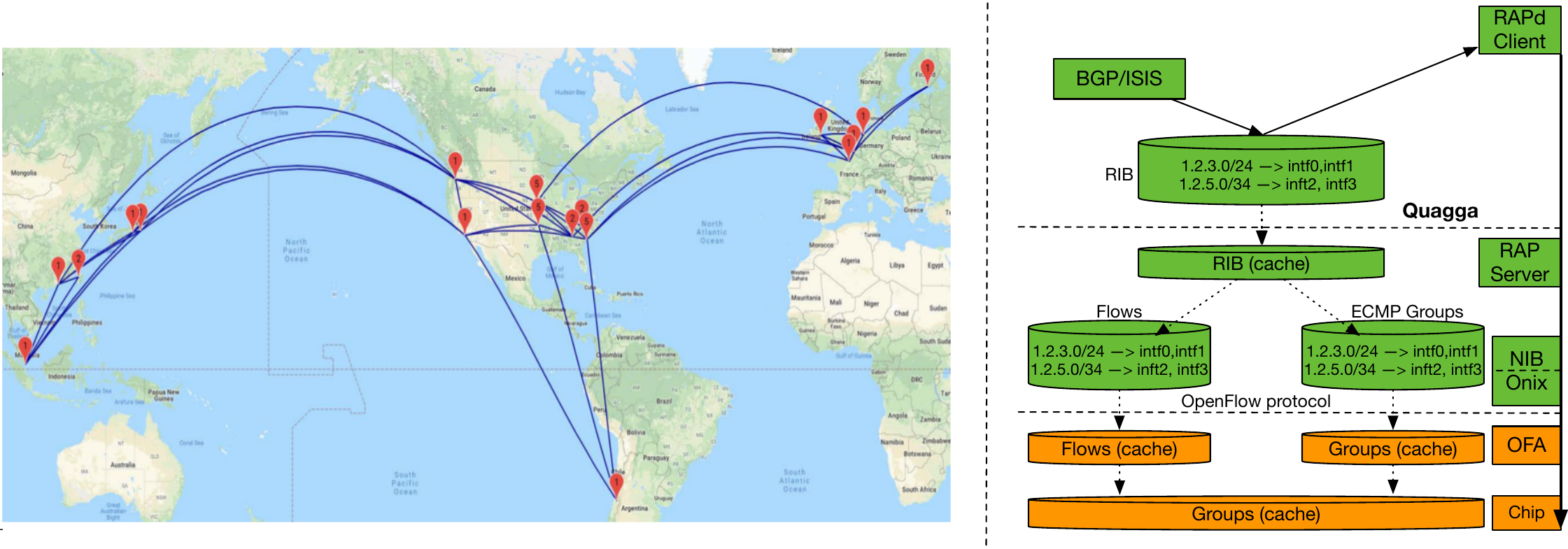}
\centering
\caption{B4 Google's Hybrid SD-WAN. (a) B4 Topology as of January, 2018; (b) OpenFlow Integration with Legacy Routing Protocols. Edited from Sources \cite{Branch:B4_After,Branch:B4}}
\label{fig:B4}
\end{figure}

\acp{wan} are in charge of providing connectivity between cloud services, \ac{saas} applications, \acp{dc} and branch locations. In the past years, \ac{saas} and cloud providers have been heavily investing in high-speed peering connections close to the enterprise \ac{wan} edge and, consequently, broadband bandwidths have raised considerably. Therefore, the overall cost of maintaining a traditional \ac{wan} with legacy protocols such as \ac{mpls}, \ac{bgp} and \ac{ospf} has greatly increased. Due to these reasons, many enterprises are migrating to a novel approach, namely, \acf{sdwan}, that allows them to highly reduce costs, augment the network flexibility and programmability and even extend existing \ac{mpls} bandwidth at branches by adding broadband connections and using \ac{pbr}. This section surveys the integration of the \ac{hsdn} paradigm in \acp{sdwan} and \ac{sdbranch}.\hfill \break
There exist two major and commonly known \ac{sdwan} deployments implemented with the \ac{hsdn} paradigm: Google’s B4 \cite{Branch:B4} and Microsoft SWAN \cite{Branch:B4_After}.\hfill \break
B4 \cite{Branch:B4} is Google’s inter-\ac{dc} \ac{sdwan} solution. It aims at customizing a self-defined \ac{hsdn} \ac{sdwan} architecture that fulfils Google’s \acp{dc} unique requirements with a two-level hierarchical control framework: \textit{(1)} the upper layer allocates a logically centralized \ac{te} server that looks over high-level \ac{te} policies and; \textit{(2)} the lower layer is a network controller located in each of the \acp{dc} that hosts and manages local control applications and site-level traffic. Google took several design decisions before implementing this massive \ac{sdwan} ranging from developing and building their own routers based on merchant switch silicon, to centralized \ac{te} and links with 100\% utilization rates. As stated in \cite[section 3.4]{Branch:B4} integrating both legacy and OpenFlow-based routing protocols, was one of the critical challenges in order to support \ac{hsdn} network deployments, in fact Google needed to develop a Routing Application Proxy (RAP) that allowed the selected routing engine, \textit{Quagga} \cite{Quagga}, to communicate and interact with the OpenFlow switches (\ac{bgp}/\ac{isis} route updates, routing-protocol packets flowing between switches and Quagga and interface updates from the switches to Quagga). Figure \ref{fig:B4} (b) shows a more detailed view of this procedure.\\
B4 is a clear example of the advantages of the migration to the \ac{hsdn} paradigm and of how it can serve to implement real global-level solutions, Hong et al. analyze in \cite{Branch:B4_After} the evolution of Google’s \ac{hsdn} \ac{wan} since its beginning up to 2018, when B4 was already comprised of 33 sites (see Figure \ref{fig:B4}) and supports services with a  required availability of 99.99\% and with links utilization rates near to 100\%.

SWAN \cite{Branch:SWAN} is Microsoft’s \ac{sdwan} inter-\ac{dc} centralized traffic engineering solution. It is primarily focused on solving the shortcomings of today’s most used \ac{te} practices with \ac{mpls} in \acp{wan}: \textit{(1)} Poor Efficiency, low volumes of traffic are carried if compared with the total capacity of the links; this leads to a poor link utilization rate (e.g., the average utilization of half the links is under 30\% and of three in four links is under 50\% in an average production inter-\ac{dc} \ac{wan}); \textit{(2)}, poor sharing, in traditional inter-\ac{dc} \acp{wan} there is a limited support for flexibly resource allocation. To achieve these goals, SWAN comprises a global network view implementing network agents with the FloodLight OpenFlow controller \cite{Ctrl:floodlight}, which helps to find globally optimized bandwidth path assignments and uses fine-grained policy rules to carry more high-priority traffic (e.g., interactive traffic) while maintaining the overall fairness among services of the same class. As an example, SWAN establishes a 5-minute period before service brokers exchange data that allows to predict the forthcoming traffic and, therefore, use the theoretically “free” bandwidth for other types of traffic (in their paper, they distinguish between three types of services/traffic: interactive, elastic and background). Consequently, the remaining resources are used more efficiently and effectively. \\
In the initial design of SWAN, only \ac{of}-enabled Switches were used (Arista 7050Ts and IBM Blade G8264s), however, Hong et al. stated in their paper that \textit{“(…) We use OpenFlow switches, though any switch that permits direct programming of forwarding state (e.g., MPLS Explicit Route Objects [3]) may be used.”}. This statement leaves room to a \ac{hsdn} \ac{sdwan} implementation. Besides, the initial design was slightly smaller than Google’s approach, with just 5 \acp{dc} all around three continents (China, USA and EU); nonetheless, SWAN is able to carry 60\% more traffic if compared to \ac{mpls} \ac{te} link utilization rates and, likewise, only a 4\% of the flows deviate over 5\% from their fair share. In \ac{mpls} \ac{te}, 20\% of the flows deviate by that much, and the worst-case deviation is much higher. \\
The main differences between these two \ac{hsdn} \ac{sdwan} inter-\ac{dc} implementations draw from the premise that they are facing different challenges. B4 aims at developing custom switches and procedures to directly integrate legacy routing protocols with a pure \ac{sdn} environment, while SWAN develops mechanisms for the congestion-free data plane updates and for effectively using the limited forwarding table capacity of commodity switches. 

Besides, the appearance of \ac{iot} environments inside the enterprise branches, the early adoption of \ac{mmtc} changes and the shift of enterprise workloads to cloud providers has led to big changes in traditional traffic patterns and to \ac{vlan} sprawl and an increase in network complexity. These facts require a new concept beyond \ac{sdwan}, a concept that includes the enterprise branch routing, security and \acp{lan} as well. This concept is known as \ac{sdbranch}. 

The \ac{sdbranch} concept has not yet fully impacted in the academic community but, in the enterprise world there are a few solutions starting to arise \cite{Branch:Aruba,Branch:CISCO,Branch:FortiNet,Branch:Versa}. Aruba's \ac{sdbranch} solution \cite{Branch:Aruba} includes \ac{sdwan} gateways supporting broadband, \ac{lte} and \ac{mpls} \ac{wan} connections, and a cloud-based management \ac{gui} for \ac{wan} and security configuration. The branch security is enforced through a stateful firewall based on \ac{dpi}, IPSec \acp{vpn}, Web content classification and reputation and cloud security integration, among others. Cisco’s Network Function Virtualization Infrastructure Software (NFVIS) \cite{Branch:CISCO} is a full operating system for the \ac{sdbranch}. NFVIS provides data flow acceleration for the \acp{vnf} used to implement routing, \ac{sdwan} and security branch services. Fortinet's Secure \ac{sdbranch} \cite{Branch:FortiNet} is focused on providing a secure and manageable branch for enterprises. It integrates several security-specific modules such as firewall, secure wireless access points, secure Ethernet and network access control.  Versa's \ac{sdbranch} \cite{Branch:Versa} relies on the FlexVNF operating system to execute routing, \ac{sdwan} and security functions on Versa boxes (white boxes are also supported). The proposed architecture enables the deployment of automated services for \ac{wan} and branch scenarios.\\
In summary, the SD-Branch concept is still young and needs to be adopted and researched by the academic community as a promising technology for future flexible networks. Moreover, B4 and SWAN represent a clear path of migration from legacy \acp{wan} to hybrid \acp{sdwan} and they demonstrate the capabilities and advantages of the \ac{hsdn} paradigm.

\section{Simulation Tools and Testbeds}\label{sec:simulation-testbed}
As in any other technological domain, researchers looking into the hybrid \ac{sdn} paradigm require Simulation Tools, Emulation Tools and testbeds that allow them to evaluate several \ac{hsdn} scenarios and its efficiency and behavior over specific conditions. If pure \ac{sdn} experiments are to be carried-out, the most commonly known tools/platforms are: Mininet, EstiNet, NS-3 and, more recently, MiniNEXT \cite{Sim:mininet, Sim:estinet, Sim:ns-3, Sim:mininext}. The most used platform is MiniNet as it implements a lightweight kernel that enables the creation of multiple custom network topologies that integrate diverse devices over a simple laptop. MiniNet defines its \ac{sw} as a “network emulation orchestration system” that orchestrates several Open vSwitch \cite{openvswitch} instances and connects them as defined by the user.\\
Nonetheless, from a \ac{hsdn} perspective, these tools lack some crucial functionalities that are key-enablers in order to be able to simulate a hybrid legacy and \ac{sdn} scenario. Rashid et al. analyze some of these challenges in their \ac{hsdn} survey \cite{Surveys:HSDN_approaches}, stating that the main lacking functionalities are related to the implementation of the required protocols for the communication between the \ac{sdn} controller and the legacy devices and to the network policy implementation. Moreover, Habib et al. \cite{TestBeds:SDNetkit} does an extensive research and comparison over the existing \ac{sdn} simulation tools and testbeds and classify them into three categories depending on how the solution is deployed/implemented: centralized, distributed and remote testbeds.
This subsection surveys the state of art in \ac{sdn}/\ac{hsdn} simulation tools and testbeds. 

\par

\begin{table}[h]
\centering
\caption{Simulation Tools and Testbeds Comparison}
\resizebox{\linewidth}{!}{%
\begin{tabular}{|c|c|c|c|c|c|c|c|c|} 
\hline
\textbf{Platform}                     & \textbf{Family} & \textbf{First Release Year} & \textbf{Open Source} & \textbf{Architecture} & \textbf{Type}                                                                   & \textbf{Legacy Protocols Support} & \textbf{Virtualization Technology}                                             & \textbf{Scalability}  \\ 
\hline
\textbf{\textit{MiniNet} \cite{Sim:mininet}}             & Simulation Tool & 2010                        & \checkmark            & Virtual               & Centralized                                                                     & None                            & KVM                                                                            & Low                   \\ 
\hline
\textbf{\textit{MiniNet CE} \cite{Sim:mininet_ce}}          & Simulation Tool & 2013                        & \checkmark            & Virtual               & Centralized                                                                     & None                              & KVM                                                                            & Medium                \\ 
\hline
\textbf{\textit{Distributed MiniNet} \cite{Sim:mininet_dist}} & Simulation Tool & 2013                        & \checkmark            & Virtual               & Distributed                                                                     & None                              & KVM                                                                            & High                  \\ 
\hline
\textbf{\textit{EstiNet} \cite{Sim:estinet}}             & Simulation Tool & 2013                        & x           & Virtual               & Centralized                                                                     & Low                              & \begin{tabular}[c]{@{}c@{}}None (kernel re-entering~\\simulation)\end{tabular} & High                  \\ 
\hline
\textbf{\textit{NS-3} \cite{Sim:ns-3}}                & Simulation Tool & 2006                        & \checkmark             & Virtual               & Centralized                                                                     & Low                            & KVM                                                                            & Medium                \\ 
\hline
\textbf{\textit{MiniNEXT} \cite{Sim:mininext}}            & Simulation Tool & 2014                        & \checkmark            & Virtual               & Distributed                                                                     & Medium                               & KVM                                                                            & Low                   \\ 
\hline
\textbf{\textit{MaxiNet} \cite{Sim:maxinet}}             & Simulation Tool & 2014                        & \checkmark            & Virtual               & Centralized                                                                     & None                               & KVM                                                                            & High                  \\ 
\hline
\textbf{\textit{DOT} \cite{TestBeds:DOT}}                 & TestBed         & 2014                        & \checkmark            & Virtual               & \begin{tabular}[c]{@{}c@{}}Distributed (centralized\\Orchestrator)\end{tabular} & High                               & \begin{tabular}[c]{@{}c@{}}Containers (Docker), \\QEU/ KVM\end{tabular}        & High                  \\ 
\hline
\textbf{\textit{ONE} \cite{TestBeds:ONE}}                 & Both            & 2018                        & \checkmark            & Virtual               & Remote                                                                          & High                              & Containers (Docker)                                                            & High                  \\ 
\hline
\textbf{\textit{RISE} \cite{TestBeds:RISE}}                & TestBed         & 2013                        & x                    & Hybrid                & Remote                                                                          & Medium                            & Local - kernel                                                                 & Low                   \\ 
\hline
\textbf{\textit{Ofelia} \cite{TestBeds:ofelia}}              & TestBed         & 2014                        & x                    & Hybrid                & Remote                                                                          & Medium                            & Xen-based VMs                                                                  & Medium                \\ 
\hline
\textbf{\textit{OTG} \cite{TestBeds:otg}}                 & TestBed         & 2017                        & \checkmark            & Physical              & Centralized                                                                     & Low                               & None                                                                           & Low                   \\ 
\hline
\textbf{\textit{SDNetkit} \cite{TestBeds:SDNetkit}}            & Simulation Tool         & 2017                        & \checkmark            & Virtual                &                Centralized                                                                 & High         & NetKit VMs                                                                     & Medium                \\ 
\hline
\textbf{\textit{NextLAB} \cite{TestBeds:NextLab}}             & TestBed         & 2019                        & \checkmark            & Hybrid                & Centralized                                                                     & High         & KVM                                                                            & Low                   \\ 
\hline
\textbf{\textit{Kentucky U.} \cite{TestBeds:Kentucky}}         & N/A             & 2019                        & x                    & Hybrid                & N/A                                                                             & High         & KVM                                                                            & Medium                \\
\hline
\end{tabular}
}

\label{Table:simTools_Tesbeds}
\end{table}

\textit{Table} \ref{Table:simTools_Tesbeds} shows a classification of the most significant \ac{hsdn}/\ac{sdn} simulation tools and testbeds up to the date. We classify these solutions by:
\begin{enumerate}
    \item \textbf{Family:} Distinguishes between “Simulation Tool” and “Testbed” to have a clear view of what kind of operations is the user going to be capable to perform on them.
    \item \textbf{First Release Year:} Clarifies the maturity and evolution of each platform.
    \item \textbf{Open Source:} Distinguishes between OpenSource platforms and commercial or non-OpenSource platforms.
    \item \textbf{Architecture:} Separates platforms into two categories: 
    \begin{itemize}
        \item \textbf{Virtual:} Platforms without a physical network deployment.
        \item  \textbf{Hybrid:} Platforms with a physical network implementation combined with virtual network elements.
    \end{itemize}
    \item \textbf{Type:} We have used the same classification as in \cite{TestBeds:SDNetkit}:
    \begin{itemize}
        \item \textbf{Centralized:} Platforms running locally.
        \item \textbf{Distributed:} Platforms running on a distributed environment. 
        \item \textbf{Remote Testbeds:} Physical networks that offer the possibility of experimenting \ac{sdn}/\ac{hsdn} using virtual overlay networks.
    \end{itemize}
    \item \textbf{Legacy Protocols Support:} In \ac{hsdn}, it is important to be able to communicate with Legacy network elements. We define four levels:
    \begin{itemize}
        \item \textbf{None:} No support at all of Legacy communication protocols.
        \item \textbf{Low:} Only one Legacy communication protocol is supported i.e. EstiNet only supports \ac{stp}.
        \item \textbf{Medium:} More than one Legacy communication protocol is supported but the integration is not fully compatible.
        \item \textbf{High:} All or almost all the legacy communication protocols are supported and the integration is fully compatible.
    \end{itemize}    
    \item \textbf{Virtualization technology:} Currently this parameter is highly meaningful depending on the kind of experiment and the amount of available resources that a researcher has in order to carry-out its research.
    \item \textbf{Scalability}: Clarifies the degree of scalability of the platform.

\end{enumerate}

\par

As explained above, MiniNet \cite{Sim:mininet} is the most widely known and used tool for \ac{sdn} simulation. However, its latest stable version (\textit{MiniNET v.2.3.0d6}) has no support for hybrid \ac{sdn} scenarios where OpenFlow and Legacy routing protocols coexist. MiniNet Cluster Edition (CE) \cite{Sim:mininet_ce} and Distributed MiniNet \cite{Sim:mininet_dist} platforms enable the execution of multiple MiniNet instances in one single system and to execute MiniNet over a distributed scenario but, as the original SW, both of them lack the functionalities to run other than OpenFlow-controlled devices.\\
Moreover, EstiNet is the only Commercial platform in Table \ref{Table:simTools_Tesbeds} but, even though it is capable of supporting several \ac{sdn} controllers as Floodlight, the only accepted Legacy routing protocol is \ac{stp}. Its leading capability is the usage of the kernel re-entering technique to simulate scenarios and its high scalability (see \cite{Sim:estinet} for further comparison between EstiNet and MiniNet). NS-3, in its latest stable version, enables certain degree of interaction between the simulated components and the real topology-enabled network components (see \cite{Sim:ns-3} for  further comparison between NS-3 and MiniNet). MaxiNet \cite{Sim:maxinet} was created as a MiniNet plugin with the aim of emulating large-scale network with a high density of devices e.g. \ac{dc} networks, but, as a MiniNet complement, it inherits the lack of \ac{hsdn} simulation capabilities from the original platform. At last, MiniNext \cite{Sim:mininext} is the only Simulation platform that supports, to some degree, legacy routing protocols. It was designed as an extension layer for MiniNet in order to allow the creation of more complex networks and the integration of traditional routing engines (e.g. Quagga , BIRD…\cite{Quagga,BIRD}) and NAT and network management components (e.g. DHCP). The main issue of MiniNext is that is no longer supported and it only works with MiniNet v.2.1.0.

Arup Raton et al. \cite{TestBeds:DOT} created one of the first \ac{hsdn} distributed testbeds, focused on improving and upgrading MiniNet scalability. The Distributed OpenFlow testbed (DOT) provides the possibility to create an emulated topology over a cluster of multiple physical machines, establishing an orchestration hierarchy in which one of the machines is the so called \textit{DOT Manager}, in charge of managing the whole emulation process, and the rest of them are the \textit{DOT Nodes}. The big advantage of this testbed, compared with the previous Simulation Tools, is that it allows the usage of two different kind of virtualization techniques (Linux Containers and Kernel Virtual Machines (KVM)) and adds full support for hybrid-networks, allowing the emulation of non-OpenFlow Switches as well as allowing the inclusion of physical switches, which are key-enablers for \ac{hsdn} environments.

The Open Network Emulator (ONE) \cite{TestBeds:ONE} is a Microsoft platform that was open-sourced in 2018 which, originally, was a collaboration project between Microsoft Azure and Microsoft Research teams called CrystalNET \cite{TestBeds:CrystalNEt} (named after a fortuneteller’s crystal ball, as it was going to be able to “reveal” the future of a network) focused on the scope of emulating large-scale cloud networks. Currently, it is an Azure \cite{Azure} product that has been integrated as an \ac{iaas} platform that allows to deploy realistic test environments for building network automation tools and testing the network behavior. ONE is highly scalable as it allows to create the virtual resources using \acp{vm} and Containers technology and is highly flexible as it supports the usage of a huge number of networks devices SW images from a wide range of different vendors. ONE also accepts real hardware devices and, hence, legacy routing protocols. These capabilities make this testbed a trustworthy platform for those researchers that are looking to develop \ac{hsdn} experiments without owning a physical infrastructure.

\begin{figure}[ht]
\includegraphics[scale=0.63]{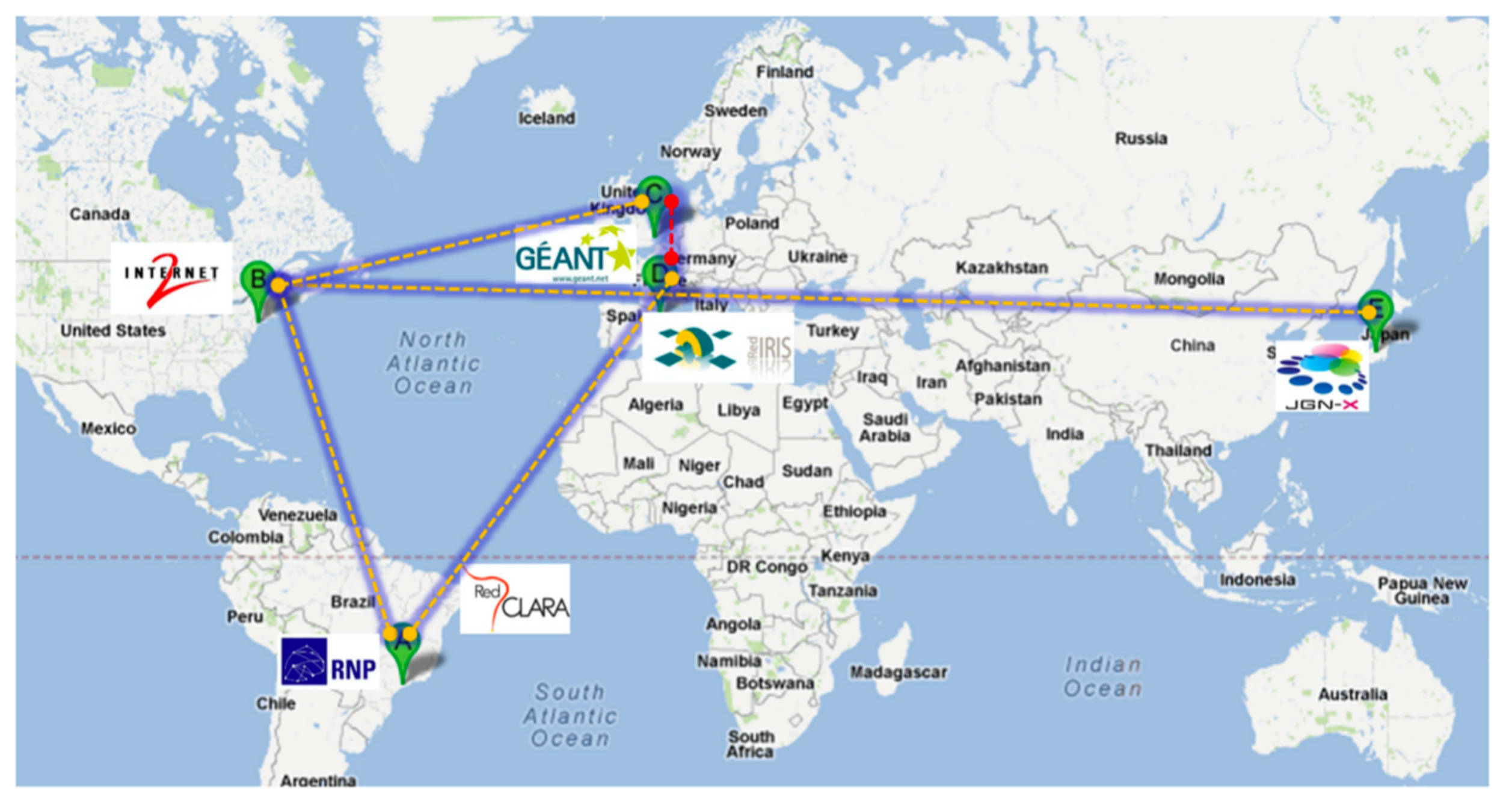}
\centering
\caption{Ofelia worldmap connectivity scheme (source \cite{TestBeds:ofelia})}
\label{fig:ofelia_net}
\end{figure}

RISE, proposed by Kanaumi et al. in \cite{TestBeds:RISE}, is a \ac{wa-ofn} hybrid Japanese testbed built on top of the JGN2plus and Japan GigaBit Network (JGN-X) \cite{JGN-X} networks that aims at allowing multiple users/customers to create WA-OFNs over existing real networks (RISE features FlowVisor \cite{sherwood2009flowvisor} to manage and control the underlying networks). Although its implementation over existing real networks may seem a good approach to obtain reliable network metrics and behavior, it is technologically limited as some of the devices within these networks do not support MAC-in-MAC and \ac{mpls}-based tunneling. RISE users are allowed to create their own wide-area hybrid network topologies in this testbed, namely \textit{Existing Virtual Networks}, and it also enables the isolation of a user  resources by assigning them a logical slice of the network (Network Slice). Some of the conducted demonstrations on the RISE testbed are shown in \cite{Testbeds:RISE_1,TestBeds:RISE_2}.\\
The RISE project achieved a federation of networks with other national-level OpenFlow testbeds such as Ofelia (EU) \cite{TestBeds:ofelia} and other European, USA and South-American projects (See Figure \ref{fig:ofelia_net}). Furthermore, based on the collected knowledge and experiences from this testbed, a new testbed was proposed to launch \ac{ict} experiments in a distributed, large-scale environment called JOSE \cite{TestBeds:JOSE}.

OFELIA \cite{TestBeds:ofelia} is a pan-European collaborative OpenFlow testbed that connects individual testbeds by a common Layer-2 infrastructure. It is comprised of seven different facilities (iMinds, U. of Bristol, ETHZ, i2CAT, TUB, CreateNet and CNIT) each of them consisting on a self-defined infrastructure (OpenFlow switches, OpenFlow controllers, legacy switches and computing resources) and with an experimentation specialty ranging from large-scale emulation and NetFPGA farms to Optical OpenFlow Experiments. It incorporates three significant tools: \textit{(1)} Web-Based GUI to manage user authentication, experiment onboarding requests and network status; \textit{(2)} Virtualization Technology Aggregate Manager \cite{TestBeds:Ofelia_Mgmt} responsible of the system control; and \textit{(3)} the FlowVisor Aggregate Manager which uses FlowVisor as an agent to provision resources on the OPN and uses a northbound API to communicate with the Web-Based GUI. These tools enable the three fundamental pillars of the OFELIA testbed: resource diversity and flexibility for the experimenters, federation extensibility and ease of experiment orchestration, control and management. Notwithstanding, it is not fully capable of supporting legacy devices.

Joshua A. et al. propose the \ac{sdn} On-The-Go (OTG) \ac{hsdn} physical testbed in \cite{TestBeds:otg}. Unlike the previous testbeds, OTG main goal is to create a portable and economical solution that fulfills the requirements of those academic researchers willing to implement \ac{sdn}/\ac{hsdn} experiments that transcend the Simulation Tools capabilities (precise and repeatable metrics, concise real-world behavior, etc.). To that end, the OTG testbed is comprised of a reduced set of components, namely: \textit{four \ac{sdn} Zodiac FX switches, four RaspberryPi 3 hosts and a Kangaroo+ mini-pc as the \ac{sdn} Controller}. Thanks to this implementation, this testbed converges on a low-cost (full set cost less than \$1000), portable and standalone \ac{hsdn} testbed capable of generating several topologies e.g. Four switches and two hosts partial mesh, Four switches and four hosts hybrid fattree with direct WAN access, etc. As demonstrated in \cite{TestBeds:otg} OTG shows a more stable performance when compared with MiniNet if running the same experiments. Additionally, as it is a modular testbed, it can be expanded and various \ac{hw} extensions can be added to the initial set of components. The big shortcoming of the OTG testbed is that it does not explicitly integrate \ac{hsdn} support, therefore, if required, it has to be manually added.

\begin{figure}[h!]
\includegraphics[scale=0.57]{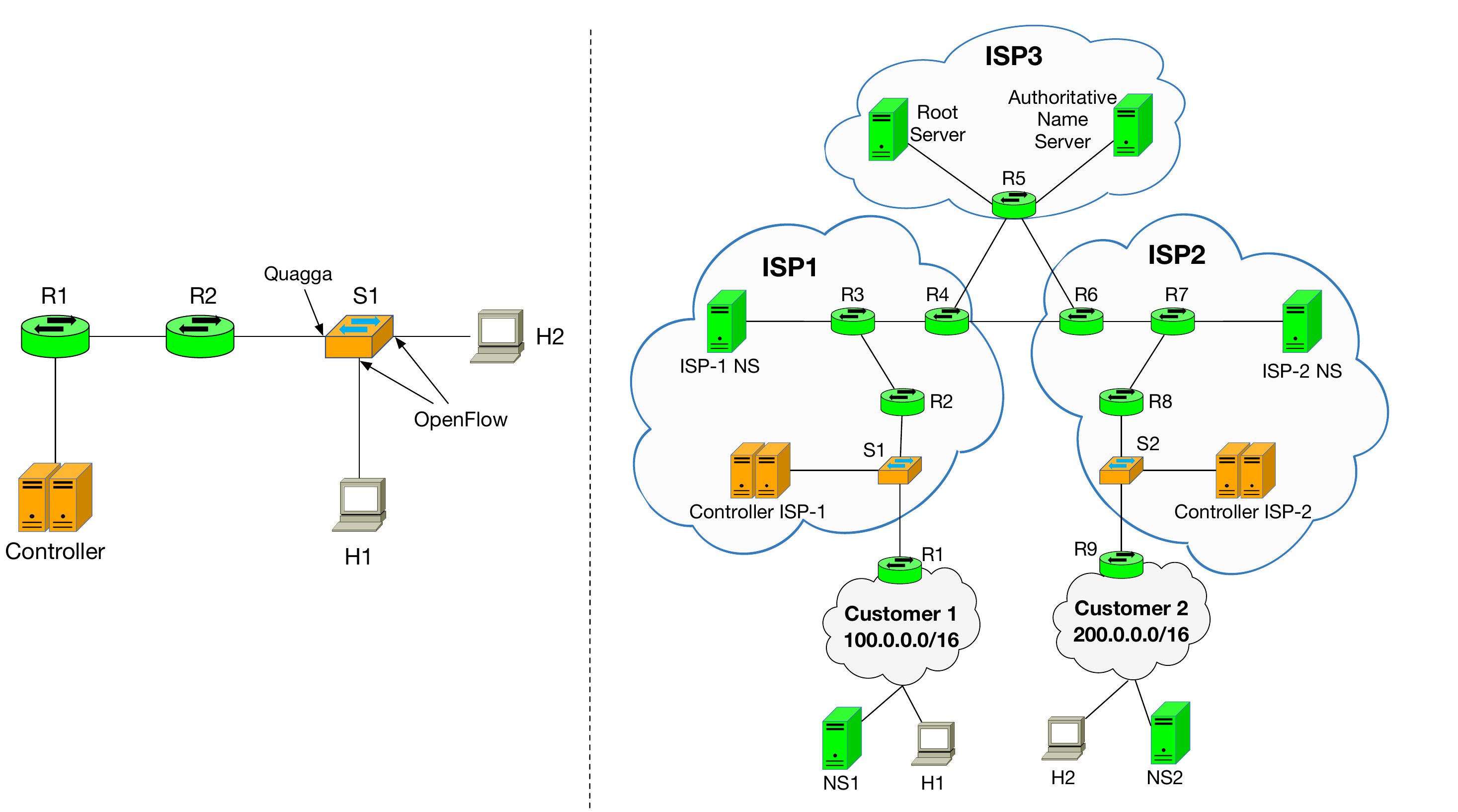}
\centering
\caption{SDNetkit Hybrid Network Topologies examples. Left: Hybrid node topology. Right: ISP complete Hybrid Network Topology. Edited from source \cite{TestBeds:SDNetkit}}.
\label{fig:SDNetkit_Topologies}
\end{figure}

SDNetkit is presented in \cite{TestBeds:SDNetkit} as a solution built over Netkit \cite{NetKit}, a freely OpenSource lightweight network emulator based on User-Mode Linux.
The main difference between this Simulation/Emulation solution and the previous ones is that it extends the Legacy networking and routing protocols of the Netkit Network Emulator in order to be \ac{sdn}-capable (it was not conceived as an independent \ac{sdn} tool). SDNetkit supports \textit{OpenFlow 1.3} and various routing engines such as \textit{Quagga} \cite{Quagga}, enabling the creation of interoperability hybrid scenarios where legacy and SDN-enabled devices and protocols coexist. More specifically, this emulation platform is capable of executing Open vSwitch \cite{openvswitch} instances with the RYU \cite{RYU} framework as the controller in parallel with legacy devices.\\
It is important to remark that SDNetkit is a modular platform that allows to extend its own functionalities by adding new SW to its repositories: “(…) e.g. it is possible to add other \ac{sdn} frameworks for implementing custom controllers, other SDN-enabled switch implementations, or general software for testing \ac{sdn}”.

\par

SDNetkit inherits the virtualization technology used by the Netkit emulator which, basically, is a form of \acp{vm} that require no administrative priviliges and consume a small amount of computing resources on the host system. As it can be observed in Figure \ref{fig:SDNetkit_Topologies}, SDNetkit can create multiple different \ac{hsdn} topologies ranging from: a network topology that uses one or more hybrid nodes (nodes that \textit{simultaneously} run OpenFlow and legacy routing protocols) to a complete hybrid network topology where legacy devices coexist with SDN-enabled devices. Consequently, all these functionalities, extension possibilities and legacy/\ac{sdn} support make this platform one of the best choices if \ac{hsdn} scenarios are to be emulated.

\par

The Nextlab testbed \cite{TestBeds:NextLab}, on the contrary, is focused on creating a platform that allows its users to share their resources in order to build a global \ac{hsdn} community and testing environment. It is conceived with a layered architecture (see Figure 1 in \cite{TestBeds:SOTE}) composed of four main components: \textit{(1)} Web-GUI, based on OvS-Mesh tool (similar to the GUI of MiniNet); \textit{(2)} Topology Creator, takes the input topology generated by the user in the Web GUI and translates it into a pre-configured PicOS script that can be executed on Pica8 switches; \textit{(3)} Open Networking Operating System (ONOS), as the \ac{sdn} Controller and \textit{(4)} Pica8 Switch, hosts the virtually created topologies. \\
NextLab brings, with this architecture, a set of advantages that enable the configuration of hybrid topologies (it uses Open vSwitch instances for the virtual elements)and \ac{qos} policies. Besides, due to the usage of ONOS as the \ac{sdn} controller, users can share their computing resources and add them to the NextLab global network. As a consequence, NextLab users can develop \ac{hsdn} experiments with minimum cost and contribute to a community where similar experiments are being carried out. The two initial nodes of this tesbed were located at Vitry-sur-Seine (France) and at Hanoi (Vietnam). Chu et al. \cite{TestBeds:NextLab} paper, execute different tests to compare the performance of NextLab and MiniNet. NextLab consumes less CPU in all scenarios. \\

There are also other \ac{hsdn} testbeds built with more specific objectives rather than to be used to deploy and test \ac{hsdn} environments. \ac{sote} by Guo et al. \cite{TestBeds:SOTE} proposes a heuristic algorithm formulated as a \ac{minpm} which minimizes the \ac{mlu}, if compared to traditional routing algorithms, comprising \ac{ospf} \cite{RFC:2328} link weight configuration with the traffic splitting ratio in \ac{sdn} nodes. To demonstrate the capabilities of their algorithm they implement an \ac{hsdn} testbed with a single \ac{sdn} controller and several \ac{hsdn} switches (see Section 6.5 in \cite{TestBeds:SOTE}). Furthermore, the authors of \cite{TestBeds:NA_HAS} propose a Network-assisted HTTP Adaptive Streaming algorithm to increase the \ac{qoe} in \ac{hsdn} networks. To do so, they simulate a \ac{hsdn} network using MiniNet \cite{Sim:mininet} , the FloodLight Controller \cite{Ctrl:floodlight} and an Open vSwitch \cite{openvswitch} operating in standalone mode to simulate a legacy switch.\\
Shi et al. \cite{TestBeds:Kentucky}, as detailed in section \ref{Arch:Island_Based}, have created a \ac{hsdn} testbed as a real application case over a campus-like topology in order to demonstrate how to integrate the \ac{sdn} paradigm with a real Legacy Network and all its challenges. In \cite{TestBeds:manet} Poularakis et al. describe the development of the testbed using a commercial \ac{sdn} dataplane, ONOS \cite{Ctrl:ONOS}, and Open vSwitch combined with in order to study the control of Mobile Ad Hoc Networks (MANET) and demonstrate its feasibility. Less remarkable but worth to mention, other \ac{hsdn} testbeds and tools are HybNET, Telekinesis and ClosedFlow respectively \cite{TestBeds:HybNET, TestBeds:Telekinesis,TestBeds:Closedflow}. To end, if interested in a deeper view over the current state of art of cross-technological \ac{sdn} and network emulator testbeds and tools, see Piccialli et al. survey \cite{Surveys:Network_Emu}.

\section{Industry and Standardization Perspectives} \label{sec:HSDN_Standardization}

White papers by many companies and documents of standardization groups have supported deployment
of \ac{hsdn}. This section briefly describes current state of deployment, white papers for \ac{hsdn}
deployment and documents from standardization groups \ac{onf} and \ac{ietf}.

According to IDC, the North American \ac{sdwan} market was \$145 million in 2017 and will grow to
\$1.8 billion in 2022~\cite{idc}. According to 2019 Cisco Global Networking Trends Survey~\cite{cisco1}, over 58\% of organizations globally have already deployed \ac{sdwan} in some form, and over 94\% of respondents believe they will deploy some form of basic or more advanced \ac{sdwan} implementation within the next two years.

Major networking equipment manufacturers, network consulting firms and network operators have supported deployment of \ac{hsdn}. In 2012 Huawei launched the industry's first \ac{hsdn} controller and \ac{sdn}-capable router~\cite{zheng2013huawei}. Huawei was also the first to pass the \ac{of} 1.2 interoperability tests managed by the Open Networking Foundation.
Entuity reports that many organizations do not adopt \ac{sdn} due to costs including retraining staff, \ac{hw}
costs, \ac{sw} licenses and hidden costs of business continuity during initial deployment and advocates
incremental deployment of \ac{hsdn}~\cite{roper2018software}.
Allied Telesis proposes a migration path to \ac{sdn} in an enterprise network~\cite{alliedtelesis}. It recommends
migration in three phases. The first phase achieves unified management of the network. The second phase uses
\acl{of} hybrid where some nodes support \acl{of} while others don't. The third phase is optimum \ac{sdn} deployment.
It indicates that transition from a hybrid network to \ac{sdn} may be gradual or never completed.
NEC describes various \ac{hsdn} introduction models that satisfy a diverse range of customer needs~\cite{nec}.
It indicates that by introducing \ac{sdn} to only the advantageous area in an existing network,
the effects and benefits of \ac{sdn} can be leveraged without affecting the existing network.

In the \ac{sdn} hybrid network architecture designed by HP~\cite{hp}, the hybrid \ac{sdn} controller retains control over all packet
forwarding on the data plane and chooses to delegate forwarding decisions to controlled switches in order to  reduce complexity of forwarding decisions controller makes and to reduce the amount of traffic on the control plane between the switches and controller. \\
The Verizon \ac{sdn}/\ac{nfv} reference architecture~\cite{verizon} describes a range approaches to \ac{sdn} deployment
and supports \ac{hsdn} as a practical solution. In this \ac{hsdn} architecture, \ac{sdn} controllers cooperate with
existing forwarding boxes and vendor-specific domain controllers.
AT\&T advocates \ac{hsdn} networks for cloud networks and big data applications. They indicate that the transition to \ac{sdn} will be a multi-phased process and having a clear technology road-map will help enterprises transition in small increments with minimal disruption to their business goals.

As businesses continue to deploy new applications and move to virtualized environments, they will
need to connect them to legacy applications and data stores. Businesses should base their decisions on network architecture that provide the features and capabilities they need now as well as in the future. They will demand a network architecture that is easy to integrate with \ac{sdn} controllers. They will also prefer open standards and interoperability between their environments and network devices.
Several ways of integrating \ac{sdn} into the \acl{dc} using Juniper \ac{sdn} gateways are described in a Juniper white
paper~\cite{juniper} on \ac{sdn} integration. 

IBM developed a reference architecture designed to help organizations compare and contrast various \ac{sdn} products~\cite{ibm}.
IBM has defined many use cases (such as micro-segmentation) and more than 100 requirements to help adapt \ac{sdn} to different environments. Requirements include the need to align network security policies to the new \ac{sdn}-driven capabilities, perhaps leveraging metadata from virtual machines in addition to
the standard IP address segmentation rules.

Recent improvements in the Intel micro-architecture enable network service providers to gain more flexibility and control over customer offerings through the use of \ac{sdn}~\cite{intel}. By virtualizing network functions on Intel architecture, network service providers can employ techniques such as deep packet inspection, geographic \ac{lb}, and power management to optimize bandwidth.
Practical implementation issues of \ac{sdn} and \ac{nfv} in \acs{wan} are described in a white paper from Wind River~\cite{windriver}. An Intel/HP/Wind River reference design demonstrates that performance of a standard open source Open vSwitch can improve by an order of magnitude by fusing Intel's Data Plane Development Kit with Wind River's Open Virtualization Profile and running the software on industry-standard hardware supplied by HP.

A Gigamon white paper~\cite{gigamon} indicates that, while some organizations may have the resources for a new
network deployment of \ac{sdn}, the majority will likely deploy hybrid networks based on traditional infrastructure side by side with \ac{sdn} capable routers and switches and dynamically provisioned workloads. For most organizations, their networking infrastructure vendors will have offerings that will allow a gradual migration to \ac{sdn}.

Cisco white paper~\cite{cisco2} describes \ac{sdn} and network programmability use cases for the defense and intelligence communities. Use cases include latency-based routing, dynamic \ac{qos}, diversion networking, central policy management for access control lists and call admission control in multiple security domain networks.

An SEL white paper~\cite{sel} examines the benefits of using \ac{sdn} 
to easily interconnect and manage traffic on Ethernet networks that communicate using IEC 61850 technology. A case study from the Itaipu Dam in South America, one of the world's largest hydroelectric facilities, is used to illustrate these benefits.

A Lumina white paper~\cite{lumina} describes an extensible software platform that enables the automation of legacy network elements using model driven frameworks, shielding the complexity of underlying south bound interfaces and enabling northbound applications. Lumina's platform uses a micro-services architecture to extend the capabilities of the OpenDayLight based \ac{sdn} controller to enable better integration with business layers.

There is limited work on \ac{hsdn} standardization. \ac{onf} document~\cite{onf1} describes various scenarios and issues to be
addressed to ensure successful migration from \ac{hsdn} to \ac{sdn}. The discussion includes traditionally controlled and \ac{sdn} controlled ports, traditional underlay with \ac{sdn} overlay, explicit invocation of traditional forwarding from \ac{sdn} and
control plane connectivity for \ac{hsdn}. RFC 7149~\cite{rfc7149} indicates that \ac{sdn} will be deployed incrementally 
and raises interesting questions such as how \ac{sdn} affects the lifetime of legacy systems. Legacy systems may be obsolete rapidly due to their \ac{hw} and \ac{sw} limitations compared with \ac{sdn}.
\section{Open Research Challenges in \texorpdfstring{\ac{hsdn}}{Hybrid SDN (hSDN)}} \label{sec:open_issues}

Based on our in-depth survey, we identified the following areas where there is limited work or satisfactory solutions are not available.

\begin{itemize}[wide, labelwidth=\leftmargin, labelindent=0pt]
    \item \textbf{P4-capable devices:} The P4 language enables programmers to define a variety of protocols and data plane behaviours to be implemented in P4-enabled nodes. Of course, in \ac{hsdn} deployments, these custom features and protocols must be carefully designed and implemented to avoid disrupting the switching and routing operations of legacy \ac{l2} and \ac{l3} devices.
    \item \textbf{\ac{qos} in \ac{hsdn}:} This is closely related to node discovery and more challenging than node discovery. Any \ac{qos} approach needs to discover all the nodes and allocate resources on the nodes. Having limited control over the nodes render \ac{qos} provisioning challenging. Existing work covers the topic superficially, with some initial contributions related to discovery protocols \cite{alvarez2020hddp}, and to resource allocation in specific scenarios such as hybrid \ac{mpls}/\ac{of} networks \cite{tajiki2019sdn}.
    \item \textbf{\acf{cpp} \acp{poc}:} As stated in section \ref{subsec:Controller_placement} this problem can lead to critical issues in \ac{sdn} and \ac{hsdn} architectures and, therefore, it needs to be properly addressed and studied. To the best of our knowledge, there are only two works that analyze this problem proposing solutions for \ac{hsdn} environments \cite{yuan2019latency,guo2019joint}. We identify two main open challenges for the \ac{cpp} in hybrid legacy/\ac{sdn} networks:
    \begin{itemize}
        \item \textit{Deployment model}. The \ac{cpp} problem should be tackled by taking into account the deployment model (cf. Section \ref{sec:deployment_strategies}) and execution (either static or progressive deployment of \ac{sdn} devices within a legacy network). In particular, a progressive deployment might require adaptive solutions of the problem, as done in dynamic contexts such as vehicular networks \cite{toufga2020towards}, or the so-called Software-Defined Drone networks \cite{9013799} and Software-Defined Satellite Networking \cite{wu2018dynamic}. 
        \item \textit{Legacy middleboxes}. The impact of legacy devices on the control channel should be also carefully addresses. Thus, in the formulation of the \ac{cpp}, the position and the type (e.g., router, firewall, \ac{ids}, etc.) of the legacy devices the control traffic must traverse should be also taken into account. Indeed, these factors might affect the end-to-end latency on the control channel, hence the optimal placement of the \ac{sdn} controller. 
    \end{itemize}
    
    \item \textbf{Security:} This topic is of vital importance in any kind of network but specially in \ac{hsdn} and \ac{sdn} networks as the attack surface might be increased. Despite all the surveyed works in Section \ref{sec:HSDN_Sec_and_Privacy}, there still exists a lack of research on the following challenges:
    \begin{itemize}
        \item \textit{Security by design}. \ac{hsdn} deployments integrate legacy and \ac{sdn} technologies, each of them with its specific security-related issues and benefits. A service-based deployment model (cf. Section \ref{sec:deployment_strategies}) focused on security services can be adopted for a secure network design. However, it might contrast the fulfillment of other requirements (e.g., \ac{qos}, budget) that might impose different deployment approaches.    
        \item \textit{Integrated security solutions}. As discussed in Section \ref{sec:HSDN_Vulnerabilities}, there are a number of vulnerabilities that affect \ac{hsdn} networks. To reduce the attack surface of \ac{hsdn} deployments, integrated solutions that combine the benefits of hardware appliances and \ac{sdn} programmability are required. 
        However, only a limited number of solutions in this regard have been proposed so far (cf. Section \ref{sec:security_frameworks}). Approaches based on the \ac{nfv} paradigm look promising ~\cite{shameli,pess2}, however the proposed placement models are limited to \ac{nfv}-enabled devices, without considering security hardware appliances when provisioning security services. 
    \end{itemize}
    \item \textbf{Emergent Technologies:} A wide range of innovative technologies have been addressed in this survey. Due to their emergent nature there are few works developing \ac{hsdn} solutions for each of these technologies. Thereupon, several key challenges for \ac{hsdn} applied on these technologies remain unresolved:
    \begin{itemize}
        \item \textit{\ac{hsdn} 5G architectures.} As demonstrated in Section \ref{sec:HSDN_Emergent_Networks} there are a couple of works approaching this challenge but there is still a lot to be done in order to achieve a full \ac{hsdn} 5G architecture or even to know if this approach brings more advantages than potential disadvantages. An integral \ac{hsdn} 5G Architecture \ac{poc} or testbed would be a promising solution as it would help to clarify the above-addresed issues. Therefore, an in-deep study of the advantages that \ac{hsdn} brings to 5G Architectures would help to establish the base to develop these further 5G-\ac{hsdn} architecture testbeds.
        \item \textit{Cloud-native \ac{hsdn} architectures with high energy efficiency.} Cloud-native environments are becoming a requirement for many technologies and services i.e. 5G architecture is designed with a cloud-native approach. As shown in section \ref{subsec:Emergent_CC_DCN} and in Table \ref{Table:CC_Surveys}, there has been a lot of research on this topic for pure \ac{sdn} but it is still an innovative research area that needs to analyze the energy efficiency management in \acp{dcn} with \ac{hsdn}. \ac{hsdn} cloud-native proposals with a focus in energy efficiency \cite{paliwal2019effective} demonstrate the potential within this combination of technologies, however these approaches are conditioned to a limited range of \acp{dcn} topology types.
        \item \textit{\ac{hsdn} and \ac{mec}-based architectures.} \ac{mec}/edge is a disrupting technology with a high interest for Communication Providers and Operators as it allows to reduce application latencies and the overall connection speeds. This technology is highly dependant of virtualization technologies and, consequently, \ac{sdn} and \ac{hsdn} are an important approach. At the time of writing, there were no works combining this two technologies. Besides, some \acp{sdo} such as \ac{etsi} are looking for \acp{poc} related to \ac{mec}/edge since 2016 \cite{ETSI:Mec_PoCs}.
        \item \textit{\ac{hsdn} budget solutions for \ac{iox} environments.} Day by day, the use cases of \ac{iox} scenarios is increasing but the number of works relating \ac{hsdn} with these technologies is just starting to arise. We think that there is a big research/opportunity gap within this topic as the benefits obtained from combining these technologies, as shown in Section \ref{subsec:IoT}, are numerous.
        \item \textit{\ac{sdbranch} \ac{hsdn} opensource solutions.} \ac{sdbranch} is a young concept that begins to have an increasing interest for the industry and that can be highly exploited by the research community as a promising technology for flexible networks. Therefore, an opensource solution that is combined with \ac{hsdn} would be an innovative approach as there are no analogous platforms up to the date.
    \end{itemize}
    \item \textbf{Simulation Tools and Testbeds:} Although there is a clear emergence and development of pure \ac{sdn} simulation tools and testbeds that support multiple legacy protocols, there is still a lack of open dedicated simulation tools that allow to experiment with \ac{hsdn}. DOT and ONE \cite{TestBeds:DOT,TestBeds:ONE} may be considered the first step towards resolving this particular challenge.
    \item \textbf{Resilience and Traffic Engineering:} There is limited work on resilience and fault-tolerance of legacy routers in \ac{hsdn} mostly due to lack of adequate protocols for discovering legacy routers. Current traffic engineering work focuses on providing \ac{sdn} benefits using a small set of \ac{sdn} capable routers. However, these approaches fail to provide the foundations for services such as \ac{qos}. Proposed traffic engineering approaches need to be improved to provide better resilience, fault-tolerance and improved estimation error to enable \ac{qos} based services.
    
    \item \textbf{Standardization of \ac{hsdn}:} Standardization efforts for \ac{hsdn} are limited considering the size of \ac{hsdn} deployments. Standardization on various issues such as traffic measurement interfaces, \ac{qos} primitives supported, ways for handling legacy devices, algorithms for node discovery, primitives for provisioning security services will enable faster deployment of \ac{hsdn} by promoting interoperable hardware and software by various vendors.
    
    As already discussed in Section \ref{sec:deployment_strategies}, there is not yet an accepted standard for \ac{nbi} between the controller and \ac{sdn} applications and network services that utilize the controller~\cite{LATIF2020102563}. This forces application developers to either choose a specific controller platform, which might not meet their needs completely, or to modify their applications to work on different controllers. In this regard, the progressive integration of \ac{sdn} technologies in legacy networks, hence the increasing need for common interfaces, is pushing for an advancement in the direction of standard \acp{nbi}~\cite{7958527,moeyersons2020pluggable}. Nevertheless, despite a few scattered initiatives, vendor specific and ad-hoc \acp{nbi} are still a major barrier to the development of portable \ac{sdn} applications.
\end{itemize}

\section{Conclusions} \label{sec:Conclusions}

In contrast to the large and well-known network providers who have already adopted \ac{sdn}, there are many organizations and enterprises who simply cannot afford a transition from legacy networking to the \ac{sdn}. Keeping the traditional network infrastructure and updating it incrementally to an \ac{sdn} infrastructure seems an affordable and technically feasible solution. The so-called \acf{hsdn} is now considered as a widely accepted transition phase and a compromise solution which have the advantages of both \ac{sdn} and legacy networking solutions. 

This paper focuses on presenting a comprehensive state-of-the-art survey of hybrid \ac{sdn}. It first studies \ac{hsdn} models in both control and data planes along with control plane optimization in terms of placement and scalability problems. The related security and privacy issues and existing vulnerabilities and threats are another investigated domain. The investigation is then supported by exploring possible threats detection and mitigation approaches and reviewing recent security modules and cyber defence frameworks. \\
This paper also discusses \ac{hsdn} network management. More specifically, it probes network update in both legacy and \ac{sdn} settings. Network automation with respect to network telemetry for management plane and network auto configuration are other topics which are reviewed elaborately. Besides, reliability, resiliency, fault-tolerance, and load balancing are also surveyed. The paper then  goes into effect traffic engineering topic and inquire into traffic measurement, traffic management and quality of service routing. The implementation and deployment of \ac{hsdn} is another topic which is studied in this paper.\\
We observed there are few topics which did not receive enough attention in the related body of literature on \ac{hsdn}. Correspondingly, recent \ac{hsdn} use cases in  5G mobile networks, cloud and data center networking, \ac{iot} connectivity, Blockchain, \ac{sdwan}, and finally \ac{sdbranch} are presented in detail. Existing simulation tools and public testbeds are another overlooked topics which reviewed and compared in this paper. Current status of standards and business issues in \ac{hsdn} along with future research directions are the last topics inspected in this paper. 

In conclusion, we believe that \ac{hsdn} has come a long way. However, there are few domains which need further investigation. While different \ac{hsdn} models and architectures, security, and traffic engineering solutions have been studied in depth, the existing literature on managing hybrid networks, privacy issues, and new \ac{hsdn} standards and business models are sparse and need further investigations.

\bibliographystyle{IEEEtran} 
\bibliography{Bib/Hybrid-SDN-Bib}

\begin{acronym}[SD-Branch]\itemsep0pt
    \acro{3gpp}[3GPP]{3rd Generation Partnership Project}
    \acro{5gppp}[5GPPP]{5G Infrastructure Public Private Partnership}
    \acro{acll}[ACL]{Access Control List} 
    \acro{ai}[AI]{Artificial Intelligence} 
	\acro{api}[API]{Application Programming Interface}
	\acro{arp}[ARP]{Address Resolution Protocol}
    \acro{as}[AS]{Autonomous System}
    \acro{bddp}[BDDP]{Broadcast Domain Discovery Protocol}
    \acro{bgp}[BGP]{Border Gateway Protocol}
    \acro{cam}[CAM]{Content-Addressable Memory}
    \acro{cpp}[CPP]{Controller Placement Problem}
    \acrodef{crc}[CRC]{Cyclic Redundancy Check}
    \acro{dc}[DC]{Data Center}
    \acro{dcn}[DCN]{Data Center Network}
    \acro{ddos}[DDoS]{Distributed Denial of Service}
    \acro{dhcp}[DHCP]{Dynamic Host Configuration Protocol}
    \acro{dos}[DoS]{Denial of Service}
    \acro{dpdk}[DPDK]{Data Plane Development Kit}
    \acro{dpi}[DPI]{Deep Packet Inspection}
    \acro{ebgp}[eBGP]{External Border Gateway Protocol}
    \acro{embb}[eMBB]{enhanced Mobile Broadband}
    \acro{epc}[EPC]{Enhanced Packet Core}
    \acro{etsi}[ETSI]{European Telecommunications Standards Institute}
    \acro{fib}[FIB]{Forwarding Information Base}
    \acro{fpga}[FPGA]{Field Programmable Gate Array}
    \acro{gui}[GUI]{Graphical User Interface}
    \acro{hal}[HAL]{Hardware Abstraction Layer}
    \acro{hsdn}[hSDN]{Hybrid SDN}
    \acro{hw}[HW]{Hardware}
    \acro{iaas}[IaaS]{Infrastructure as a Service}
    \acro{ibgp}[iBGP]{Internal Border Gateway Protocol}
    \acro{icn}[ICN]{Information Centric Network}
    \acro{ict}[ICT]{Information and Communication Technology}
    \acro{ietf}[IETF]{Internet Engineering Task Force}
    \acro{igp}[IGP]{Inter Gateway Protocol}
    \acro{ilp}[ILP]{Integer Linear Programming}
    \acro{iot}[IoT]{Internet of Things}
    \acro{iov}[IoV]{Internet of Vehicles}
    \acro{iox}[IoX]{Internet of Everything}
    \acro{ids}[IDS]{Intrusion Detection System}
    \acro{ips}[IPS]{Intrusion Prevention System}
    \acro{isis}[IS-IS]{Intermediate System to Intermediate System}
    \acro{isp}[ISP]{Internet Service Provider}
    \acro{itu}[ITU]{International Telecommunication Union}
    \acro{itur}[ITU-R]{International Telecommunication Union Recommendation}
    \acro{kpi}[KPI]{Key Performance Indicator}
    \acro{l2}[L2]{Layer 2}
    \acro{l3}[L3]{Layer 3}
    \acro{lan}[LAN]{Local Area Network}
    \acro{lb}[LB]{Load Balancing}
    \acro{lldp}[LLDP]{Link Layer Discovery Protocol}
    \acro{lsa}[LSA]{Link State Advertisements} 
    \acro{lte}[LTE]{Long-Term Evolution}
    \acro{mano}[MANO]{Management and Orchestration}
    \acro{mec}[MEC]{Multi-Access Edge Computing}
    \acro{mimo}[MIMO]{Multiple-Input and Multiple-Output}
    \acro{minpm}[MINPM]{Mixed Integer Non-Linear Programming Model}
    \acro{mitm}[MITM]{Man in the Middle}
    \acro{ml}[ML]{Machine Learning}
    \acro{mlu}[MLU]{Maximum Link Utilization}
    \acro{mmtc}[mMTC]{massive Machine-Type Communications}
    \acro{mpls}[MPLS]{Multi-protocol Label Switching}
    \acro{nbi}[NBI]{North-Bound Interface}
    \acro{netconf}[NETCONF]{Network Configuration Protocol}
    \acro{nfv}[NFV]{Network Function Virtualization}
    \acro{nfvi}[NFVI]{Network Function Virtualization Infrastructure}
    \acro{nfvo}[NFVO]{Network Function Virtualization Orchestrator}
    \acro{ngmn}[NGMN]{Next Generation Mobile Networks}
    \acro{ns}[NS]{Network Slicing}
    \acro{of}[OF]{OpenFlow} 
    \acro{ofn}[OFN]{OpenFlow Network}
    \acro{ofdp}[OFDP]{OpenFlow Discovery Protocol}
    \acro{onf}[ONF]{Open Networking Foundation}
    \acro{ospf}[OSPF]{Open Shortest Path First}
    \acro{ovs}[OvS]{Open Virtual Switch}
    \acro{paas}[PaaS]{Platform as a Service}
    \acro{pbr}[PBR]{Policy Based Routing}
    \acro{p4}[P4]{Programming Protocol-Independent Packet Processors}
    \acro{poc}[PoC]{Proof of Concept}
    \acro{qoe}[QoE]{Quality of Experience}
    \acro{qos}[QoS]{Quality of Service}
    \acro{ran}[RAN]{Radio Access Network}
    \acro{rest}[REST]{Representational State Transfer}
    \acro{rib}[RIB]{Routing Information Base}
    \acro{rpc}[RPC]{Remote Procedure Call}
    \acro{rt}[RT]{Real-Time}
    \acro{sa}[SA]{Service Assurance}
    \acro{saas}[SaaS]{Software as a Service}
    \acro{sbi}[SBI]{South-Bound Interface}
    \acro{sdn}[SDN]{Software-Defined Networking}
	\acro{sdbranch}[SD-Branch]{Software-Defined Branch}
	\acro{sdo}[SDO]{Standard Definition Organization}
	\acro{sdr}[SDR]{Software Defined Radio}
	\acro{sdwan}[SD-WAN]{Software-Defined WAN}
	\acro{sfc}[SFC]{Service Function Chaining}
	\acro{sitn}[SITN]{Ships in the Night}
	\acro{sla}[SLA]{Service-Level Agreement}
	\acro{snmp}[SNMP]{Simple Network Management Protocol}
    \acro{sote}[SOTE]{SDN/OSPF Traffic Engineering}
    \acro{sr}[SR]{Segment Routing}
    \acro{stp}[STP]{Spanning Tree Protocol}
    \acro{sw}[SW]{Software}
    \acro{tcam}[TCAM]{Ternary Content-Addressable Memory}
    \acro{te}[TE]{Traffic Engineering}
    \acro{tls}[TLS]{Transport Layer Security}
    \acro{tm}[TM]{Traffic Matrix}
    \acro{tr}[TR]{Technical Recommendation}
    \acro{ttl}[TTL]{Time To Live}
    \acro{urllc}[uRLLC]{Ultra-reliable and Low Latency Communications}
    \acro{v2v}[V2V]{Vehicle to Vehicle}
    \acro{vim}[VIM]{Virtual Infrastructure Manager}
    \acro{vm}[VM]{Virtual Machine}
    \acro{vnf}[VNF]{Virtual Network Function}
    \acro{vnfm}[VNFM]{Virtual Network Function Manager}
    \acro{vpp}[VPP]{Vectorized Packet Processing}
	\acro{vsnf}[VSNF]{Virtual Security Network Function}
    \acro{vlan}[VLAN]{Virtual LAN}
    \acro{vpn}[VPN]{Virtual Private Network}
    \acro{wan}[WAN]{Wide Area Network}
    \acro{wa-ofn}[WA-OFN]{Wide Area OpenFlow Network}
    \acro{wlan}[WLAN]{Wireless LAN} 
    \acro{xml}[XML]{Extensible Markup Language}

\end{acronym}

%

\end{document}